\documentclass[twocolumn,a4paper,superscriptaddress,floatfix,longbibliography]{revtex4-1}

\usepackage[english]{babel}
\usepackage[utf8]{inputenc}
\usepackage{amsmath}
\usepackage{graphicx}
\usepackage{amssymb,epsfig,color,textcase}
\usepackage{hyperref}
\usepackage{xcolor} 

\usepackage{chemformula} 

\AtBeginDocument{%
    \newwrite\bibnotes
    \def\bibnotesext{Notes.bib}
    \immediate\openout\bibnotes=\jobname\bibnotesext
    \immediate\write\bibnotes{@CONTROL{REVTEX41Control}}
    \immediate\write\bibnotes{@CONTROL{%
    apsrev41Control,author="08",editor="1",pages="1",title="0",year="1"}}
     \if@filesw
     \immediate\write\@auxout{\string\citation{apsrev41Control}}%
    \fi
}%

\def\t2g{$t_{2g}$}

\def\tildej{\tilde\jmath}

\providecommand{\U}[1]{\protect\rule{.1in}{.1in}}

\makeatletter
\renewcommand{\p@subsection}{}
\renewcommand{\p@subsubsection}{}
\makeatother

\newcommand{\nocontentsline}[3]{}
\newcommand{\tocless}[2]{\bgroup\let\addcontentsline=\nocontentsline#1{#2}\egroup}

\begin{document}

\title{Orbital effects in solids: basics, recent progress and opportunities}

\author{Daniel I. Khomskii}
\affiliation{II. Physikalisches Institut, Universit\"at zu K\"oln,
Z\"ulpicher Stra\ss e 77, D-50937 K\"oln, Germany}

\author{Sergey V.~Streltsov$^*$}
\affiliation{Institute of Metal Physics, S. Kovalevskoy St.\ 18, 620990 Ekaterinburg, Russia}
\affiliation{Department of theoretical physics and applied mathematics, Ural Federal University, Mira St.\ 19, 620002 Ekaterinburg, Russia}
\email{streltsov@imp.uran.ru}

\date{\today}

\begin{abstract}

The properties of transition metal compounds are largely determined by nontrivial  interplay of different degrees of freedom: charge, spin, lattice, but also orbital ones. Especially rich and interesting effects occur in systems with orbital degeneracy. E.g. they result in the famous Jahn--Teller effect leading to a plethora of consequences, in static and in dynamic properties, including nontrivial quantum effects. In the present review we discuss the main phenomena in the physics of such systems, paying central attention to the novel manifestations of those. After shortly summarising the basic phenomena and their description, we concentrate on several specific directions in this field. One of them is the reduction of effective dimensionality in many systems with orbital degrees of freedom due to directional character of orbitals, with concomitant appearance of some instabilities leading in particular to the formation of dimers, trimers and similar clusters in a material. The properties of such cluster systems, largely determined by their orbital structure, are discussed in detail, and many specific examples of those in different materials are presented. Another big field which acquired special significance relatively recently is the role of relativistic spin--orbit interaction. The mutual influence of this interaction and the more traditional Jahn--Teller physics is treated in details in the second part of the review. In discussing all these questions special attention is paid to novel quantum effects in those.
\end{abstract}

\maketitle
\tableofcontents

\tocless\section{Abbreviations}

\begin{center}
\begin{table}[h!]
   {\renewcommand{\arraystretch}{1.2}
   \begin{tabular}{l l}
   AFM & antiferromagnetic \\
   DFT & Density functional theory \\
   DMFT & Dynamical mean field theory \\
   GKA   & Goodenough--Kanamori--Anderson \\
   FCC   & Face centred cubic (lattice) \\
   FM & ferromagnetic \\
   JT  & Jahn-Teller \\
   L   & Ligand \\
   ML$_6$   & Metal-ligand octahedra \\
   PCO  & Plaquette charge order \\
   TM or M   & Transition metal \\
   \end{tabular}
   }
\end{table}
\end{center}

\section{Introduction~\label{Sec:intro}}
When treating transition metal (TM) compounds, one has to discuss first which degrees of freedom determine their properties. First, there are charge degrees of freedom, of electrons and ions. Electrons can have, generally speaking, two different states --- they can be itinerant (in chemistry this corresponds to molecular orbital picture) and localized.  Then, there exist electron spins --- especially important for localized electrons. All strong magnets are of this type. Of course all the phenomena in solids occur on a background of the lattice, i.e.\ one has to take into account the interaction with the ions. But, besides these, there exist also orbital degrees of freedom.  Transition metal ions have $d$ electrons with orbital moment $l=2$, i.e.\ in isolated TM atoms or ions these $d$ states are 5-fold degenerate, $l=2$, $l_z=(+2, +1, 0, -1, -2)$. Thus the orbital degrees of freedom and the effects connected with those are very important ingredients of the physics and chemistry of TM compounds.

In this review we will first, in Secs.~\ref{sec:el-in-solids}--\ref{Sec:OO}, present the basic notions used in describing orbital effects in solids. In particular, special attention will be paid to quantum effects in orbital physics, both at the level of a single site as well as in concentrated solids. After that we will concentrate on a novel development in orbital physics, focusing primarily on two main directions: In closely related Secs.~\ref{Sec:D-reduction} and \ref{Sec:CMI} we discuss the special properties connected with the directional character of orbitals, leading to the Peierls phenomenon with the formation of well-defined clusters (``molecules in solids''), and consider orbital-selective effects in cluster materials in general. Then in Sec.~\ref{SOC} we present some results of the very hot nowadays topic --- the role of spin--orbit coupling in systems with orbital freedom, mostly in $4d$ and $5d$ compounds, but not only those. Finally we would like to mention that there exist many books and reviews covering some of these topics --- e.g.\ \cite{Goodenough,Gehring1975,KK-UFN,Bersuker1989,Kaplan1995,Tokura2000,Bersuker2006,khomskii2014transition,Streltsov-UFN,Oles2017a}; one can find more details, especially as to the ``classical'' topics, in these publications. For the basic notions we mostly follow the presentation in \cite{KK-UFN,Streltsov-UFN}.

\section{~Electrons in solids\label{sec:el-in-solids}}
\begin{figure}[b]
   \centering
  \includegraphics[width=0.45\textwidth]{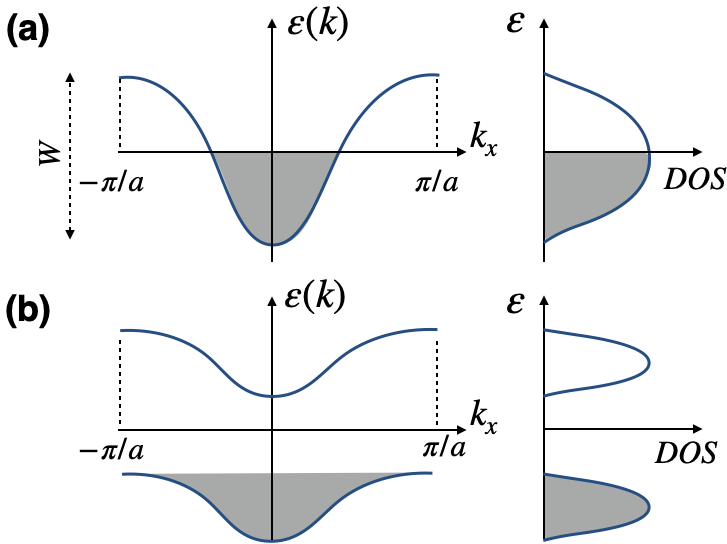}
  \caption{\label{Band-insulator}Typical band spectrum, $\varepsilon(k)$, along one direction in reciprocal space for a metal (a) or for a band insulator (b). Occupied states are shown by a grey shade. Right part of each plot schematically demonstrates corresponding density of states (DOS).}
\end{figure}

First we shortly discuss the description of different states of electron in solids, on the simple example of non-degenerate electrons (e.g.\ $1s$ electrons). In solid state physics we describe such electrons in a crystal by including, first, the kinetic energy of electrons due to hopping of electrons from site to site,
\begin{eqnarray}
\label{kin-energy-R}
\hat H = -t \sum_{\langle ij \rangle \sigma} c^{\dagger} _{i\sigma} c_{j\sigma},
\end{eqnarray}
where the operators $c^{\dagger} _{i\sigma}$, $c _{i\sigma}$ denote the processes of creation and annihilation of an electron at site~$i$ with spin~$\sigma $ (this representation is called in quantum mechanics the method of second quantization; a simple presentation of it is given e.g.\ in the Appendix~B of \cite{khomskii2014transition}). Thus this term describes the hopping, i.e.\ the transfer of an electron from site~$j$ to site~$i$: electron is annihilated at site~$j$ and recreated at site~$i$ with amplitude~$t$, called a hopping parameter. If we make a Fourier transform of this term, we get
\begin{eqnarray}
\label{kin-energy-k}
\hat H =  \sum_{ {\bf k} \sigma} \varepsilon ({\bf k}) c^{\dagger} _{{\bf k} \sigma} c_{{\bf k} \sigma},
\end{eqnarray}
where $\varepsilon ({\bf k})$ is the one-electron spectrum, which of course depends on the type of the lattice and the values of hopping parameters. E.g., for a cubic lattice with nearest neighbour hopping $\varepsilon ({\bf k})$ has the form
\begin{eqnarray}
\varepsilon ({\bf k}) = -2t \left( \cos (k_xa) +  \cos (k_ya) + \cos (k_za) \right),
\end{eqnarray}
see Fig.~\ref{Band-insulator}(a); here $a$ is the lattice constant. This description of electrons in solids is called a tight binding approximation. 

We have to fill this band by a certain number of electrons according to the Pauli principle (two electrons per state), so that e.g.\ for one electron per site this band will be half-filled, and the system would be metallic, with itinerant electrons, Fig.~\ref{Band-insulator}(a).  If we have several ions in the unit cell or if we have not $s$, but, say, $p$ or $d$ ions, then where would be several bands, and if there is a band gap between completely filled and empty bands, as shown in Fig.~\ref{Band-insulator}(b), then we have a band insulator. This is the situation in e.g.\ MgO or HfO$_2$, where $p$ or $d$ states of a metal are completely filled or empty.

In transition metal compounds the $d$ band of metals is typically not completely filled and according to the band theory these materials should be metals, while many of them do not conduct electricity; for example the energy gap in NiO, with Ni ions having electronic configuration~$3d^8$, is $\sim4\,\rm eV$~\cite{Sawatzky1984}. In fact what is missing in the treatment present above is the Coulomb repulsion between electrons. In the simplest case we can include only the repulsion at the same site, characterized by the parameter~$U$. For one orbital per site (e.g.\ $s$ atoms) this term can be written as
\begin{eqnarray}
\label{Uterms}
\hat H =  U \sum_i  n_{i\uparrow} n_{i\downarrow},
\end{eqnarray}
where $n_{i\sigma} = c^{\dagger}_{i\sigma} c_{i\sigma}$ is the number of electrons with spin $\sigma$ at site~$i$.
\begin{figure}[t]
   \centering
  \includegraphics[width=0.35\textwidth]{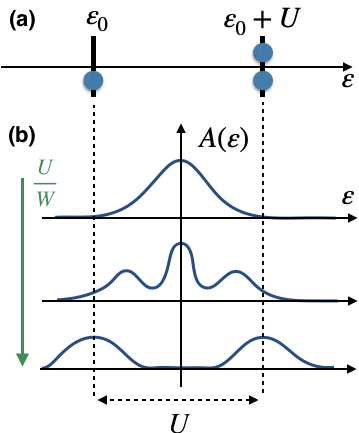}
  \caption{\label{Hubbard}Sketch illustrating the formation of Hubbard bands. (a) shows the energy levels of a quasi-isolated ion with one or two electrons (experiencing repulsion~$U$). In (b) we show how the spectral function changes with increase of the ratio of the on-site electron--electron repulsion $U$ and the bandwidth~$W$. For small $U/W$ we have a metal described by the Fermi-liquid theory, while for large $U/W$ the lower and upper Hubbard bands are formed, and for one electron per site the lower Hubbard band would be filled and the upper one empty, so that the system turns out to be a (Mott--Hubbard) insulator. In (b) $U$ increases from top to bottom.}
\end{figure}

The combination of the kinetic (band) energy \eqref{kin-energy-R} and the on-site interaction \eqref{Uterms} will give us what is called the Hubbard model. It is clear that for one electron per site and for strong interaction $U \gg t$ the electrons would prefer to stay at the same site and become localized, so that, instead of a metal, the material will be an insulator --- the Mott, or Mott-Hubbard insulator (these ideas were first proposed by Schubin and Vonsovskii in 1934 \cite{Schubin1934}, then by Peierls and finally elaborated by Mott, see \cite{Mott1937} and the Appendix A.1 in \cite{khomskii2014transition}). In this state the electrons are localized at sites, and correspondingly there will exist local magnetic moments --- in this case just with the spin $S=1/2$, but with larger spins for a general situation with more localized electrons per ion. Due to different types of exchange interactions such localized spins will finally give rise to all the plethora of different magnetic states in TM compounds (itinerant electrons can also produce some magnetic states, but usually with much smaller magnetic moments). If we change the electron occupation $n$, or make electron hopping $t$ or the corresponding bandwidth $W=2zt$  (where $z$ is the number of nearest neighbours) larger than the interaction, $W>U$, the electrons would become itinerant and the system would be metallic --- there would occur the insulator--metal, or Mott transition.

The electronic structure of materials, where electron motion becomes correlated, e.g.\ due to on-site Coulomb repulsion, is very different from what we have in conventional band metals or insulators. It is easy to demonstrate what would happen in this case on the example of quasi-isolated atoms, where electrons interact with repulsive energy~$U$. If we have a single electron on a level with energy $\varepsilon_0$, then adding another electron we obtain the state with energy $\varepsilon_0+U$, as shown in Fig.~\ref{Band-insulator}(a). In real solids these atomic levels become bands, and in the limit of $U \gg W$ we have, instead of these atomic levels, what is called the lower and upper Hubbard bands --- see the lower panel of Fig.~\ref{Band-insulator}(b). In the opposite limit of small $U$ the band metal picture is restored, see the upper part of Fig.~\ref{Band-insulator}(b).  The intermediate regime is of course the most interesting one. According to calculations performed within the Dynamical Mean-Field Theory (DMFT) in this situation we have both Hubbard bands because of localized electrons, and also a quasi-particle peak at the Fermi energy (zero energy in Fig.~\ref{Band-insulator}(b)) leading to heavy (theorists say ``dressed by the interaction'') metallic electrons at the Fermi level~\cite{Georges-96}. The full many-particle theory describing the transition from a Mott insulating to a metallic regime taking into account various non-local contributions is still to be developed, but some very interesting ideas concerning a possible topological nature of this transition have recently been formulated~\cite{Irkhin2019}.

\section{Single-site effects\label{Sec:single-site-effect}}
\begin{figure}[t]
   \centering
  \includegraphics[width=0.49\textwidth]{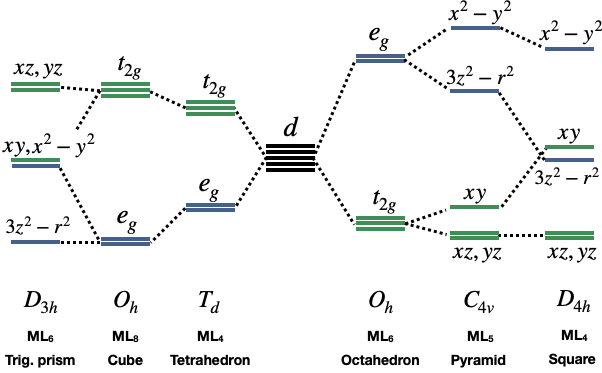}
  \caption{\label{CFS}Diagram of the crystal field splittings of the $d$ levels for different surroundings of the transition metal ion.}
\end{figure}

\subsection{Crystal-field splitting\label{Sec:CFS}}
When we consider real TM systems, we also have to include orbital degrees of freedom.
In isolated TM atoms or ions with full spherical symmetry five $d$ levels with $l=2$ are degenerate. When a TM ion is put in a solid, this degeneracy is lifted because of interaction with the surroundings --- mainly with nearest neighbouring ligands, e.g.\ oxygen, sulphur, chlorine etc.\ (for simplicity we will often speak below about oxides, although the main conclusions are mostly equally valid for other anions). The most typical situation is the six-fold coordination --- a TM ion in a ligand octahedron. In this case the five-fold degenerate $d$ levels are split into a lower $t_{2g}$ triplet and a higher-lying $e_g$ doublet, Fig.~\ref{CFS}, with the wave functions~\cite{Abragam} 
\begin{eqnarray}
\label{eg}
&&e_g: 
\begin{cases}
|3z^2-r^2 \rangle = | l_z=0 \rangle \sim 3z^2 - r^2, \\
|x^2-y^2 \rangle = \frac 1 {\sqrt 2} \left(  | l_z=2 \rangle +  | l_z=-2 \rangle \right) \sim x^2 - y^2,
\end{cases}\\
\label{t2g}
&&t_{2g}: 
\begin{cases}
|xy \rangle = -\frac {\rm i} {\sqrt 2} \left( | l_z=2 \rangle -   |l_z=-2 \rangle \right) \sim xy, \\
|xz \rangle = -\frac {1} {\sqrt 2} \left( | l_z=1 \rangle -   |l_z=-1 \rangle \right) \sim xz, \\
|yz \rangle = \frac {\rm i} {\sqrt 2} \left( | l_z=1 \rangle +   |l_z=-1 \rangle \right) \sim yz.
\end{cases}
\end{eqnarray}
Note that ``$xy$'', ``$3z^2-r^2$'' etc.\ are not just labels, but real mathematical expressions for the angular part of the corresponding wavefunctions (in what follows we will often denote  $|3z^2-r^2\rangle$ as $|z^2\rangle$ for shortness). 

\begin{figure}[t]
   \centering
  \includegraphics[width=0.49\textwidth]{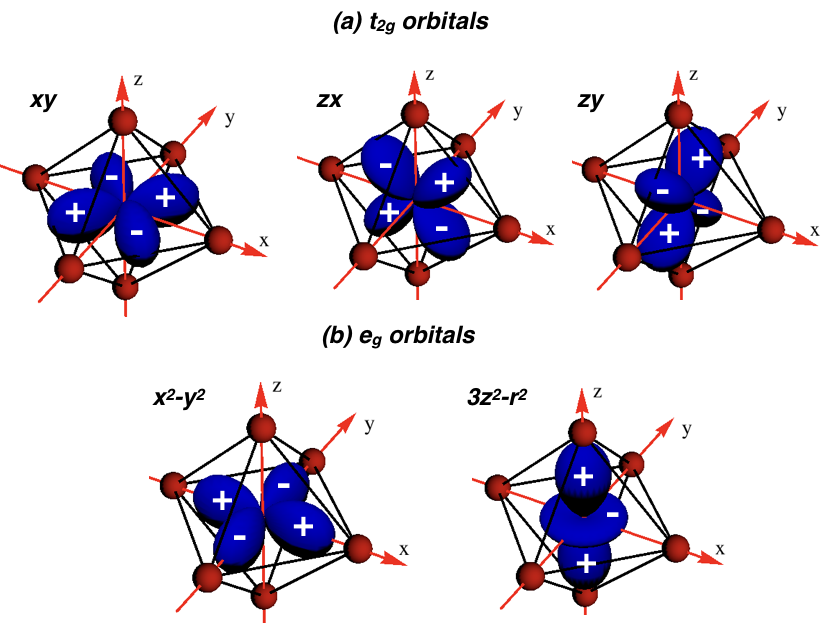}
  \caption{\label{cubic-harmonics}Cubic harmonics corresponding to $d$ orbitals (``$+$'' and ``$-$'' denote the sign of the wavefunction in a given region).}
\end{figure}

These orbitals are illustrated in Fig.~\ref{cubic-harmonics}. We see, first, that a particular orbital has a very specific direction in space --- this will be very important for many effects discussed in this review. The occupation of a particular orbital makes the original ``spherical'' ion non-spherical, with its shape being ellipsoidal --- i.e.\ the real order parameter for eventual orbital ordering would be in fact a quadrupole moment --- a symmetric second-order tensor. But in many cases one can reduce the description to a simpler one --- see Sec.~\ref{Sec:OO} below.

From the expressions (\ref{eg}), (\ref{t2g}) one can see that in the $e_g$ states the orbital moment is quenched, so that the spin--orbit coupling 
\begin{eqnarray}
\label{eq:full-SOC}
\hat H_{SOC} = \lambda \hat {\mathbf L} \cdot \hat {\mathbf S}
\end{eqnarray}
is in the first approximation zero for the $e_g$ states.  However, it can act on the $t_{2g}$ states, which form a triplet. In fact it can be shown that matrix elements of the spin--orbit coupling \eqref{eq:full-SOC} for the $t_{2g}$ wavefunctions are exactly the same as for the $p$ orbitals (also forming a triplet with $l=1$) if we substitute $\lambda \to - \lambda$~\cite{Sugano-book}. Therefore it is customary to describe the $t_{2g}$ states by an effective orbital moment $\tilde l =1$ --- with only, in the simplest case, the change of sign of the spin--orbit coupling for this effective moment.  
The eigenfuctions of the $z$ component of this orbital moment $ \tilde l$ have the form~\cite{Abragam}:
\begin{eqnarray}
&& | \tilde l^z_0 \rangle =   |xy \rangle, \nonumber  \\
&& | \tilde  l^z_{\pm 1} \rangle =   -\frac {1} {\sqrt 2} \left( {\rm i}|  xz \rangle \pm    |yz \rangle \right).
\end{eqnarray}           

The $t_{2g}$--$e_g$ crystal field splitting $\Delta_{CF} = 10Dq$ for $3d$ oxides is typically $\sim1.5$--$2\,\rm eV$, for $4d$ systems it is $\sim2.5-3\,\rm eV$, and for $5d$ oxides it is $\sim 3$--$4\,\rm eV$. Note, that the $t_{2g}$--$e_g$ crystal field splitting is typically much larger than the spin-orbital parameter $\lambda$, which does not exceed 0.5 eV even for heavy $5d$ elements, see Sec.~\ref{SOC} for more detailed information. The crystal-field splitting is caused by a combined action of $d$--$p$ hybridization with ligands, and also by a Coulomb repulsion of electronic charge of a particular orbital with the negatively charged ligands --- e.g.\ O$^{2-}$: it is clear that the stronger repulsion of the $e_g$, say $3z^2-r^2$ orbital directed towards ligands would make the energy of $e_g$ states higher than for the $t_{2g}$ orbitals with the electron density pointing in between the ligands.

Octahedral coordination, though the most common, is not the only possible coordination of TM ions in solids. In Fig.~\ref{CFS} we collected most typical situations. Those shown on the right-hand part of Fig.~\ref{CFS}  are in fact variations of the octahedral geometry --- due to lower symmetry the $t_{2g}$ and $e_g$ levels get split, while on the left-hand part the order of states is inverted so that the $e_g$ levels lie lower. The  $t_{2g}$--$e_g$ splitting also changes; thus for a tetrahedron $\Delta_{CF}$ is 4/9 of that for an octahedron. 
\begin{figure}[t]
   \centering
  \includegraphics[width=0.49\textwidth]{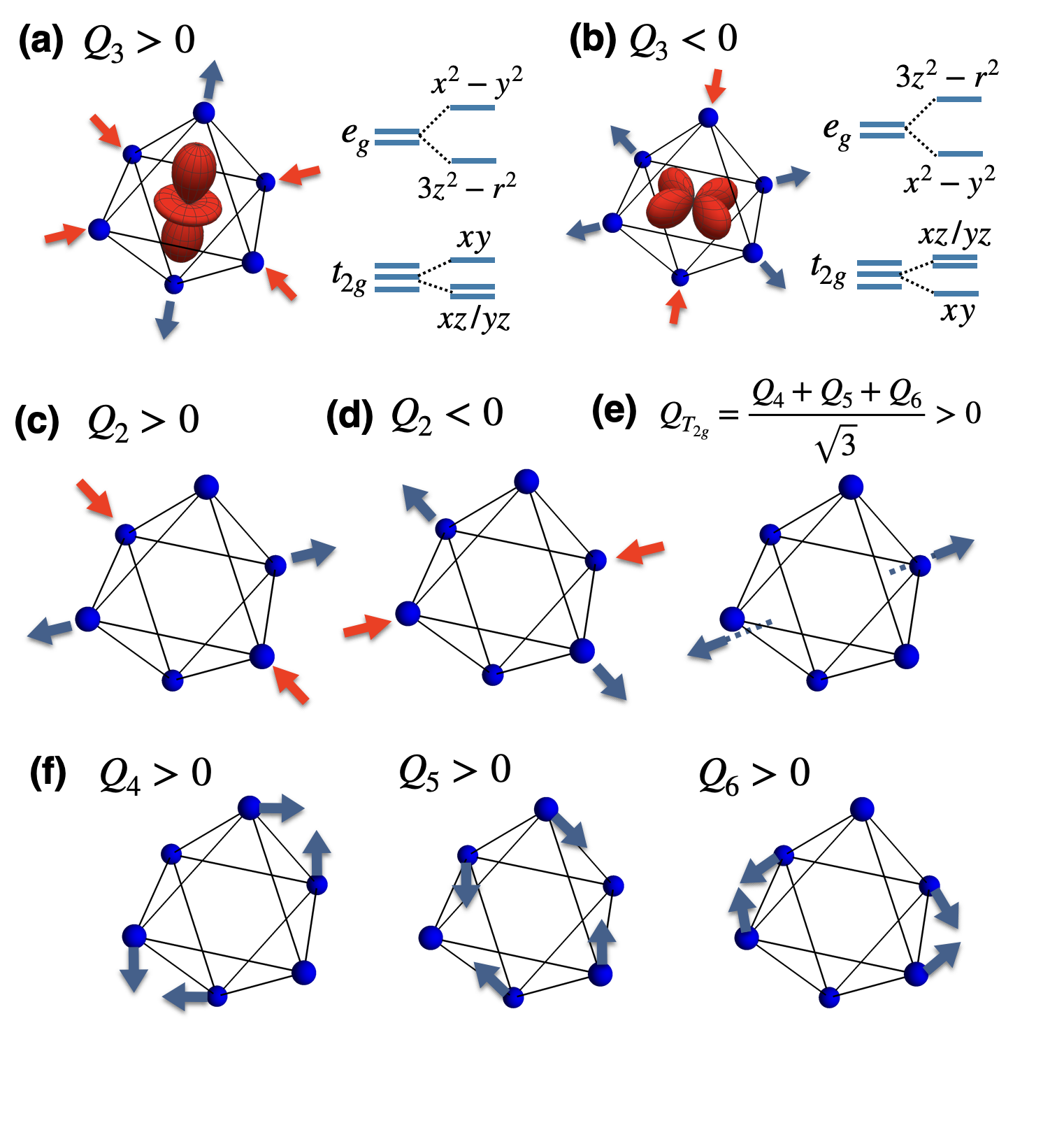}
  \caption{\label{JT-Q}Different possible symmetrical vibrations. (a--d) Two $E=\{Q_2,Q_3\}$ modes (for each we show distortions of each signs), (f) three $T=\{Q_4,Q_5, Q_6\}$ modes, and (e) symmetric $T_{2g}$ mode corresponding to trigonal elongation. For tetragonal distortions  we also show electronic level diagrams.}
\end{figure}

Further distortions can also split the octahedral crystal-field levels. Tetragonal elongation, so-called $Q_3$ mode \cite{VanVleck1939}, leads to additional splitting of electronic levels shown in Fig.~\ref{JT-Q}(a). Tetragonal distortion of opposite sign gives level ordering of Fig.~\ref{JT-Q}(b). Similarly, trigonal distortions (elongation or contraction along one of the [111] axes, which is a linear combination of $Q_4$, $Q_5$, and $Q_6$ modes, shown in Fig.~\ref{JT-Q}(f)) split the $t_{2g}$ levels, but do not split the $e_g$ levels. More details are presented in the next section.

When we know the crystal field scheme, we can understand the state of an ion with several electrons. One just has to fill these levels from below by electrons. In doing that one has to keep in mind the Hund's rule coupling which for isolated atoms or ions states that, first of all, the system chooses the state with maximum possible spin and for that one --- the state with maximum possible orbital moment. These rules lead to a unique state for isolated ions, but for ions in a crystal there exist different possibilities. For octahedral case, which we will mainly discuss in the following, the first, second and third electrons occupy the lowest $t_{2g}$ states, with the total spins respectively $\frac12$, 1 and~$\frac32$. But for the fourth electron the situation is {\it a priori} not clear. It can occupy one of the $e_g$ states with spin up, giving the total spin $S=2$; this is what we call the high spin (HS) state. We gain by that the Hund's interaction energy of this spin with the three spins on the $t_{2g}$ levels, $3J_H$, but since we have to put this electron on a higher-lying $e_g$ level this costs us the energy $\Delta_{CF} = 10Dq$. If $10Dq$ is less than $3J_H$, this will be the state formed. But if $10Dq > 3J_H$ it is better to put this fourth electron on one of the $t_{2g}$ levels, but then necessarily with the opposite spin. This is the low spin (LS) state with $S=1$ (in oder to calculate gain due to intra-atomic exchange one needs to count number of pairs having the same spins~\cite{Streltsov-UFN}, thus $E_{Hund}^{LS}=-3J_H$ and  $E_{Hund}^{HS}=-6J_H$). The Hund's coupling is $\sim0.8$--1~eV for $3d$ electrons~\cite{Vaugier2012}, $\sim0.6$--0.7~eV for $4d$ electrons~\cite{Vaugier2012} and $\sim0.5$~eV for $5d$ electrons~\cite{Yuan2017}. As the crystal-field splitting $10 Dq$ has the opposite tendency, the result is that for $3d$ ions the HS situation is the most typical (although there are important exceptions --- Co$^{3+}$ and Fe$^{2+}$ with the configuration $3d^6$ may also be in a LS state), while for $4d$ and especially for $5d$ systems the LS situation realizes nearly exclusively.

It is interesting to note that the crystal-field effects can sometimes favor a certain type of ligand polyhedra   surrounding a transition metal ion. E.g. it is well known that in molybdenum and niobium disulfides (and related materials) Mo and Nb ions are prone to have prismatic, not octahedral coordination. Indeed, in this situation we gain a lot of energy by putting two $d$ electrons of Mo$^{4+}$ ions in MoS$_2$ or a single $d$ electron of Nb$^{4+}$ ions in NbS$_2$ onto $3z^2-r^2$ orbital, see Fig.~\ref{CFS}, instead of leaving them on one of the $t_{2g}$ orbitals in octahedra. While this example illustrates that the orbital effects can affect crystallochemistry, there are of course also many other important factors (such as ionic radii, degree of covalency etc.) so that in oxides Mo and Nb are typically in octahedra.

\subsection{The Jahn--Teller effect\label{sec:JT}}
We saw in the previous section that in the high spin case for four $d$ electrons (in octahedral field), the fourth electron comes to one of the $e_g$ states. But, as these states are degenerate in the case of regular octahedra (cubic symmetry), there is an extra freedom: this electron can occupy any of the $e_g$ states, $z^2$ or $x^2-y^2$, or any of their linear combination. Thus in this case we have an extra (orbital) degeneracy, besides the usual Kramers degeneracy (spin up or down). This situation leads to the phenomenon known as the Jahn--Teller effect~\cite{Jahn1937,Bersuker2006}. In simple terms it tells us that the symmetric state with degenerate ground states (excepting Kramers degeneracy) is unstable with respect to distortions reducing this symmetry and leading to a splitting of these levels and to a decrease of total energy. Another, equivalent formulation of the Jahn-Teller theorem is that the symmetric state of a molecule or an ion in a crystal (except  linear molecules) with the degenerate ground state is not a minimum of the energy; there always exist some distortions which lift this degeneracy and decrease the total energy.

\begin{figure}[t]
   \centering
  \includegraphics[width=0.5\textwidth]{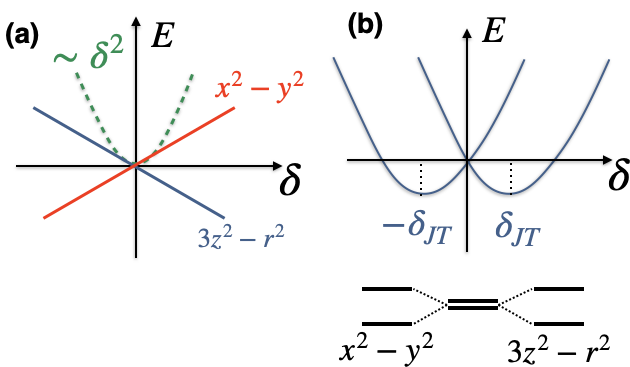}
  \caption{\label{JT}Sketch illustrating the essence of the Jahn--Teller effect. (a) blue and red lines show level splitting due to tetragonal distortion $\delta$ ($\delta>0$ corresponds to elongation, $\delta<0$ to compression), the first term in \eqref{JT-simplest}. Green dashed line is the elastic energy in harmonic approximation, i.e.\ the Hooke's law, the second term in \eqref{JT-simplest}. (b) shows the sum of these two contributions, which results in the formation of two Jahn--Teller minima.}
\end{figure}

For a simple doubly-degenerate level with one electron or hole on it (such as in the low spin Ni$^{3+}$ $(t_{2g}^6 e_g^1)$,  high spin Mn$^{3+}$ $(t_{2g}^3 e_g^1)$ or Cu$^{2+}$ $(t_{2g}^6 e_g^3)$) which could be split by one type of distortion, e.g.\ tetragonal distortion, the energy can be written as
\begin{eqnarray}
\label{JT-simplest}
E_{JT} = \pm g |\delta| +\frac {B\delta^2}2,
\end{eqnarray}
where the first term describes the splitting of the $e_g$ levels with the distortion $\delta$ (similar to the Zeeman splitting in the case of magnetic field), and the second term is the elastic energy, see Fig.~\ref{JT}(a).  The sign in the first term is responsible for the ``sign''  of distortion, i.e.\ contraction or elongation. The sum of these two terms is shown in Fig.~\ref{JT}(b). One sees that the undistorted state $(\delta=0)$ with degenerate levels does not correspond to an energy minimum. There are two minima at $\pm \delta_{JT}$ with one sign of distortion stabilising one orbital, and the opposite distortion the other one.

\begin{figure}[t]
   \centering
  \includegraphics[width=0.5 \textwidth]{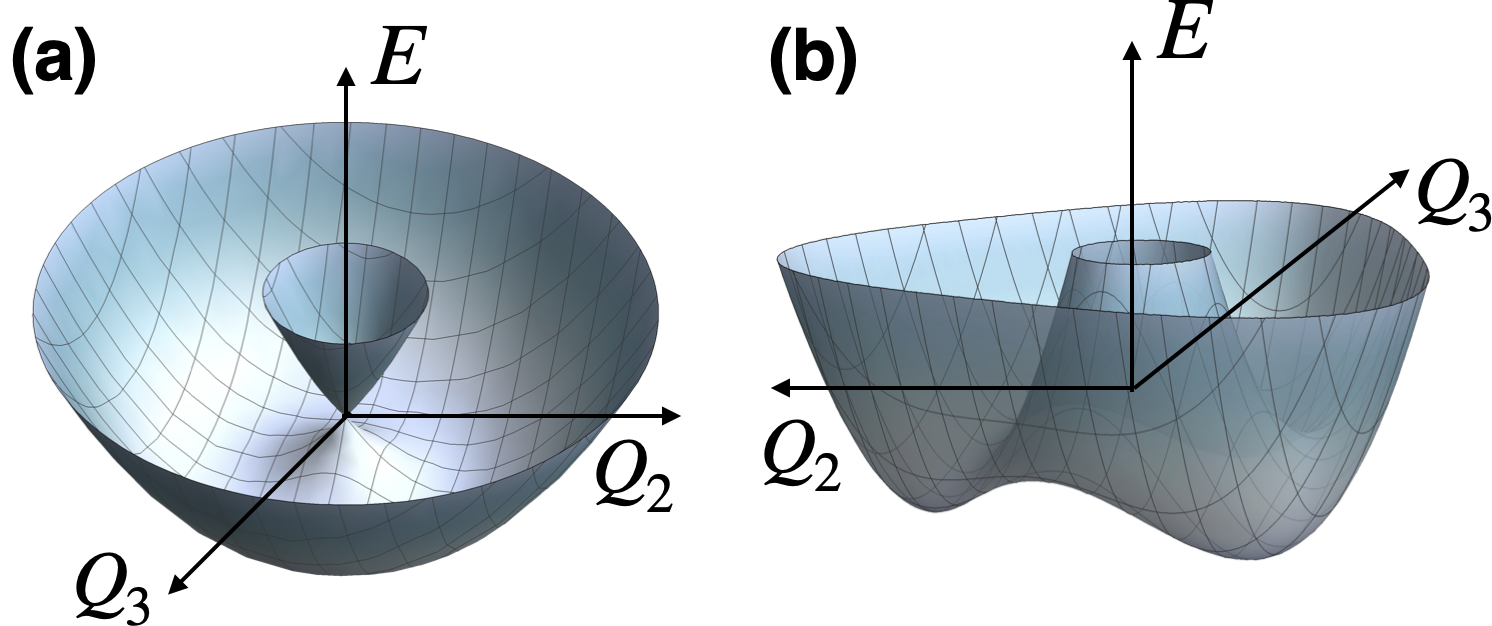}
  \caption{\label{JT-Mexican-hat}(a) ``Mexican hat'' energy surface for the $e \otimes E$ problem ($e_g$ electronic states and $E = \{Q_2,Q_3\}$ vibrations) in case of linear Jahn Teller effect and in harmonic approximation, see \eqref{JTeg}. (b) Effect of anharmonicity or of higher-order Jahn--Teller coupling resulting in warping (corrugation) of the trough at the bottom of the Mexican hat potential, leading typically to stabilization of elongated octahedra.}
\end{figure}

\begin{figure}[t]
   \centering
  \includegraphics[width=0.35 \textwidth]{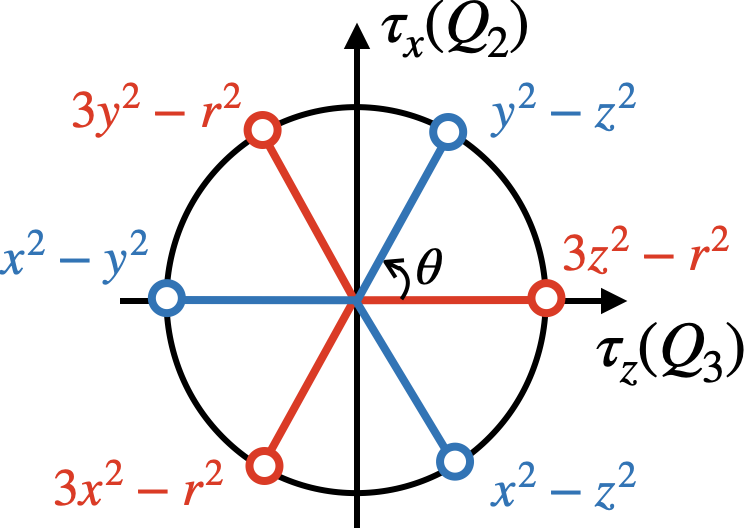}
  \caption{\label{Magic-circle} This sketch shows that one can describe both the Jahn--Teller distortions $E=\{ Q_2, Q_3\}$ and corresponding occupied orbital, characterized by the pseudospin $\tau$, by a single variable, the angle $\theta$. Explicit expression of the ground state electronic wavefunction and distortions are given by Eqs.~\ref{theta1} and \ref{theta2}. The plot corresponds to the case of a single electron in the $e_g$ orbitals (situation of e.g.\ Mn$^{3+}$), for a single hole (e.g.\ Cu$^{2+}$) one needs to change $\theta \to \pi - \theta$.}
\end{figure}

Several points should be mentioned here. First of all, for Jahn--Teller ions with $e_g$ degeneracy the splitting of degenerate levels can be caused not only by tetragonal distortions, Fig.~\ref{JT-Q}(a--b), but also by orthorhombic ones. These two are $E$ vibrations. They are shown in Fig.~\ref{JT-Q}(c--d), and they also split $e_g$ levels into two singlets (with the wavefunctions which are a mixture of $|x^2-y^2\rangle$  and $|z^2\rangle$ orbitals of Eq.~\eqref{eg}).  The orthorhombic  distortions, $Q_2$ mode, also split $t_{2g}$ levels into three singlets. The coupling to these two vibrations, $Q_3$ (tetragonal) and $Q_2$ (orthorhombic), is the same; this actually follows from the group symmetry analysis. In effect the $e_g$ levels can be split by coupling both with $Q_3$ and $Q_2$ modes, and the total energy surface, instead of Fig.~\ref{JT}, takes the form of Fig.~\ref{JT-Mexican-hat}(a), known as the ``Mexican hat''  (obtained by rotating the curves in Fig.~\ref{JT}(b) around the $z$-axis). In effect not just the two states, the tetragonal compression and tetragonal elongation shown  in Fig.~\ref{JT-Q} (a) and~(b) are equivalent in this approximation, but all the states at the bottom of the trough of the ``Mexican hat'' have the same energy. One can describe these states by a mixing angle $\theta$, shown in Fig.~\ref{Magic-circle} \cite{Kanamori1960,KK-UFN}: the distortions (i.e.\ nuclei part of the wavefunction) can be parametrized as
\begin{eqnarray}
\label{theta1}
|\theta \rangle = \cos (\theta) Q_3 + \sin ( \theta) Q_2,
\end{eqnarray}                                                                            
while corresponding electron wavefunction (spinor) is
\begin{eqnarray}
\label{theta2}
|\theta \rangle = \cos (\theta/2) |z^2\rangle + \sin (\theta /2) |x^2-y^2\rangle.                                           
\end{eqnarray}                                                                           
If we take into account the interaction of the $e_g$ electrons with both $Q_3$ and $Q_2$ distortions, the effective Jahn--Teller interaction, instead of \eqref{JT-simplest}, would now have the form
\begin{eqnarray}
\hat H_{JT}^E = &-&\frac g2 \left ((n_{z^2} – n_{x^2-y^2})Q_3  +  (c^{\dagger}_{z^2} c_{x^2-y^2} + {\rm h.c.}) Q_2 \right)  \nonumber \\
    &+& \frac B2 (Q_3^2 +Q_2^2),          
\label{JTeg}
\end{eqnarray}   
see e.g.\ \cite{Gehring1975},  where $n_{z^2}= c^{\dagger}_{z^2}c_{z^2}$, and the same for the $x^2-y^2$ orbital. Corresponding expression for Jahn-Teller coupling of $t_{2g}$ electrons can be found e.g. in Ref.~\cite{StreltsovPRX}.

It is very convenient to introduce here the pseudospin notation for doubly-degenerate $e_g$ orbitals: similar to real spin $\frac12$ one can describe them by a pseudospin $\tau=\frac12$, so that $\tau^z=+\frac12$ corresponds to the $z^2$ orbital, and $\tau^z = -\frac12$ to the $x^2-y^2$ orbital. Then the Hamiltonian \eqref{JTeg} acting in the subspace of the $z^2$ and $x^2-y^2$ orbitals would take the form
\begin{eqnarray}
\label{tau-Q}
\hat H_{JT}^{E} = -g (\hat \tau_z Q_3  +  \hat \tau_xQ_2 )   + \frac B2 (Q_3^2 +Q_2^2).          
\end{eqnarray}

One interesting remark should be made here~\cite{Khomskii2000c,Khomskii2000a}. The pseudospin operators $\hat { \pmb \tau}$ in many respects behave similar to the usual electron spin 1/2 operators, they can be represented by the same Pauli matrices and have the same algebra, they obey exactly the same commutation relations. But different components of $\hat {\pmb \tau}$ have different transformation properties. The usual spin 1/2 operators describing Kramers doublets are odd with respect to time inversion. But for $\tau$-matrices describing the $e_g$ orbitals the situation is different: $\tau_z$ and $\tau_x$ are even for time inversion, and they actually describe the nonspherical charge distribution of quadrupolar type. Only $\tau_z$ and $\tau_x$ couple to the lattice and enter the Jahn--Teller Hamiltonian \eqref{tau-Q}. On the other hand the $\tau_y$-matrix is imaginary, and the eigenstates of $\tau_y$ are complex combinations of $|z^2\rangle$ and $|x^2-y^2\rangle$,  $|\tau_y = \pm 1\rangle = \frac 1{\sqrt 2} (|z^2\rangle \pm  {\rm i} |x^2-y^2\rangle)$. Such states have spherical (or rather cubic) distribution of charge density, i.e.\ they do not correspond to the usual Jahn--Teller distorted states with the conventional orbital ordering. On the other hand, as all complex wave function in quantum mechanics they are transformed to complex conjugate with time inversion, i.e.\ these states are actually magnetic. However it is not a usual magnetic ordering, these states do not correspond to the states with nonzero magnetic dipoles; actually these states have nonzero magnetic octupoles~\cite{Khomskii2000a}. The ordering of higher order multipoles was invoked to explain the properties of a number of real materials, notably rare earths systems, but also some transition metal compounds, see e.g.\ \cite{Jackeli2009a,Kuramoto2009,Santini2009,Chen2010a}.

There are several important implications of the interaction (\ref{JTeg},\ref{tau-Q}). First of all, we see that, contrary to the simple picture described at the beginning of this section, the Jahn--Teller effect for an isolated TM site, or an isolated molecule with degenerate ground state, does not simply lead to a particular distortion with the corresponding unique occupation of one particular orbital: many such distortions --- in this approximation actually an infinite number of those --- are degenerate, having the same energy. Thus one should expect significant quantum effects in this case: there should be strong quantum fluctuations between these states even at zero temperature. This is indeed the case. The motion of the system along the trough of Fig.~\ref{JT-Mexican-hat}(a) should be quantized as the standard quantum rotator. The presence of the conical intersection in the ``Mexican hat'' also has profound implications: its presence strongly influences the character of excited states, optical properties, dynamics of the system, etc. Because of that the phase of electron wavefunction changes sign when we move the system in the configuration space around this point; this is a clear manifestation of the geometric, or Berry phase (and in fact this very notion first appeared in physics in the context of the Jahn--Teller effect \cite{Longuet-Higgins1,Herzberg1963,Longuet-Higgins1975}, long before the famous paper by Sir Michael Berry \cite{Berry1984}).  This is a very important and actively studied field nowadays, especially in the study of dynamics of molecules, etc., see e.g.~\cite{Bersuker1989}.

One relatively straightforward consequence of this treatment is the strong violation of the adiabatic approximation. Usually in treating electron systems in molecules and in solids  we first consider fast electronic motion on the background of a fixed, static lattice, composed of heavy ions. For a doubly-degenerate state one would then write the total wavefunction as 
\begin{eqnarray}
|\Psi_i\rangle =  |\psi_i\rangle  |\phi\rangle,       
\end{eqnarray}
where $\Psi$ is the total wavefunction, $\psi$ is the electronic part of it, and $\phi$, the same for both states in the doubly-degenerate $e_g$ problem discussed above, describes the static or frozen lattice. In case of a Jahn--Teller state, as we see e.g.\ from  the comparison of Fig.~\ref{JT-Q}(a) and~(b),  different lattice distortions correspond to different electronic states, i.e.\ for each state $i$ we have
\begin{eqnarray}
|\Psi_i\rangle =  |\psi_i\rangle  |\phi_i\rangle.
\end{eqnarray}                                 
This leads for example to the suppression of nondiagonal matrix elements of electronic operators:
\begin{eqnarray}
\langle \Psi_i | \hat A |\Psi_j\rangle =   \langle \psi_i| \hat A |\psi_j\rangle \langle\phi_i|\phi_j \rangle.
\end{eqnarray}
I.e.\ such matrix element is suppressed by the scalar product of different distortions $\langle \phi_i |\phi_j \rangle$, which can be much smaller than~1. This is called the Ham's reduction factor~\cite{Ham1965}; it is a very clear manifestation of the quantum nature of single-site (or single molecule) Jahn--Teller effect.  (In condensed matter physics the analogous phenomenon for the motion of electrons interacting with the (polar) lattice is known as the polaron band narrowing~\cite{alexandrov1995}.)  The state of the system described by \eqref{tau-Q}, in which every electronic state has its own distortion, is called the vibronic state. In treating vibronic physics it is essential to treat both electron and lattice degrees of freedom quantum-mechanically --- in contrast to the approximation most often used in considering for example different electronic or magnetic phase transitions in bulk solids, for which we usually treat electronic and magnetic degrees of freedom quantum-mechanically, but most often consider the lattice quasiclassically. Usually it is justified by the applicability in most cases of the adiabatic approximation. To which extent the vibronic or quantum effects, often crucial for single-site or molecular systems, should be taken into account in considering concentrated solids, is an open and a very important question.

Now, the full degeneracy of all $|\theta \rangle$ states (\ref{theta1}), (\ref{theta2}) exists when we treat the lattice in a harmonic approximation and use the lowest-order, i.e.\ linear Jahn--Teller coupling, see the Hamiltonians (\ref{JT-simplest},\ref{JTeg},\ref{tau-Q})  (in this approximation the energy does not depend on the mixing angle~$\theta$ (\ref{theta1}), (\ref{theta2}), since it contains it as $ \cos^2 \theta + \sin^2 \theta=1$). When we include higher-order Jahn--Teller coupling and take into account lattice anharmonicity, we have to add the term  $k \cos (3 \theta)\delta^3$ to the total energy coming from~(\ref{JTeg},\ref{tau-Q})\cite{Opik1957a,Kanamori1960}. The coefficient $k$ in this expression is usually negative, see e.g.\ \cite{Khomskii2000a}, so that the angles $\theta =0$, $\pm 2\pi/3$ are preferred. According to \eqref{theta1} this means that the elongation in the $z$ direction ($\theta=0$) is more preferable than compression ($\theta=\pi$). Other states with the same energy, $\theta =2\pi/3$ and $-2\pi/3$, also correspond to local elongations, but along $x$ and $y$ axes: in the original system the axes $x$, $y$, $z$ were equivalent, and this equivalence is preserved in the Jahn--Teller distorted case. This leads to the warping of the trough (the bottom of the ``Mexican hat''), so that the states $\theta = 0$, $\pm 2 \pi/3$ on the circle in the $(Q_2, Q_3)$-plane become local minima, see Fig.~\ref{JT-Mexican-hat}(b). There are many concentrated systems with $e_g$ electrons, which were first claimed to be compressed, but finally were found to have elongated octahedra \cite{Prodi2004,Streltsov2014d,Prodi2014,Kruger2010,Streltsov12Cu,Well2019}. The situation with isolated Jahn--Teller sites is in this sense very different: the system can tunnel between minima, so that quantum effects are preserved in this case. This can lead, in particular, to the preservation of the doublet as the ground state, but to a singlet state for very strong nonlinear coupling~\cite{Koizumi1999}. 
\begin{figure}[t]
   \centering
  \includegraphics[width=0.3\textwidth]{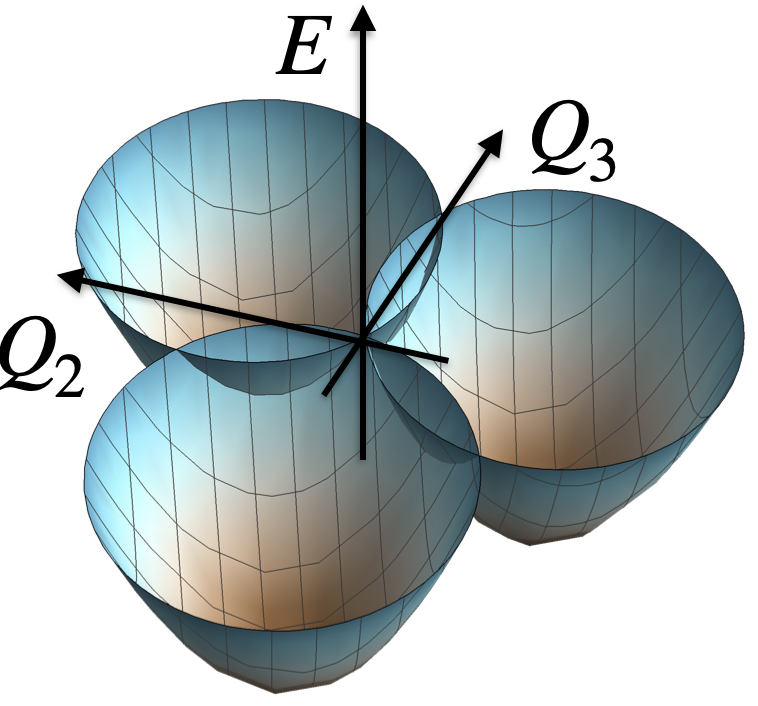}
  \caption{\label{JT-t2g}Energy surface of the $t \otimes E$ Jahn--Teller problem, one electron on the $t_{2g}$ orbitals in the presence of tetragonal ($Q_3$) and orthorhombic ($Q_2$) distortions.}
\end{figure}

Thus the $e_g$ levels in octahedra can be split by doubly-degenerate $E_g$ deformations $Q_2$,~$Q_3$ (this situation is known as the $e \otimes E$ problem). The same $E_g$ distortions also lead to a splitting of the $t_{2g}$ levels, see Fig.~\ref{JT-Q}(a,b), the $t \otimes E$ problem. But for these levels the situation is rather different. Here the resulting energy surface in case of one electron in the triply-degenerate $t_{2g}$ levels has the form of three paraboloids, shown in Fig.~\ref{JT-t2g}.  The minima correspond here to local contractions of the ML$_6$ octahedra along three different metal--ligand bonds, one of such compressions is shown in Fig.~\ref{JT-Q}(b). At such contraction the singlet level, here $xy$, goes down by energy~$E_{JT}$, but the doublet $zx$, $yz$ goes up by $E_{JT}/2$ (so that the centre of gravity of these levels remains at the same place). It is because of this factor that for $t_{2g}$ case the degeneracy between tetragonal elongation and tetragonal compression is already lifted in the simplest approximation of linear Jahn--Teller coupling and harmonic lattice. Still, some degeneracy remains: the octahedra can be compressed along $z$, $x$ or $y$ axes --- three equivalent paraboloids in Fig.~\ref{JT-t2g}. But, in contrast to the case of $e_g$ electrons, there is no tunnelling between these minima and no conical intersection, as was the case for the Mexican hat of Fig.~\ref{JT-Mexican-hat}, thus the quantum effects in this case ($t \otimes E$ problem) are much weaker than in the $e_g$ case. As we discuss below in Sec.~\ref{JTplusSOC}, this will change when we include real relativistic spin--orbit  coupling (SOC), which is potentially very important for $t_{2g}$ systems with unquenched orbital momenta.

As was mentioned in the previous section, $t_{2g}$ electrons are also split by trigonal $T_{2g}$ vibrations (the $t \otimes T$ problem). This problem however cannot be solved analytically and is more complicated. Many results in this field, and also in treating $t_{2g}$ electrons interacting both with $T_{2g}$ and $E_g$ vibrations, are presented in the monograph~\cite{Bersuker2006}.

\section{Cooperative Jahn--Teller effect, orbital ordering and magnetism\label{Sec:OO}}

Important effects appear when we consider concentrated systems such as TM compounds --- oxides, chlorides, sulphides etc. First of all, there appears an interaction between orbital degrees of freedom at different sites, so that for example the Jahn--Teller effect becomes cooperative Jahn--Teller effect. Nowadays we speak more often about orbital ordering --- at such  cooperative lattice distortion a particular orbital is occupied at each centre. But these terms actually describe exactly the same situation --- just stressing different possible mechanisms of such cooperative ordering. 
\begin{figure}[t]
   \centering
  \includegraphics[width=0.5\textwidth]{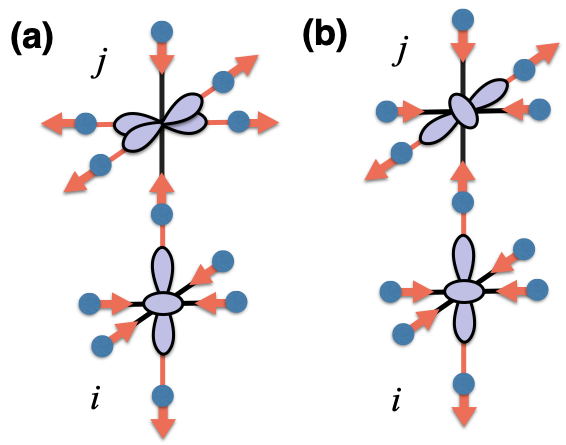}
  \caption{\label{OO-two-centers}Orbital ordering in the case of a single electron in the $e_g$ shell induced by lattice distortions. Nonlinear effects typically stabilize this pattern (b).}
\end{figure}

\begin{figure}[t]
   \centering
  \includegraphics[width=0.5\textwidth]{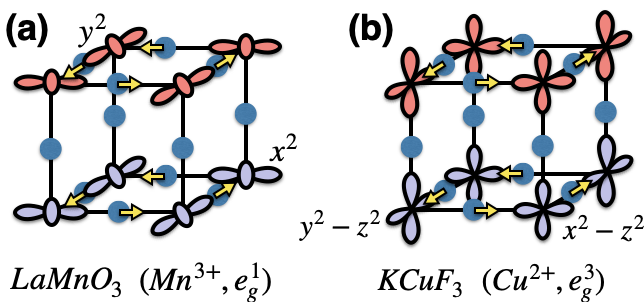}
  \caption{\label{LaMnO3-KCuF3}Orbital ordering pattern, which is realized in LaMnO$_3$ and KCuF$_3$ (for one of two polytypes). In both cases the metal--oxygen octahedra are locally elongated. The arrows show the directions of shifts of oxygen and fluorine ions (small circles on the bonds). This results in the occupation of the $x^2$ (i.e.\ $3x^2-r^2$) and $y^2$ (i.e.\ $3y^2-r^2$) electron orbitals in LaMnO$_3$ and the $x^2-z^2$ and $y^2-z^2$ hole orbitals in KCuF$_3$. Different colours correspond to different (opposite) spins --- the so called A-type magnetic structure.}
\end{figure}

The simplest mechanism of coupling between different distortions and respective orbital occupation at neighbouring sites is illustrated in Fig.~\ref{OO-two-centers}(a). In order not to induce global change of the crystal volume it is favourable to alternate distortions, e.g.\ to locally elongate MO$_6$ octahedra at site~$i$, and to contract them at site~$j$. Correspondingly, for one $e_g$ electron the orbital occupation would be ``antiferro'': the $z^2$ orbital would be occupied at site~$i$, and $x^2-y^2$ at site~$j$.  If we have strong nonlinear effects, then, as discussed above, local elongations are always preferred at each site. With this factor taken into account, the system may prefer to order as shown in Fig.~\ref{OO-two-centers}(b): there will be an elongation along $z$ at site~$i$ and along $x$ (or $y$) at site~$j$. Such type of ordering apparently is met in the parent compound of colossal magnetoresistance manganites LaMnO$_3$, see Fig.~\ref{LaMnO3-KCuF3}(a), where the oxygen distortions and the corresponding ordering of the $x^2$ and $y^2$ orbitals occupied by an $e_g$-electron is shown; or in KCuF$_3$, see Fig.~\ref{LaMnO3-KCuF3}(b), where the same distortion corresponds to alternating occupation of the hole orbitals $x^2-z^2$ and $y^2-z^2$.

Actually this mechanism of cooperative Jahn--Teller effect and orbital ordering due to electron--lattice coupling is always present and in many cases gives a dominant contribution to the ordering. It is important to note that the Jahn--Teller distortion at one site causes strain in the crystal which is actually long-range (decaying as $1/r^3$), which may also have different signs depending on the direction in the crystal \cite{Khomskii2001,Khomskii2003}. This strain would be felt by other Jahn--Teller ions in the crystal, thus in fact such orbital--lattice (or Jahn--Teller) mechanism of intersite interaction and of cooperative Jahn--Teller effect and orbital ordering is actually long-range. It is interesting that such simple treatment turns out to be quite successful in explaining many features of orbital ordering~\cite{Khomskii2003}.

One can formally describe this interaction, leading to a cooperative Jahn--Teller effect and orbital ordering, by proceeding from the Jahn--Teller interaction of the type (\ref{JTeg}), (\ref{tau-Q}), generalized for the case of many sites, i.e.\ adding the site index $i$ both to the electronic or pseudospin operators and to distortions; one can find the details e.g.\ in \cite{Gehring1975,KK-UFN}. Excluding the distortions, or phonon operators $Q_2$, $Q_3$, one finally gets the effective interorbital, or pseudospin interaction having the form of an exchange interaction, schematically written as
\begin{eqnarray}
\label{KK}
\hat H = \sum_{i \ne j} J^{\alpha \beta}_{ij} \hat \tau^{\alpha}_i \hat \tau^{\beta}_j,
\end{eqnarray}
where the pseudospin exchange $J_{ij} \sim g^2/B$ is in general long-range, and where the product of pseudospins can in principle be anisotropic. In any case, if we treat this interaction as we are used to do in the physics of magnetism, we can obtain that there would be no orbital (pseudospin) ordering at high temperatures, but typically there would be some type of ordering at lower temperatures --- be it ferro-orbital (ferro pseudospin), or antiferro, or some more complicated type of ordering. As is clear from our presentation, this orbital ordering is simultaneously an ordering of lattice distortions, i.e.\ a structural phase transition. The cooperative Jahn--Teller transition and the corresponding structural phase transition cannot exist without one another. Actually this situation is rather unique in the physics of solids. There are many different types of structural transitions in solids, and very rarely does one really know the true microscopic origin of those; it is often very difficult to predict whether a particular crystal would display such transition with changing temperature, pressure etc. Modern {\it ab initio} calculations can provide this information, but the simple physics of these transitions often still remains obscure. Cooperative Jahn--Teller transitions present a rare case when the microscopic nature of the transition is very clear: having a chemical formula of a compound, and just by looking at the electron occupation one can say that if, for example, in a symmetric situation there is an orbital degeneracy, then such system would definitely experience a structural transition leading to a reduction of symmetry of a crystal (e.g.\ cubic-tetragonal transition) and correspondingly to a lifting of orbital degeneracy.

From the form of the Hamiltonian \eqref{KK}, which actually has the form of magnetic exchange interaction, one can immediately suspect that, similar to magnetic systems, there may exist in orbital ordering significant quantum effects, as they are often observed in magnetic systems (and note that such quantum effects are different or at least only weakly related to vibronic effects considered in Sec.~\ref{sec:JT}). The situation, see more detailed discussion below, is however not so straightforward. The main problem here is that, as discussed above, the orbital ordering is intrinsically connected with lattice displacements, but the lattice is ``heavy'', so that many quantum effects existing for spin case may be suppressed for pseudospins, i.e.\ for orbitals.  
\begin{figure}[t]
   \centering
  \includegraphics[width=0.5\textwidth]{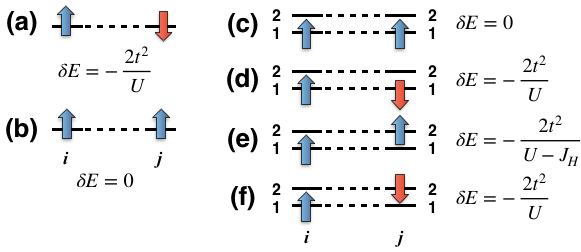}
  \caption{\label{2t2U}Illustration of interplay between orbital and magnetic structures. For nondegenerate case, if there is overlap only between half-filled orbitals, cases (a) and (b), then the AFM spin ordering takes place, since only then we gain $\delta E$ due to a virtual electron hopping from site to site (Pauli principle forbids such hopping for parallel spins, case (b)). For degenerate orbitals, (c--f),  if e.g.\ there is overlap only between the same orbitals, $t_{11}=t_{22}=t$, $t_{12}=0$,  the maximal energy gain due to electron hopping will be achieved in the situation with antiferro orbital, but ferro spin ordering, case~(e).}
\end{figure}

Besides electron--lattice, or Jahn--Teller mechanism of intersite interaction, there exists also a purely electronic, or exchange mechanism, also leading to orbital (and spin) ordering. This phenomenon, sometimes called Kugel--Khomskii mechanism, can be illustrated by simple considerations presented in Fig.~\ref{2t2U}. The typical mechanism of exchange interaction between localized electrons, described for example by the nondegenerate Hubbard model~\eqref{Uterms}, is shown in Fig.~\ref{2t2U}(a,b). For strong interaction, $U \gg t$, the electrons are localized at sites, but their spins are undetermined. Coupling between spin orientations at neighbouring sites is provided by virtual hopping of electrons to neighbouring sites and back.  As always, one would decrease the energy if this hopping is allowed (this is in fact the uncertainty principle of quantum mechanics). This process, however, is forbidden for parallel spins of Fig.~\ref{2t2U}(b) by the Pauli exclusion principle. But for antiparallel spins it is allowed, and leads (in second order of perturbation theory in hopping $t/U\ll 1$) to an energy gain $\sim t^2/U$. That is, in this case it is preferable to have electrons with antiparallel spins at neighbouring sites, which typically produces antiferromagnetic ordering for a concentrated system. The effective Hamiltonian describing such interaction is a Heisenberg exchange interaction 
\begin{eqnarray}
\label{Heisenberg}
\hat H = \sum_{i \ne j} J_{ij} \hat {\mathbf S}_i \cdot \hat {\mathbf S}_j
\end{eqnarray}
with the isotropic exchange
\begin{eqnarray}
\label{2t-over-U}
J_{ij}= \frac {2t_{ij}^2}U.
\end{eqnarray}
The virtual hopping processes leading to the exchange and shown in Fig.~\ref{2t2U}(a) are actually of two types: One is when  the ``blue'' electron from the left site hops to the right and then back. In the result the spin states of both sites do not change. This gives the Ising term $\sim \hat S^z_i \hat  S^z_j$ in the Hamiltonian \eqref{Heisenberg}. But there exist another process: when at the second step not the  ``blue'' electron returns to its original place, but the ``red'' electron from the right site moves to the left. In effect electrons, and spins at these two sites exchange their places. This is the real exchange process, and it is described by the terms of the type $\hat S^-_i \hat S^+_j$  (reversal of spins at both sites). Such terms  finally combine to ($\hat S^x_i \hat S^x_j + \hat S^y_i \hat S^y_j$), which, together with the first term  $\hat S^z_i \hat S^z_j$ (hopping of the same electron back and forth) combine to the Heisenberg exchange $J\left( {\mathbf S}_i \cdot {\mathbf S}_j\right)$. There can be situations, however, when the real exchange processes described above are impossible and we end up with the Ising-like Hamiltonian, see the discussion after Eq.~\eqref{Kitaev-constant}.

This is actually the main mechanism of exchange interaction in magnetic insulators (without complicated orbital structure, see below). If hopping occurs via ligand $p$ states, this is called superexchange interaction, and instead of direct hopping between $d$ orbitals, $t_{dd}$, one needs to use the effective one:
\begin{eqnarray}
\tilde t_{dd} \sim \frac{t_{pd}^2}{\Delta_{CT}},
\end{eqnarray}
where $t_{pd}$ is the $p$--$d$ hopping and  $\Delta _{CT}$ is the charge transfer energy --- the energy needed to transfer electron from the filled $p$ shell of O$^{2-}$ to TM, i.e.\ the energy of the process (``reaction'') $d^n p^6 \to d^{n+1}p^5$.

One can apply the same arguments for example to doubly degenerate orbitals, Fig.~\ref{2t2U}(c--f). Assuming for simplicity only diagonal hopping (hopping between the same orbitals, $t_{11}=t_{22}=t$, $t_{12}=0$), one would get now four situations shown in Fig.~\ref{2t2U}: (c)~same orbitals -- same spins, (d)~same orbitals -- opposite spins, (e)~different orbitals -- same spins, and (f)~different orbitals -- opposite spins. One sees that in case (c) the hopping is forbidden by the Pauli principle, thus we don't gain any energy in this situation. For the cases (d) and (f) the energies are $-2t^2/U$. But in case (e) in the intermediate state, when we transfer an electron from the orbital~2 of site~$j$ to the same orbital~2 at site~$i$, the spins of this excited virtual state would be parallel. But one has to remember that in atomic physics there is a Hund's rule interaction, which tells us that the energy of states with parallel spins is lowered by the Hund's energy~$J_H$. That is, in the denominator in the expressions of the energy in this case we will have not just $U$ --- the repulsion energy of two electrons at the same site --- but $U-J_H$. We see that in this case the system would prefer to have at neighbouring sites opposite orbitals but the same spin --- i.e.\ this mechanism would lead simultaneously {\it to both orbital and spin ordering}, in this case antiferro orbital and ferro spin (which may occur at different temperatures, but which are still intrinsically connected). The effective Hamiltonian describing this situation, instead of the simple Heisenberg exchange \eqref{Heisenberg}, would in this case  schematically have the form
\begin{eqnarray}
\label{KK2}
\hat H = \sum_{i \ne j} J^S_{ij} \hat {\mathbf S}_i \cdot \hat {\mathbf S}_j +J^{\tau}_{ij} \hat {\mathbf \tau}_j \hat {\mathbf \tau}_j + J^{S\tau}_{ij} (\hat {\mathbf S}_i \cdot \hat {\mathbf S}_j)( \hat {\mathbf \tau}_i \hat {\bf \tau}_j), 
\end{eqnarray}
i.e.\ it describes the interaction and coupling of two degrees of freedom, spin and orbital ones. The form of spin exchange here is the simple Heisenberg exchange (${\mathbf S}\cdot{\mathbf S}$), but 
because the electron hoppings of different orbitals are very different, see e.g. Fig.~\ref{OO-two-centers}, the resulting orbital terms may be rather anisotropic, see e.g.\ \cite{KK-JETP,KK-UFN}.
\begin{figure}[t]
   \centering
  \includegraphics[width=0.35\textwidth]{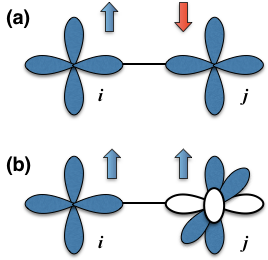}
  \caption{\label{GKA-direct}Illustration of the Goodenough--Kanamori--Anderson rules. If there is an overlap between two half-filled orbitals (i.e.\ for one electron per orbital), then the spins order in an antiferromagnetic fashion~(a). In contrast the overlap between half-filled and empty orbitals results in ferromagnetic ordering~(b).}
\end{figure}

Specifically, directional character of  orbitals, with the lobes of, for example, particular $e_g$  orbitals directed along respective bonds, cf. Fig.~\ref{OO-two-centers}, can lead to bond-dependent orbital interaction. Thus to optimise the electron hopping  in cubic lattice like that of  perovskites it might be favourable to occupy for the metal-metal bond parallel to $z$ the $z^2$ orbitals, but for similar bonds in $x$ direction -- the orbitals $x^2$ (i.e. $3x^2 - r^2$), and similarly for $y-$bonds. The resulting interaction would then have the form of the ``compass'' model
\begin{eqnarray}
\label{compass-model}
\hat H = J \left( 
\sum_{\langle i,j \rangle_x} \tau_i^x \tau_j^x +
\sum_{\langle i,j \rangle_y} \tau_i^y \tau_j^y +
\sum_{\langle i,j \rangle_z} \tau_i^z \tau_j^z 
\right)
\end{eqnarray}
introduced in~\cite{KK-UFN}, see Eq. 34 in \cite{KK-UFN}.  Such bond-dependent interaction results in very unusual properties, including strong quantum effects, see e.g.~\cite{Nussinov2015}. This compass model, naturally emerging in orbital physics, is the prototype of the popular nowadays Kitaev model, see Sec. \ref{sec:SOCexchange} below.



The connection between orbital structure and magnetic ordering is actually well known in the field of magnetism. It is known there as the Goodenough--Kanamori, or Goodenough--Kanamori--Anderson (GKA) rules. The simplest cases, closely related to the simple model presented above, are shown in Fig.~\ref{GKA-direct}. In the first case illustrated in Fig.~\ref{GKA-direct}(a) the electrons can hop between sites only to the already half-filled orbital, which is possible only for antiparallel spins, as has been already explained in Fig.~\ref{2t2U}(a--b). This would give rise to a strong AFM exchange $J_{\it AFM} \sim t^2/U$. In the second case the half-filled orbitals are orthogonal, and an electron from site~$i$ can only hop to the empty orbital at site~$j$. In a first approximation these hoppings are possible for any orientation of the spins $S_i$ and $S_j$. But the same Hund's rule exchange tells us that the energy of the  intermediate state  with two electrons at site~$j$ would be lower for parallel spins. In effect we have here a ferromagnetic exchange, but reduced in comparison to the first case:
\begin{eqnarray}
\label{FMexchange}
J_{\it FM} \sim -\frac {t^2 J_H}{U^2}. 
\end{eqnarray}
A very good manifestation of this rule is presented by the perovskite KCuF$_3$ with the strong Jahn--Teller ion Cu$^{2+}$. Its orbital structure (one of two possible modifications) is shown above in Fig.~\ref{LaMnO3-KCuF3}(b). As we see, the occupied (by one hole) $e_g$ orbitals are orthogonal (i.e.\ only hopping between a half-filed and an empty orbital is possible, similar to Fig.~\ref{2t2U}(e)), which according to the GKA rules gives weak ferromagnetic interaction. On the other hand the lobes of theses orbitals along the $c$-axes are directed towards each other, which leads to a very strong antiferromagnetic coupling in this direction. As a result this material, with practically cubic structure, is magnetically one of the best quasi-one-dimensional antiferromagnets~\cite{Kadota1967}!

Before we come to other geometries it should be mentioned that these very simple arguments actually explain why most transition metal oxides with localized electrons are antiferromagnetic; and even if some of them turn out to be ferromagnetic, the Curie temperature in them are typically much smaller than the N\'eel temperature. This is because of very different dependence of exchange constants on the largest energy scale in such systems --- the Hubbard~$U$. One may see that the AFM exchange in \eqref{2t-over-U} is inversely proportional to~$U$, while the FM exchange is smaller, typically $\sim 1/U^2$, see \eqref{FMexchange}. 

\begin{figure}[t]
   \centering
  \includegraphics[width=0.35\textwidth]{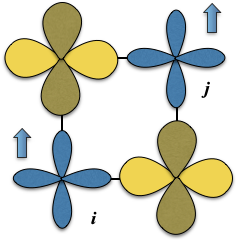}
  \caption{\label{GKA-90}Situation of 90$^{\circ}$ metal--ligand--metal geometry (e.g.\ with common edge of two octahedra), when half-filled $d$ orbitals (blue) overlap with two different ligand (oxygen, fluorine etc.)\ $p$ orbitals (dark and bright yellow). This results in a ferromagnetic exchange.}
\end{figure}

\begin{figure}[t]
   \centering
  \includegraphics[width=0.5\textwidth]{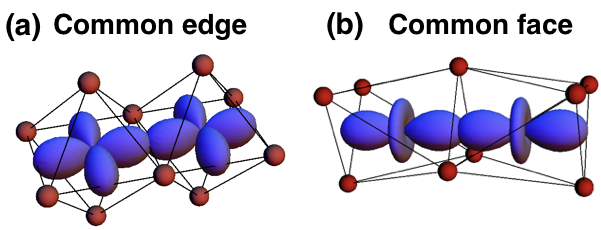}
  \caption{\label{Different-sharing}Two types of structures (in case of metals in octahedral coordination) where we have strong direct $d$--$d$ hopping.}
\end{figure}

The situation with orbital ordering and with the resulting exchange strongly depends on the geometry of the lattice. Often people cite the rule obtained above and illustrated in Fig.~\ref{GKA-direct}:  ``Ferro orbitals -- antiferro spins, antiferro orbitals -- ferro spins''. This is indeed generally true for systems with octahedra with common corner and with $\sim180^{\circ}$ metal--oxygen--metal bonds.  (Actually the hopping between $d$ orbitals in compounds such as TM oxides rarely occurs via direct $d$--$d$ overlap and hopping --- although such cases do exist and will be very important for us later, in Secs.~\ref{Sec:D-reduction} and \ref{Sec:CMI}. Most often these hoppings occur via intermediate ligands, e.g.\ along the Mn--O--Mn or Cu--F--Cu straight bond in LaMnO$_3$ and KCuF$_3$ mentioned above.) But the situation can be very different for example for octahedra with a common edge, sharing two common oxygens, with approximately 90$^{\circ}$ $M$--O--$M$ bonds, see Figs.~\ref{Different-sharing}(a) and~\ref{GKA-90}. As one sees from latter figure, in this case the same ``ferro'' ordered $x^2-y^2$ orbitals overlap with orthogonal $p$ orbitals of a ligand. Thus the virtual hopping of electrons (actually hoppings of $p$ electrons of a ligand to two TM sites and back) would be through (or rather from) these two orthogonal $p$ orbitals, and the Hund's exchange of two $p$ holes on oxygen (which is really not small at all, $J_H^{\rm oxygen} \sim 1.6$~eV \cite{Mazin1997} --- even bigger than the corresponding Hund's exchange on TM) would finally again make the spins of TM ions parallel. That is, in this case we have ferro orbital, but also ferro spin ordering --- but again with weaker exchange, reduced by the factor $J_H^{\rm oxygen}/\Delta_{CT}$. This is in fact the second GKA rule: in most cases the exchange for 90$^{\circ}$ bonds is ferromagnetic but weak.  

There are many specific situations for this geometry. For some orbital occupations the exchange for 90$^{\circ}$ bonds may still be antiferromagnetic and relatively strong.
It may also change for the case of strong spin--orbit interaction: e.g.\ for ions such as Ir$^{4+}$ it may be strongly anisotropic, which can lead to what is called Kitaev interaction, or Kitaev model~\cite{Kitaev2006,Jackeli2009}, which will be briefly discussed in Sec.~\ref{sec:SOCexchange}.

In fact besides these two most often discussed situations --- octahedra with common corner and with 180$^{\circ}$ $M$--O--$M$ exchange (such as perovskites, etc.), and octahedra with common edge and 90$^{\circ}$ exchange (e.g.\ exchange of nearest neighbours of systems such as NiO with the NaCl-type structures, or exchange of $B$-sites in spinels), there exist a third situation, strangely enough almost ignored in the literature until recently --- the case of octahedra with a common face, i.e.\ with three common oxygens, with the $M$--O--$M$ angle $\sim 70.5^{\circ}$, Fig.~\ref{Different-sharing}(b). This situation is actually met in quite a big number of different compounds, which will be also discussed below in Sec.~\ref{Sec:CMI}. The type of spin--orbital exchange for this case was recently considered in \cite{Kugel2015,Khomskii2016}.  It actually  resembles more the case of systems with common corners than with common edge. In particular, the simplified symmetric model with only diagonal hopping $t_{11}=t_{22}=t$, $t_{12}=0$, considered above, see Fig.~\ref{2t2U}(b), is actually indeed realized for the case of a common face. Interestingly enough, in this simple case the effective spin--orbital (``Kugel--Khomskii'') Hamiltonian \eqref{KK2} takes a very symmetric form, with the Heisenberg-like interaction (scalar product) not only for the spin, but also for the orbital (pseudospin) operators, and with a definite relations between coupling constants $J^S$, $J^{\tau}$, $J^{S\tau}$:
\begin{eqnarray}
\label{SU4}
\hat H = \sum_{i\ne j} J_{ij} \left( \hat {\mathbf S}_i \cdot \hat {\mathbf S}_j + \hat {\pmb \tau}_j \cdot \hat {\pmb \tau}_j + 4(\hat {\mathbf S}_i \cdot \hat {\mathbf S}_j)(\hat {\pmb \tau}_i \cdot \hat {\pmb \tau}_j) \right),
\end{eqnarray}
which one can rewrite as
\begin{eqnarray}
\hat H = \sum_{i\ne j} J_{ij}  
\left(\frac 12 + 2\hat {\mathbf S}_i \cdot \hat {\mathbf S}_j \right) 
\left(\frac 12 + 2\hat {\pmb \tau}_j \cdot \hat {\pmb \tau}_j \right)   + const, \nonumber
\end{eqnarray}
where $\frac 12 + 2\hat {\mathbf S}_i \cdot \hat {\mathbf S}_j$ and 
$\frac 12 + 2\hat {\pmb \tau}_j \cdot \hat {\pmb \tau}_j$ are the usual permutation operators. It is clearly seen now that this Hamiltonian has a very high symmetry --- theoreticians speak in this case about SU(4) symmetry. Such situation can actually appear in some transition metal compounds~\cite{Pati1998,Kugel2015,Khomskii2016,Yamada2018,Natori2019}, and even in twisted bilayer graphene~\cite{Venderbos2018}. Such symmetric model can also lead to very unusual quantum effects, such as the formation of spin-orbital liquid ~\cite{Corboz2012}.

An interesting question arises: what is the relative importance of the two mechanisms of orbital ordering discussed above --- Jahn--Teller coupling via the lattice, (\ref{Heisenberg}),~(\ref{KK2}), and a purely electronic (exchange, or Kugel--Khomskii) mechanism \ref{KK}. In real systems of course both these mechanisms act simultaneously and usually cooperate, i.e.\ typically they lead to the same type of orbital ordering. This question was addressed theoretically in \cite{Pavarini2008,Leonov10,Pavarini2010}: in calculations one can turn off the interaction via phonons, considering a frozen lattice, and thus separate electron--lattice and exchange contributions to the ordering. The results seem to show that these two mechanisms are comparable, the Jahn--Teller mechanisms in most cases being somewhat stronger (with the ratio $\sim$ 60:40 or 70:30). But it is important to note that even for frozen lattice the purely electronic mechanisms would also lead to orbital ordering and to the connected spin ordering.   

The cooperative Jahn--Teller effect is experimentally observed in quite a lot of different materials --- almost in all those containing partial occupation of degenerate states in a symmetric situation. These are most of bulk materials containing strong Jahn--Teller ions Cu$^{2+}$ ($t_{2g}^6 e_g^3$) and Mn$^{3+}$ ($t_{2g}^3 e_g^1$). Among these are the already mentioned KCuF$_3$ and its layered analogue K$_2$CuF$_4$~\cite{Khomskii1973,Ito1976}; prototype systems such as La$_2$CuO$_4$ giving High-$T_c$ superconductivity with doping; or the prototype colossal magnetoresistance material LaMnO$_3$. These are the examples of systems with strong Jahn--Teller effect, typical for the case of $e_g$ degeneracy. But systems with partially-filled $t_{2g}$ levels can also show cooperative Jahn--Teller effect and orbital ordering, although in this case both the electron--lattice interaction and that leading to a superexchange (Kugel-Khomskii) mechanism are weaker. Also for $t_{2g}$ electrons the real relativistic spin--orbit coupling is not quenched, and it can lead to other types of ordering. These questions will be considered in more details in Sec.~\ref{JTplusSOC}.

\section{Reduction of dimensionality due to orbital degrees of freedom\label{Sec:D-reduction}}

One of the very specific effects connected with orbital degrees of freedom is the often observed reduction of effective dimensionality due to a particular orbital ordering. I.e.\ a system behaves not as one would expect from its (high temperature) crystal structure, but its dimensionality would be lower. Some examples of such materials, where orbitals are responsible for this reduction, are given in Table~\ref{Tab:Dim-reduction}. The origin of this effect is very easy to understand just by looking at orbitals shown in Fig.~\ref{cubic-harmonics}.  Thus, if in a cubic system an electron occupies the $z^2$ orbital of Fig.~\ref{cubic-harmonics}, it is intuitively clear that this electron can easily hop from this orbital to neighbouring sites in the $z$ direction, but much less likely in the $x$ and $y$ directions. Or an electron on the $x^2-y^2$ orbital of Fig.~\ref{cubic-harmonics} can move in the $xy$ plane, but practically cannot hop in the $z$ direction. This leads to strong anisotropy of many physical properties, in the limiting cases effectively reducing the dimensionality --- e.g.\ from (cubic) three-dimensional (3D)  to two-dimensional (2D) for the $x^2-y^2$ orbitals or to almost one-dimensional (1D) for the $z^2$ orbitals. We have already mentioned above in the discussion of KCuF$_3$ that because of a particular type of orbital ordering, shown in Fig.~\ref{LaMnO3-KCuF3}, this material, crystallographically cubic, magnetically becomes really a one-dimensional antiferromagnet~\cite{Iio1978}.

\begin{table}[t!]
\centering \caption{\label{Tab:Dim-reduction}Examples of materials, where effective reduction of dimensionality due to orbitals degrees of freedom is observed.}
\vspace{0.2cm}
\begin{tabular}{llcccccc}
\hline
\hline
$1D \to 0D$  & chains$ \to $dimers                  & NaTiSi$_2$O$_6$ \cite{Isobe2002,Streltsov2008}\\
$1D \to 0D$  & chains$ \to $dimers                  & TiOCl \cite{Seidel2003}\\
$2D \to 0D$  & triangular lattice $\to$ trimers & LiVO$_2$ \cite{Cardoso1988,Pen1997}\\
$2D \to 0D$  & square lattice $\to$ dimers     & La$_4$Ru$_2$O$_{10}$ \cite{Wu2006}\\
$2D \to 0D$  & depleted square lattice $\to$ tetramers     & CaV$_9$O$_9$~\cite{Starykh1996,Korotin1999}\\
$3D \to 0D$  & hollandite $\to$ tetramers       & K$_2$Cr$_8$O$_{16}$ \cite{Hasegawa2009,Toriyama2011}\\
$3D \to 0D$  & spinel $\to$ tetramers/trimers             & AlV$_2$O$_4$ \cite{Horibe2006,Browne2017}\\
$3D \to 0D$  & spinel $\to$ octamers             & CuIr$_2$S$_4$ \cite{Radaelli2002,Khomskii2005a}\\
$3D \to 1D$  & spinel $\to$ chains $\to$ dimers                 & MgTi$_2$O$_4$ \cite{Schmidt2004,Khomskii2005a}\\
$3D \to 1D$  & perovskite $\to$ chains          & KCuF$_3$ \cite{Hutchings1969}\\
$3D \to 1D$  & pyrochlore $\to$ chains          & Tl$_2$Ru$_2$O$_7$ \cite{Lee2006}\\
\hline
\hline
\end{tabular}
\end{table}

In the next two subsections we discuss possible mechanisms of effective reduction of dimensionality starting with two very different limits: in Sec.~\ref{Sec:VBC} we begin with the atomic limit, which is more appropriate for insulators with localized electrons, while in Sec.~\ref{Sec:Orbital-Peierls} the band limit is considered. We will discuss examples of some particular materials with very different initial dimensionality and show that these two pictures, being qualitatively different, rather often lead to the same conclusions, although in some situations only one of them gives a correct result explaining experimental findings.
\begin{figure}[t]
   \centering
  \includegraphics[width=0.5\textwidth]{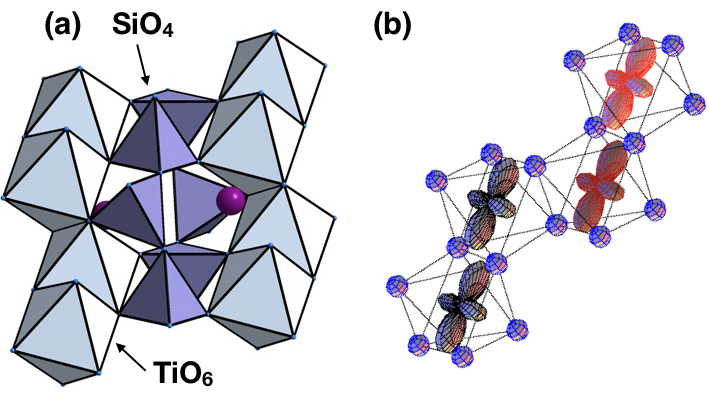}
  \caption{\label{NaTiSi2O6}(a) Crystal structure of NaTiSi$_2$O$_6$. (b) Orbital ordering ($xy$ orbitals) as realized in the low temperature phase in this material (results of the DFT${}+{}$U calculations)~\cite{Streltsov2008}.}
\end{figure}

\subsection{Orbital-dependent valence bond condensation\label{Sec:VBC}}


{\bf Chains ($\mathbf {1D}$).} Probably one of the conceptually simplest examples --- although in materials with relatively complicated crystal structure --- is met in NaTiSi$_2$O$_6$. This material belongs to pyroxenes --- silicates which are not (yet) very popular among physicists, but which are extremely important in geology: pyroxenes are among the main rock-forming minerals, constituting more than 10\% (volume) of the Earth's crust~\cite{anderson2007}. These systems have a chain-like structure composed of zig-zag chains of edge-sharing MO$_6$ octahedra, with SiO$_4$ tetrahedra in between, see Fig.~\ref{NaTiSi2O6}(a). Specifically, NaTiSi$_2$O$_6$ has Ti$^{3+}$ ($d^1$) in these octahedra, which form quasi-1D chains with localized electrons with antiferromagnetic interaction. Correspondingly, at high temperatures magnetic susceptibility has typical form for an antiferromagnetic chain of $S=1/2$ (the so-called Bonner--Fisher curve~\cite{Bonner1964}). But there occurs in NaTiSi$_2$O$_6$ a structural phase transition at $T_c \sim 210$~K, below which the material becomes practically diamagnetic~\cite{Isobe2002}. The explanation of this behavior was proposed in \cite{Konstantinovich-04,Streltsov2008}.

It turned out that, whereas at high temperatures all three $t_{2g}$ levels of Ti$^{3+}$ are more or less equally populated, below $T_c$ only one type of orbitals, $xy$, is occupied, while the other two $t_{2g}$ states become empty. Thus there occurs in NaTiSi$_2$O$_6$ below $T_c$ an orbital ordering of ferro type --- the same orbital occupied at every site, see Fig.~\ref{NaTiSi2O6}(b). But in this zig-zag geometry after this orbital ordering the chain practically becomes a collection of weakly coupled dimers, with a singlet ground state for each dimer (forming in fact something like a simple valence bond --- as in a hydrogen molecule). And indeed, {\it ab initio} calculations show that the exchange interaction inside such dimers is very strong and antiferromagnetic, $J_{\rm intra} \sim 400$~K, while the exchange between dimers is weak and most probably even ferromagnetic, $J_{\rm inter} \sim -5$~K \cite{Streltsov2008}. This is explained by the orbital orientation: in a dimer the occupied orbitals are directed towards one another and  have larger direct $d$--$d$ overlap, leading to strong AFM interaction. However there is no such overlap between dimers, which finally, according to the GKA rules discussed in Sec.~\ref{Sec:OO}, gives very weak ferromagnetic interaction. In any case, just this orbital ordering itself, even without any lattice distortion, would produce this splitting of a magnetic chain into singlet dimers. Of course lattice also plays an important role here: below 210~K the Ti atoms inside dimers move towards each other, so that the Ti--Ti distance in a dimer becomes very short, $d_{{\rm Ti}-{\rm Ti}} \sim 3.05\,$\AA~\cite{Redhammer2003}. But we want to stress that even without such lattice distortions the respective orbital ordering alone would explain the physical properties of this system. We see here a clear example of the role of the directional character of orbitals, and how orbital ordering effectively reduces the dimensionality of the system --- here from 1D chains to zero-dimensional (0D) clusters, singlet Ti--Ti dimers.
\begin{figure}[t]
   \centering
  \includegraphics[width=0.5\textwidth]{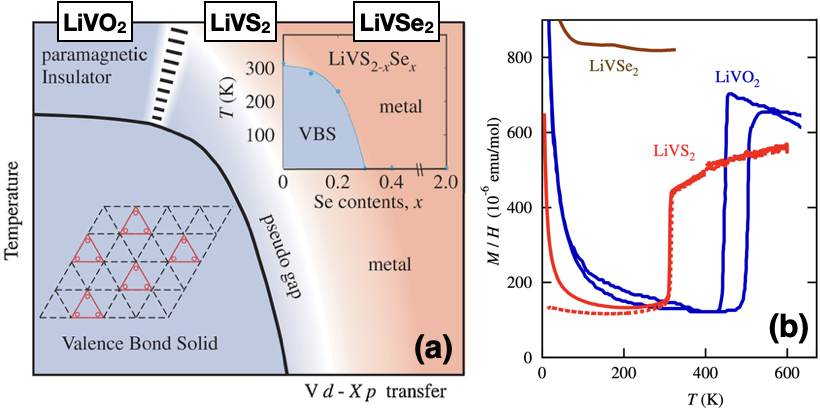}
  \caption{\label{LiVO2-exp}(a)~Phase diagram of LiV$L_2$, where $L={}$O, S, Se. (b)~Temperature dependence of magnetic susceptibility for  LiVO$_2$,  LiVS$_2$, and  LiVSe$_2$. Reproduced with permission from Ref.~\cite{Katayama2009}. Copyright 2009 American Physical Society.}
\end{figure}

{\bf Triangular lattice ($\mathbf {2D}$).} This lattice represents a special and very important class of lattices with the spin frustration. Indeed if magnetic ions are antiferromagnetically coupled on such a lattice and if there is an easy axis (Ising-like)  magnetic anisotropy  along which spins are aligned, then the ground state turns out to be highly degenerate. This is because it is impossible to have three collinear spins forming triangle antiferromagnetically ordered - at best only two of them can be antiparallel, two of three bonds can be ``satisfied'' and the last one remains ``frustrated''. 

\begin{figure}
   \centering
  \includegraphics[width=0.49\textwidth]{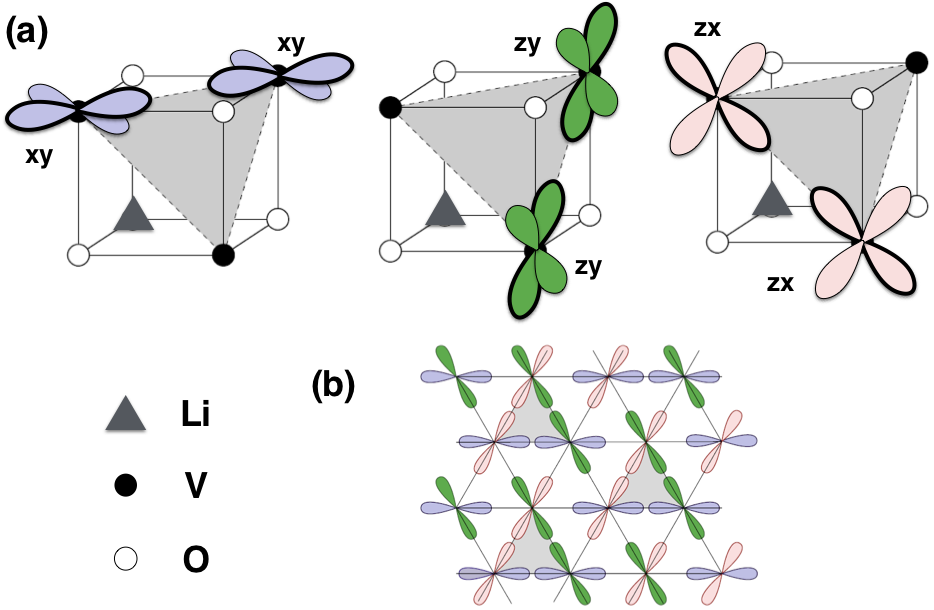}
  \caption{\label{LiVO2-orbitals}Possible mechanism of nonmagnetic ($S=0$) ground state in LiVO$_2$ due to orbital ordering. There is a strong overlap between two $t_{2g}$ orbitals for every pair of V ions forming trimers (shaded triangles). This results in the formation of three molecular orbitals ($xy-xy$, $xz-xz$, and $yz-yz$), which are occupied by all (six) available $d$ electrons.}
\end{figure}

Effects related to spin frustrations represent a special and fast developing field in modern theory of magnetism, for detailed description see e.g. \cite{Ramirez1994,Zvyagin2013,Schmidt2017,Zhou2017}, but in this review we are mainly interested in other type of effects, related to physics of orbitals. A good example of such effects is the orbital ordering is LiVO$_2$. It is a layered material with triangular layers of V$^{3+}$ ($d^2$) (it can be visualized as a rock salt VO with every other [111] layer of V replaced by Li). It is an insulator with magnetic properties (very similar to those of  NaTiSi$_2$O$_6$) shown in Fig.~\ref{LiVO2-exp}(b): the magnetic susceptibility has a Curie--Weiss behaviour at high temperatures, but below $T_c \sim 500\,$K it strongly drops and becomes very small (with a small Curie tail due to magnetic impurities --- most probably V ions at Li sites), see e.g.~\cite{Katayama2009}. Simultaneously there occurs a structural transition with the formation of a $\sqrt 3 \times  \sqrt 3$ superstructure. This phenomenon was explained in \cite{Pen1997} as a consequence of orbital ordering. The three-fold degenerate $t_{2g}$ orbitals order below $T_c$ into three sublattices shown in Fig.~\ref{LiVO2-orbitals}(b). In that process there appear tightly-bound trimers, triangles shaded in Fig.~\ref{LiVO2-orbitals}(b). The electron hopping and consequently V--V exchange is strong and antiferromagnetic inside these triangles, but much weaker and probably ferromagnetic in between those. In effect we have a singlet state of such triangles. Thus in this case the 2D system becomes actually zero-dimensional below~$T_c$, consisting of almost isolated trimers.

One can in principle think of different states of these trimers, different ways to describe them and to get a total spin singlet state. One is the molecular orbital description, with $d$ electrons forming molecular orbitals either at each valence bond separately, using direct $d$--$d$ hopping between respective orbitals, $xy$, $xz$, and $yz$  (the picture used in the figure caption of Fig.~\ref{LiVO2-orbitals}), or a total molecular orbital state of a triangle, including also $d$ electron hopping via oxygens. Three $t_{2g}$ orbitals are shown in Fig.~\ref{LiVO2-orbitals}(a). One sees that there exists a direct $d$--$d$ overlap and hopping for the $xy$ orbitals for neighbours in the $xy$-directions, the $yz$ orbitals in the $yz$-direction, and $xz$ orbitals in the $xz$-chains.  This would give strong hopping between respective occupied orbitals at the bonds of strongly bound  triangles (shaded in Fig.~\ref{LiVO2-orbitals}(b)).  The corresponding electronic structure is shown in Fig.~\ref{MO-in-clusters}. Six electrons of three V ions occupy three lowest molecular orbitals and this gives $S_{tot}=0$. (If one includes weaker electron hopping via ligands, some of these triplet levels will be split into singlets and doublets, but this would not change the conclusions).

However there exists also an alternative picture, leading to the same singlet state of a triangle: the picture with electrons localized at each V and forming there a total spin $S=1$ according to the Hund's rule. These localized $S=1$ spins would have strong AFM exchange inside triangles, so that the ground state would again correspond to a total singlet, $S_{tot}=0$. Which picture, that of delocalized electrons inside triangles forming molecular orbitals, or the one with strong electron correlations at each V, is {\it a priori} not clear.

Spectroscopic studies carried out in \cite{Pen1997} seem to agree better with the second picture, that of localized electrons with $S=1$ at each~V, with strong AFM exchange within a triangle. However below the structural transition, in a diamagnetic state, the V--V distance in trimers becomes very short, $d_{{\rm V}-{\rm V}}=  2.56\,$\AA\ --- shorter than the V--V distance in V metal, $d_{{\rm V}-{\rm V}}^{\rm metal} = 2.62\,$\AA! One could expect that in this case the molecular orbital picture should work.  Nevertheless the electron  correlation effects in vanadium ions, $3d$ ions, seem to be strong enough to  impose a different state, still with electrons more or less localized at V ions. This is an important general message. Usually one uses the qualitative criterion, according to which if the metal--metal distance in a given bond is of the order or smaller that the corresponding distance in the respective metal, then one should describe the electron state at this bond by molecular orbital picture. The list of metal--metal bonds in pure metals can be found e.g.\ in~\cite{Streltsov-UFN}. This  ``rule of thumb'' usually works qualitatively quite well, especially in $4d$ and $5d$ systems. However apparently this rule can be violated in $3d$ systems:  electron correlations in those can still be large enough to lead to a localized state of electrons, even with very short metal--metal distances. Thus e.g.\ in the trimer system Ba$_4$NbMn$_3$O$_{12}$, which, despite very short Mn--Mn distance of $2.47\,$\AA\ --- also shorter that the distance in metallic Mn --- behaves as a system with localized electrons~\cite{Streltsov2018a}, in particular it has a long-range antiferromagnetic ordering~\cite{Nguyen2019a}.

In LiVO$_2$ the magnetic--nonmagnetic transition occurs between two insulating states. Interestingly enough, when one replaces O by S, the corresponding transition in LiVS$_2$ becomes a real metal--insulator transition: the high-temperature state has an undistorted triangular lattice and is metallic, though with relatively correlated electrons, and the low-temperature state is actually the same as in LiVO$_2$ --- a diamagnetic state with V trimers~\cite{Katayama2009,Kojima2019}.  And LiVSe$_2$ remains metallic  down to the lowest temperatures~\cite{Katayama2009}. Apparently going from oxide to sulphide and then to selenide moves the system closer to the itinerant regime. It would be very interesting to carry out spectroscopic studies similar to those done for LiVO$_2$~\cite{Pen1997} to check which picture, that of molecular or  localized electrons, works better in which case.
\begin{figure}[t]
   \centering
  \includegraphics[width=0.49\textwidth]{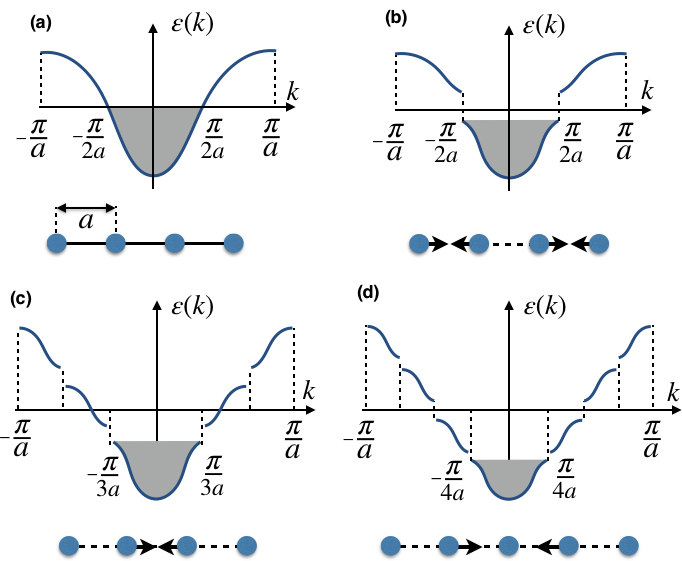}
  \caption{\label{Peierls}Sketch illustrating the Peierls effect in case of a chain with partially-filled nondegenerate (e.g.\ $s$) states. The band structure at half-filling is shown in~(a). This situation turns out to be unstable with respect to dimerization and the energy gap opens as shown in~(b). (c) and (d) demonstrate the same effect, but for band fillings 1/3 and 1/4, which leads to trimerization and tetramerization respectively.}
\end{figure}

\begin{figure}[t]
   \centering
  \includegraphics[width=0.41\textwidth]{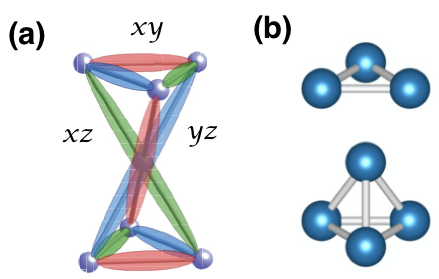}
  \caption{\label{AlV2O4}Two possible types of clusters formed by V ions in AlV$_2$O$_4$. Reproduced with permission from Ref.~\cite{Horibe2006,Browne2017}. Copyright 2006 and 2017 American Physical Society.}
\end{figure}

{\bf Spinel lattice ($\mathbf {3D}$).}  
One can also use this picture to explain even more complicated superstructures, for example the one found in the mixed valence spinel AlV$_2$O$_4$, with the average vanadium valence V$^{2.5+}$. Originally it was claimed that there appears in this system a superstructure with heptamers --- ``molecules'' consisting of seven V ions, see Fig.~\ref{AlV2O4}(a)~\cite{Horibe2006}. However in the recent pair distribution function study~\cite{Browne2017} it was found that in fact the middle V ion in a heptamer shifts toward one of the bases, so that instead of a heptamer there appears a tetrahedral cluster and a triangular one, Fig.~\ref{AlV2O4}(b). One can explain this by noting that there exist in AlV$_2$O$_4$ nominally two V$^{2+}$ ions with three $d$ electrons, and two V$^{3+}$ ($d^2$) ions. And the ions with three electrons form  tetrahedra (three bonds per V), whereas those with two electrons form triangles, two bonds per V (here one V$^{3+}$ remains ``alone'', not included in any cluster; this agrees with magnetic properties of AlV$_2$O$_4$~\cite{Horibe2006}). Our non-magnetic DFT calculations indeed confirm that such a superstructure, consisting not of regular heptamers but of tetrahedra and triangles, has the lowest energy.

\subsection{Orbitally-induced Peierls effect\label{Sec:Orbital-Peierls}}

The reduction of dimensionality is especially important because often 2D, and especially 1D systems display very unusual magnetic properties and lead, in particular, to specific novel types of phase transitions~\cite{Vasiliev2018,vasiliev-book}. Such is the famous Peierls dimerization in 1D systems, or, more generally, Peierls instability (not necessarily dimerization). In this situation the 1D metallic system, with, for example, one electron per site, forming a half-filled 1D band, would develop dimerization (alternation of short and long bonds with the respective large and small intersite hopping). This dimerization would open a gap at the Fermi-surface, so that the energies of the occupied electron states decrease, see Fig.~\ref{Peierls}(a,b). This decrease of energy is larger than the corresponding loss of elastic energy, so that formally this process should always occur in 1D metals with half-filled bands, see e.g.\ \cite{Bulaevskii:1975} for details. One can visualize this process as a first step in going from the 1D metal made for example by equally-spaced  hydrogen atoms, to the formation of hydrogen dimers --- H$_2$ molecules (in real quasi-one-dimensional materials, e.g.\ containing weakly coupled chains, some interchain coupling is usually still present, which, if strong enough, can suppress this instability). The same effect occurs not only for half-filled 1D bands leading to dimerization, but also for other fillings. Thus, e.g., 1/4-filled 1D band (one electron per two sites) would similarly lead to a Peierls distortion with tetramerization, and a 1/3-filled band would lead to trimerization, see Fig.~\ref{Peierls}(c) and~(d). It turns out that quite a few structural (and simultaneously often metal--insulator) transitions can be at least qualitatively explained by this mechanism. And orbitals, due to their directional character, can be very efficient in this: they can lead to an effective reduction of dimensionality, making the system for example electronically 1D --- after which the Peierls mechanism turns on and leads to the formation of superstructures such as dimers, trimers etc.
\begin{figure}
  \includegraphics[width=0.39\textwidth]{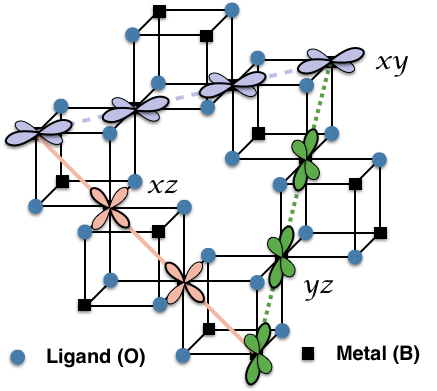}
  \caption{Crystal structure of normal spinels $AB_2L_4$, where transition metals occupy $B$ sites ($A$ ions are not shown). One can see that $t_{2g}$ orbitals should form 1D bands if one neglects hybridization via ligand $p$ states.}
  \label{Spinel} 
\end{figure}

\begin{figure}[t]
  \includegraphics[width=0.49\textwidth]{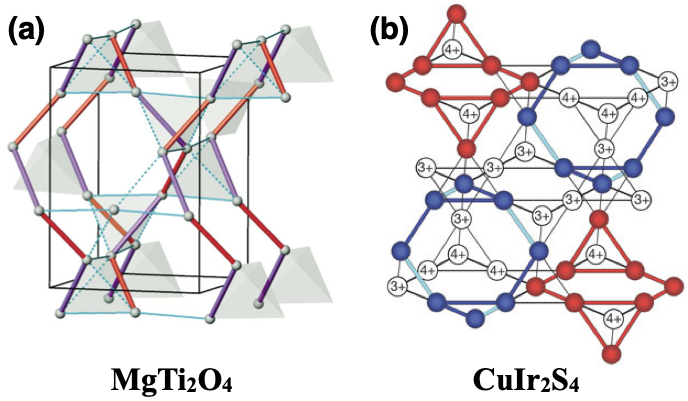}
  \caption{Crystal structures of MgTi$_2$O$_4$~(a) and CuIr$_2$S$_4$~(b) in low temperature distorted phases. (a)~Red and violet bonds  are respectively short and long bonds. (b)~Red balls are Ir$^{3+}$, blue balls are Ir$^{4+}$. Light blue bonds are Ir$_2$ singlet dimers. Reprinted by permission from Springer Nature ~\cite{Radaelli2002} and American Physical Society~\cite{Schmidt2004}. Copyright 2002 Springer Nature and 2004 American Physical Society.}
  \label{MgTi2O4-CuIr2S4}
\end{figure}

{\bf Spinels ($\mathbf {3D}$).} A very clear example of reduced dimensionality is provided by systems with local geometry resembling that in LiVO$_2$ --- in spinels MgTi$_2$O$_4$, ZnTi$_2$O$_4$, (Mg,Zn)V$_2$O$_4$ and in CuIr$_2$S$_4$. In the spinel $AB_2L_4$ (where $A$ and $B$ are two different metals, and $L$ is O, S, Se or Te) the $B$-sites, occupied in our systems by TM ions Ti, V or Ir sitting in $BL_6$ octahedra, form a lattice of corner-shared tetrahedra --- the so-called pyrochlore lattice (in real pyrochlores $A_2B_2$O$_7$ both $A$- and $B$–ions form such lattices, but the detailed connection of $AL_6$ and $BL_6$ octahedra is different). It can be seen from Fig.~\ref{Spinel} that one can visualize these cubic systems as containing a set of mutually perpendicular 1D chains running in the $xy$, $xz$, and $yz$ (and in $\bar {x}y$, $\bar {x}z$, and $\bar {y}z$) directions. This representation may seem rather artificial, but it acquires a meaning if we look at the shape of the $t_{2g}$ orbitals on this lattice, Figs.~\ref{LiVO2-orbitals},~\ref{Spinel}. One immediately sees that, similar to the case of LiVO$_2$ shown in Fig.~\ref{LiVO2-orbitals}(b), here the direct $d$--$d$ hopping becomes very anisotropic and orbital-selective. The $xy$ orbitals can hop to neighbouring sites in the $xy$ directions, i.e.\ in the $xy$ chains, the $xz$ orbitals can hop in the $xz$ chains, and the $yz$ orbitals in the $yz$ chains. In effect these materials, being really three-dimensional (they are typically cubic at high temperature), electronically become in this approximation quasi-1D\null!  And as such they can easily develop a Peierls-like distortion with the corresponding opening of the energy gap at the Fermi level in these 1D bands. This picture was proposed in \cite{Khomskii2005a} to explain phase transitions with the formation of beautiful superstructures observed in MgTi$_2$O$_4$~\cite{Schmidt2004}, and in CuIr$_2$S$_4$~\cite{Radaelli2002}. These superstructures are shown in Fig.~\ref{MgTi2O4-CuIr2S4}. In both these there occurs, with decreasing temperature, a transition from cubic metallic state to tetragonal insulating state, with extra superstructures appearing below~$T_c$. In MgTi$_2$O$_4$ short and long Ti--Ti bonds are formed, with short and long bonds forming spirals, as shown in Fig.~\ref{MgTi2O4-CuIr2S4}. An even more exotic superstructure is observed below $T_c$ in CuIr$_2$S$_4$: there are formed in it octamers made of Ir$^{3+}$ (nonmagnetic states $t_{2g}^6$, red in Fig.~\ref{MgTi2O4-CuIr2S4}b) and nominally magnetic Ir$^{4+}$ states ($t_{2g}^5$). And in Ir$^{4+}$ octamers there appears an extra distortion leading to the formation of short nonmagnetic Ir--Ir dimers (light blue bonds in Fig.~\ref{MgTi2O4-CuIr2S4}(b)).

The origin of these superstructures is not easy to understand proceeding from the cubic pyrochlore-type lattice. But it becomes immediately evident when we take into account that electronically these systems are 1D-like. In the cubic high-temperature phase there exist three sets of 1D bands, formed by the $xy$, $xz$, and $yz$ orbitals. After tetragonal transitions these levels/bands are split, and in MgTi$_2$O$_4$ with one $d$ electron per Ti the lower two $xz$ and $yz$ bands become both 1/4-filled. Correspondingly, the Peierls effect would lead to tetramerization in the $xz$ and $yz$ directions. This is exactly what happens in MgTi$_2$O$_4$: tetramerization in the $xz$ and $yz$ chains (alternation of short--intermediate--long--intermediate bonds). And in effect if we connect only short bonds with short bonds and long bonds with long bonds we would have these spirals of Fig.~\ref{MgTi2O4-CuIr2S4}. Thus this apparently strange pattern becomes quite simple if we only look at it from the right point of view, thinking about what happens in the 1D chains --- the natural building blocks of spinel structures with $t_{2g}$ electrons. Interestingly, more accurate model studies taking into account many-particle effects confirm this simple picture~\cite{DiMatteo2004,DiMatteo2005}.

Exactly the same physics explains an even more exotic octamer structure of CuIr$_2$S$_4$, Fig.~\ref{MgTi2O4-CuIr2S4}(b). In this case the upper 1D band turns out to be 3/4-filled, i.e.\ we would also get tetramerization. And indeed this is the periodicity along straight chains in octamers pattern of Fig.~\ref{MgTi2O4-CuIr2S4} (sequence is Ir$^{3+}$, Ir$^{3+}$, Ir$^{4+}$, Ir$^{4+}$).

The notion of orbitally-driven Peierls state \cite{Khomskii2005a} seems to work in these systems very well, at least qualitatively. Actually this is even somewhat surprising. This picture was based on a strongly simplified treatment --- taking into account only direct $d$--$d$ hopping and ignoring the hybridization and hopping via ligand (O, S) $p$ orbitals. But in fact this hopping is not at all small, especially in sulphides such as CuIr$_2$S$_4$. Nevertheless, even this simplified model turns out to be very successful in explaining the exotic superstructures obtained in MgTi$_2$O$_4$ and CuIr$_2$S$_4$.

\begin{figure}[t]
   \centering
  \includegraphics[width=0.47\textwidth]{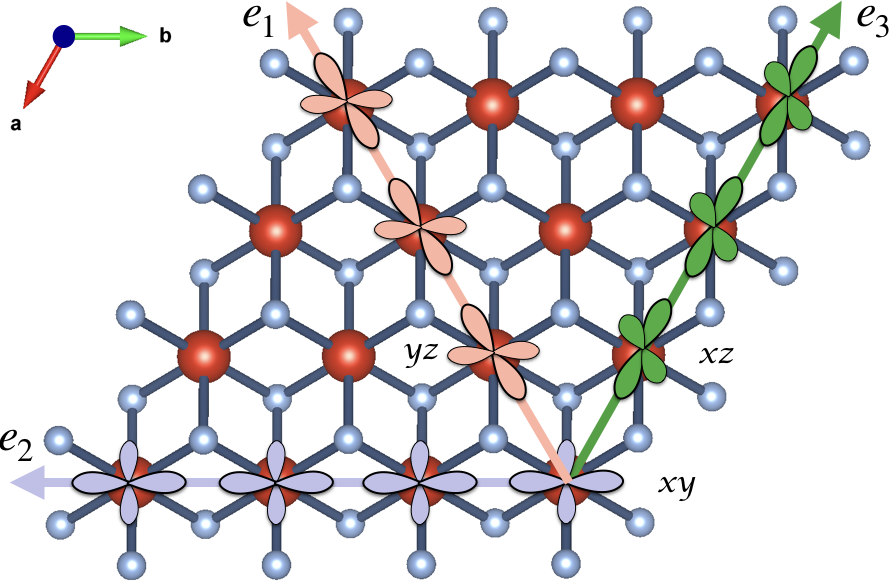}
  \caption{\label{Peierls-triang}Orbitally-driven Peierls effect in case of triangular lattice. Strong direct overlap between different $t_{2g}$ orbitals results in formation of three 1D bands, and Peierls phenomenon leads (for LiVO$_2$) to trimerization on each of
these chains (shown by different colors), cf. Fig.~\ref{LiVO2-orbitals}, where this trimerization is clearly seen.}
\end{figure}

{\bf Triangular lattice ($\mathbf {2D}$).}
Having this picture in mind we can also give a somewhat different explanation of superstructures observed in some other materials. Thus the superstructure with the formation of tightly bound triangles discussed above for LiVO$_2$ and LiVS$_2$ can also be understood from this point of view. Indeed also in triangular V layers in LiVO$_2$ shown in Fig.~\ref{Peierls-triang}  one would have for direct $d$--$d$ hopping the formation of three sets of one-dimensional bands, running in three directions $e_1$, $e_2$, and $e_3$ in a triangular layer at 120$^{\circ}$ with each other. For V$^{3+}$ 
with two electrons per site ($N_e = 2$) these 3 bands ($N_{b}=3$) would each be 1/3-filled ($N_e/(N_{b}N_{spin}) = 1/3$, since $N_{spin} = 2$), so that we should expect Peierls distortion with the trimerization in all three directions $e_1$, $e_2$, and~$e_3$. And indeed the superstructure with trimers in LiVO$_2$ shown in Fig.~\ref{LiVO2-orbitals}(b) exactly corresponds to such trimerization! Thus one would get the same superstructure proceeding both from the picture of localized electrons with orbital ordering \cite{Pen1997} as well as from the more itinerant point of view, with Peierls transition of partially-filled 1D bands.

\begin{figure}[t]
   \centering
  \includegraphics[width=0.5\textwidth]{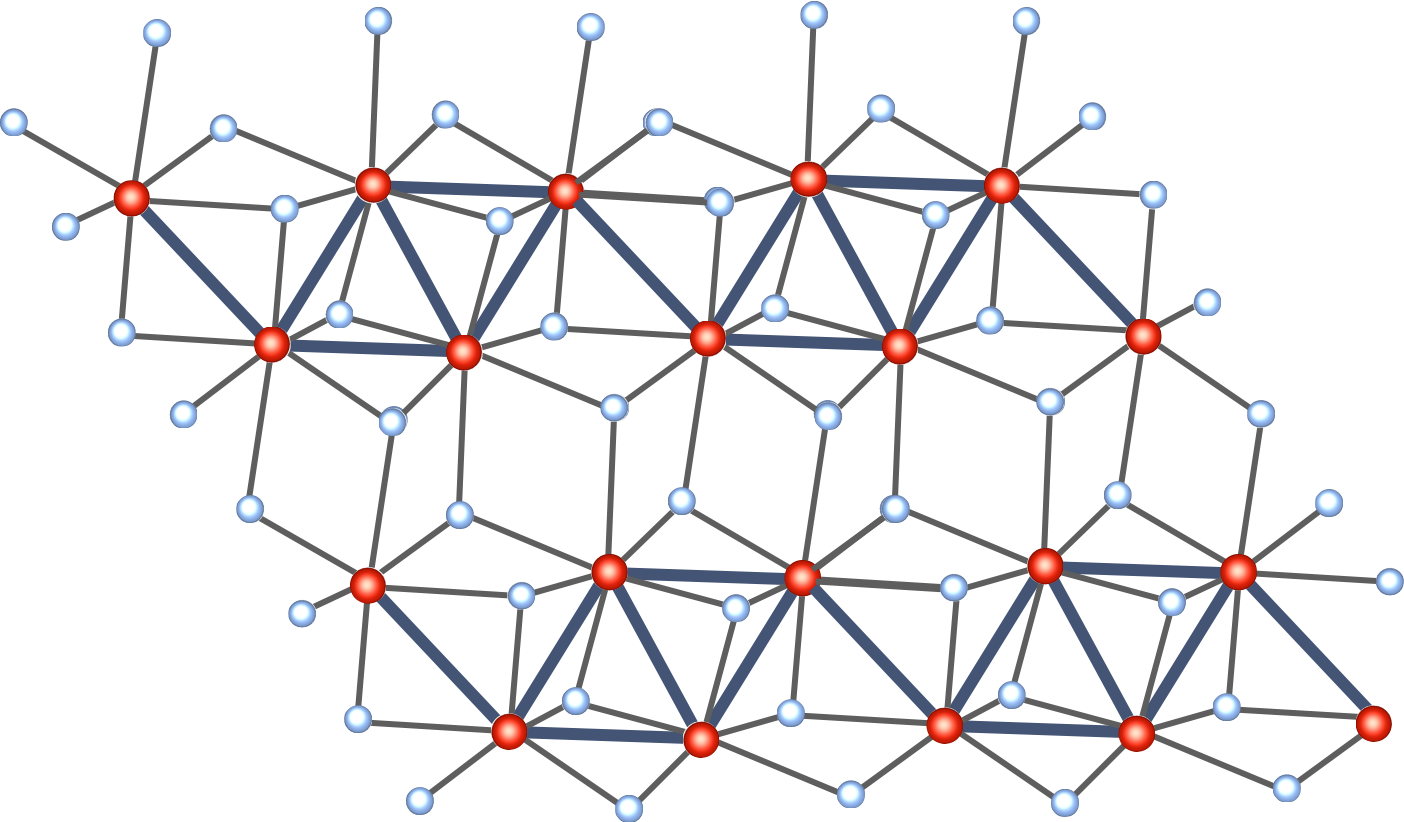}
  \caption{\label{ReS2}Diamond necklace crystal structure of ReS$_2$. Re ions are shown by red balls, S  by blue balls, and short Re--Re bonds by thick dark blue lines. }
\end{figure}

It is interesting that this treatment works also for other electron fillings. For example, various superstructures were observed in dichalcogenides~\cite{Wilson1975a}. One of the most beautiful ones was found in ReS$_2$ and ReSe$_2$~\cite{Meetsma1996}. This superstructure, shown in Fig.~\ref{ReS2}, is sometimes called ``diamond necklace''. In this case each Re, forming a triangular lattice, has not two, as in LiVO$_2$, but three $d$ electrons. There are now again three active ``1D'' bands, and for three electrons per Re each such band would be half-filled, so that one should expect dimerization in all three direction. And indeed the diamond necklace structure of Fig.~\ref{ReS2}  is exactly that --- looking at straight chains we see  the dimerization in each of them.  Of course not all superstructures observed in $MX_2$ dichalcogenides can be explained by this picture, especially those with transition metals not in octahedra but in prismatic coordination. But some of them very clearly demonstrate the physics described above.

\begin{figure}[t]
   \centering
  \includegraphics[width=0.41\textwidth]{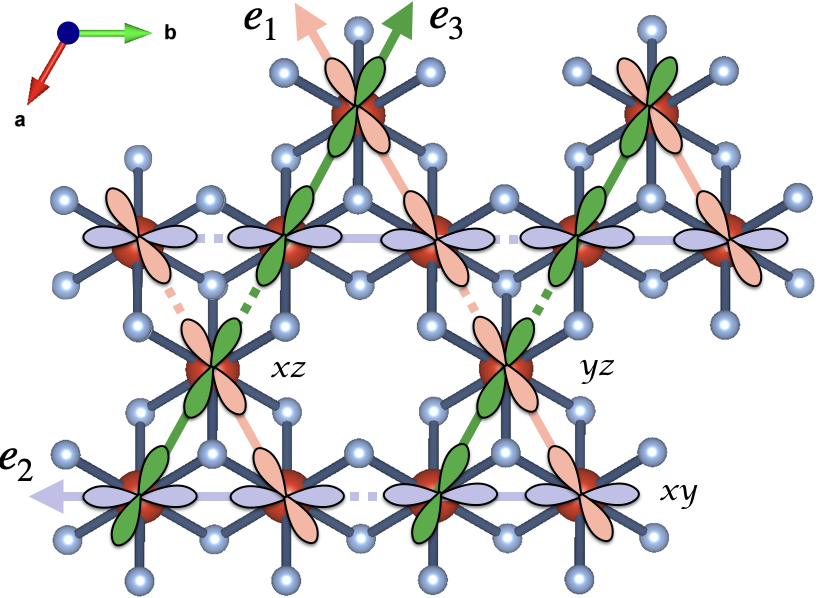}
  \caption{\label{Peierls-kagome}Formation of the breathing kagome lattice due to the orbitally-driven Peierls effect. Strong direct overlap between different $t_{2g}$ orbitals results in the formation of 1D bands, and Peierls phenomenon leads to dimerization in each of these chains. Short (long) metal--metal bonds are shown by solid (dashed) lines. Note, that in contrast to Figs.~\ref{LiVO2-orbitals}(a),~\ref{Spinel}  only the ``active'' (directed along bonds connecting metal sites) lobes of $t_{2g}$ orbitals are shown.}
\end{figure}

{\bf Kagome lattice ($\mathbf {2D}$).} Similarly one can also rationalize in this way the superstructures obtained in other systems, e.g.\ in Zn$_2$Mo$_3$O$_8$, which is a prototype of a large and very rich class of materials based on Mo$_3$O$_8$ clusters, in which such phenomena as giant optical diode effect~\cite{Yu2018}, valence-bond condensation and freezing of part of magnetic moments~\cite{Sheckelton2012,Haraguchi2015,Akbari-Sharbaf2018,Iida2019}, multiferroicity~\cite{Wang2015c,Kurumaji2017b,Streltsov2019,Solovyev2019,Tang2019} etc.\ were observed. In these systems Mo$^{4+}$ ($d^2$)  ions form basically a 2D kagome lattice, but it is a ``breathing kagome'', in which small and large triangles alternate, see Fig.~\ref{Peierls-kagome}. Each such triangle with two electrons per Mo, or again 6 per triangle, is exactly equivalent to small triangles in LiVO$_2$, Fig.~\ref{LiVO2-orbitals}(b), and these triangles are diamagnetic, with total spin $S_{tot} = 0$.   One usually takes this structure as given, not asking for the reasons for such breathing. But its origin  becomes clear  in the same picture of the Peierls  phenomenon. Also in this case the direct $d$--$d$ hopping would give 1D bands, running along all chains in the kagome lattice, see Fig.~\ref{Peierls-kagome}. But in contrast to LiVO$_2$, where each V participates in three bands (three intersecting chains) which gives rise to 1/3-filled bands and to trimerization, in a kagome lattice each site takes part only in two bands --- it lies at the intersection of two chains. In effect these two bands, with two electrons per site, would be half-filled each, so that one should expect dimerization in all chains. And indeed the breathing kagome structure corresponds exactly to that --- the period is doubled in the $e_1$, $e_2$, and $e_3$ directions. The dimerization in itself does not yet guarantee that precisely the breathing kagome structure would be realized: generally speaking the phases of different distortions, or accompanying charge density waves  in Zn$_2$Mo$_3$O$_8$ (as well as in LiVO$_2$)  could be different, which could in principle lead to different superstructures even with dimerization (or trimerization in LiVO$_2$)  in all three directions. Apparently it is higher order interactions which lock the phases of these charge density waves and lead to the formation of the observed structure.

The same physics, with small modifications, could also explain the formation of triangular molecules in another similar system with strongly breathing basic kagome lattice --- the molecular magnet Nb$_3$Cl$_8$. In contrast to Zn$_2$Mo$_3$O$_8$ here we have not 6, but 7 electrons per trimer. But apparently the strongest instability is again towards trimerization, the extra, 7th electron giving self-doping of these Nb$_3$ trimers, so that each of them, preserving basically the same structure, would have $S=1/2$. This is also what happens when one dopes Zn$_2$Mo$_3$O$_8$ for example adding Li, LiZn$_2$Mo$_3$O$_8$~\cite{Torardi1985a,Sheckelton2012}, or going to Li$_2$InMo$_3$O$_8$ or Li$_2$ScMo$_3$O$_8$~\cite{Torardi1985a,Haraguchi2015,Akbari-Sharbaf2018,Iida2019}: Mo$_3$ clusters survive although the degree of breathing may change, and the doped electron leads to the formation of a magnetic state of each triangle, with $S=1/2$ per trimer.

\subsection{General comments on effective reduction of dimensionality}
Thus, one can see that both approaches, the one based on the atomic limit and exploiting the idea of valence bond formation, and another one utilizing the concept of orbitally-induced Peierls effect, can be used to explain physical properties of real materials. We have shown on the example of LiVO$_2$ that both mechanisms may work equally well. Nevertheless, there are situations, e.g.\ AlV$_2$O$_4$, where only one of them is able to describe observed superstructures. 

Concluding this large section we would like to stress several important points concerning the influence of orbital degrees of freedom on the effective reduction of dimensionality in solids:

(i) The directional character of orbitals often leads to an effective reduction of dimensionality of systems with orbital degrees of freedom. This sometimes directly leads to the formation of well-defined clusters, ``molecules in solids'' (see the example of NaTiSi$_2$O$_6$ above), or it can make the system essentially one-dimensional, after which the Peierls effect takes place, also leading to the formation of dimers, trimers etc.

(ii) There are also other arguments, besides the ones presented in Secs.~\ref{Sec:VBC} and ~\ref{Sec:Orbital-Peierls}, to explain observed superstructures. E.g.\ a phenomenological concept of metal--metal bonds is often widely used in a chemical literature. In the geometries with ML$_6$ octahedra with common edge or common face the metal--metal distances can be relatively short, and in this case one may expect the formation of metal--metal bonds. And if we have different $t_{2g}$ orbitals at each site, due to their directional character each orbital can form a bond in its own direction. Then one can have the situation when there would be as many such (strong) bonds formed by a given site as there are active $d$ electrons on it. Indeed, triangles in LiVO$_2$ and Zn$_2$Mo$_3$O$_8$, with $d^2$ ions, are the structures with two bonds per site. Similarly, the diamond necklace of ReS$_2$, Fig.~\ref{ReS2}, corresponds to each Re ($d^3$) forming exactly three strong (short)  metal--metal bonds. 
 
A description of such clusters is not so trivial as it looks at first sight. The simple ``rule of thumb'' is to compare the metal--metal distance in a cluster with the distance in the corresponding metal. If these are comparable or if the distance in a cluster is even shorter, one often has the situation in which the electrons are delocalized within the cluster and can be described by the molecular orbital picture. But this rule, although it often works, or at least gives a good qualitative feeling of what we can expect, can sometimes break down, especially in $3d$ systems, in which correlation effects can still be strong enough to lead to electron localization. Most probably the situation in many cases is actually ``in-between'': the electrons in such clusters are half-way between being completely itinerant in a cluster and being localized at respective centres.

(iii) In the present section, discussing Peierls-like phenomena and the formation of metal clusters in solids we mostly illustrated these effects on the example of materials which became popular and were studied relatively recently. However the story actually goes rather far back. One of the first, and still very important system showing this effect is VO$_2$, in which the strong metal-insulator transition,  occurring in the very convenient temperature range, slightly above room temperature ($T_C$=340 K), is driven, or accompanied by the formation of V dimers along chains in the $c$ direction of the rutile structure~\cite{Imada1998,khomskii2014transition}. Orbital repopulation plays very important role at this transition in VO$_2$~\cite{Haverkort2005}.  Interestingly, such dimers with strong metal-metal bonds can not only form valence-bond crystals, as in most examples discussed above, but can also exist in a kind of a dimer liquid -- the famous example being the resonating valence bond state (RVB) of P.W. Anderson~\cite{Anderson1973}. Experimentally such state was discovered for example in a Magneli phase Ti$_4$O$_7$~\cite{Lakkis1976}.

(iv) Yet one more interesting effect is connected with the cluster (most often dimer) formation. As the metal-metal distance in such dimers is usually smaller than the corresponding distance in the original material, formation of such dimers is usually accompanied by the decrease of the sample volume, i.e. such dimers may be stabilized by the external pressure. And indeed, such phenomenon -- a transition from the regular to the dimer structure under pressure – was discovered in several materials: in the first heavy fermion $3d$ system LiV$_2$O$_4$~\cite{Pinsard-Gaudart2007}, in the very popular nowadays Kitaev materials (see Sec.~\ref{sec:SOCexchange}): in different forms  ($\alpha$, $\beta$) of Li$_2$IrO$_3$ \cite{Hermann2018,Veiga2019}, in $\alpha-$RuCl$_3$~\cite{Bastien2018}. In this sense these honeycomb systems under pressure become nonmagnetic, resembling a similar system with Ru instead of Ir or Rh, Li$_2$RuO$_3$. This compound is nonmagnetic at low temperature due to dimerization~\cite{Miura2007}. Above 540~K a magnetic response appears and the crystal structure as measured by X-ray diffraction looks like nondimerized, but close inspection using pair distribution function analysis shows that Ru dimers exist even at 540~K, but seems to flow over the lattice so that in average one sees the uniform structure~\cite{Kimber2013}. This interpretation is consistent with recent results of Raman and NMR spectroscopy~\cite{Arapova2017,Ponosov2019a,Ponosov2020}.  In the formation of dimerized structure in Li$_2$RuO$_3$ orbital degrees of freedom also plays crucial role~\cite{Jackeli2008,Miura2009,Pchelkina2015a}. Interestingly enough, as dimer state is often insulating whereas the starting material is metallic, this may present a rather unique situation, in which  there occurs   under pressure a transition from metal to insulator, contrary to the usual situation where pressure rather works in favour of metallic state.

(v) These phenomena usually occur in the vicinity of a Mott transition, in case when the electron interaction (e.g.\ the Hubbard~$U$) and the kinetic energy of electrons (measured by the electron hopping $t$ or the resulting bandwidth $W = 2zt$) are comparable.  In fact Mott transitions  can occur not only by transforming the whole system directly from one state, e.g.\ a Mott insulator, to a homogeneous metal, but rather ``in stages'': first the electrons are delocalized in specific small clusters (dimers, trimers, etc.), in which they behave practically as itinerant, but the whole system is still insulating because the hopping between clusters remains small. And only later, e.g.\ at higher pressures, can the system  become a homogeneous metal. Thus such spontaneous formation of clusters with itinerant electrons, described e.g.\ by the molecular-orbit picture, typically occurs when the system is close to localized--itinerant crossover, i.e.\ close to Mott transition.

(vi) The situation in this respect can be different for different orbitals: some of them, with stronger overlap, may be already ``on the itinerant side'' of the Mott transition, whereas the others can still remain largely localized. More detailed discussion of cluster Mott insulators, especially with orbital-selective delocalization,  will be given in the next section.

\section{Cluster Mott insulators or ``molecules'' in solids~\label{Sec:CMI}}
\subsection{Orbital-selective behaviour~\label{Sec:OSB}}
\begin{figure}[t]
   \centering
  \includegraphics[width=0.5\textwidth]{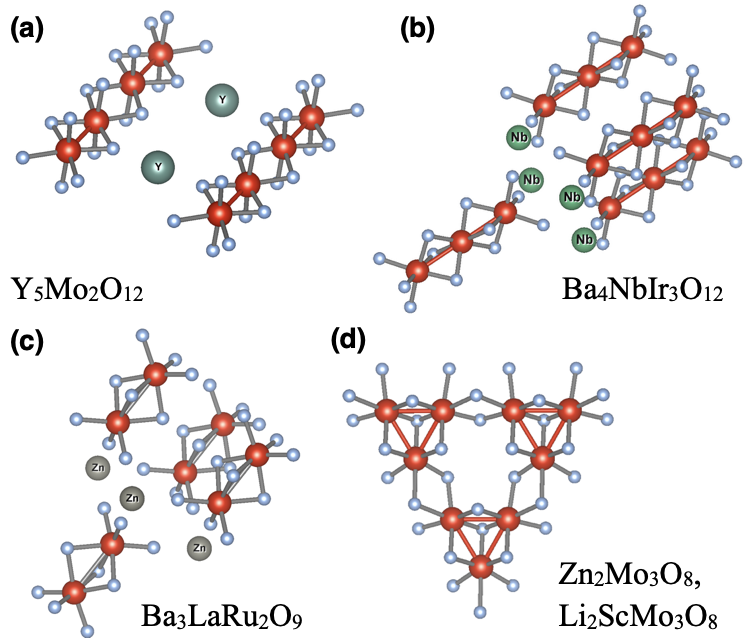}
  \caption{\label{Clusters-examples}Examples of materials with various structural clusters of transition metal ions: (a)~Mo dimers in Y$_5$Mo$_2$O$_{12}$~\cite{Torardi1985}; (b)~Ir trimers in Ba$_4$NbIr$_3$O$_{12}$~\cite{Nguyen2019b},  (c) Ru~dimers in Ba$_3$LaRu$_2$O$_9$~\cite{Senn2013a} and (c) Mo~trimers in Zn$_2$Mo$_3$O$_8$~\cite{Cotton1964} or Li$_2$ScMo$_3$O$_8$~\cite{Haraguchi2015,Akbari-Sharbaf2018,Iida2019}. Transition metal ions are shown by large red balls, while oxygens by smaller light blue balls.}
\end{figure}

In the previous section we largely discussed spontaneous formation  of different superstructures, in particular due to a reduction of effective dimensionality caused by the directional character of $d$ orbitals involved. Often this resulted in the formation of well-defined clusters --- dimers, trimers, etc. However in many cases similar clusters exist  just due to the crystal structure itself.  In Fig.~\ref{Clusters-examples} we show several such examples. One can show that also in such ``preformed'' clusters orbital degrees of freedom still  can play a significant role and can largely determine their properties. In particular, different orbitals can behave differently: orbitals directed towards neighbours, say in a dimer or a trimer, may behave as delocalized and should be described by molecular orbitals, whereas the electrons on other orbitals may still be localized.

The simplest example of such behaviour, considered in \cite{Streltsov2014,Streltsov2016b}, is that of a dimer with two orbitals: one, orbital~$a$, with the lobes along the metal--metal bonds and with very strong inter-site hopping~$t$,  and another, orbital~$b$, with lobes perpendicular to the bond and with weak hopping~$t'$ (for simplicity $t'$ can be taken as~0). In the case of a common edge geometry $a$ are $xy$ orbitals, while $b$ can be $xz$ or $yz$ orbital. For a common face $a=a_{1g}$ and $b$ is one of the $e_g^{\pi}$ orbitals, see Fig.~\ref{Different-sharing}. 

\begin{figure}[t]
   \centering
  \includegraphics[width=0.49\textwidth]{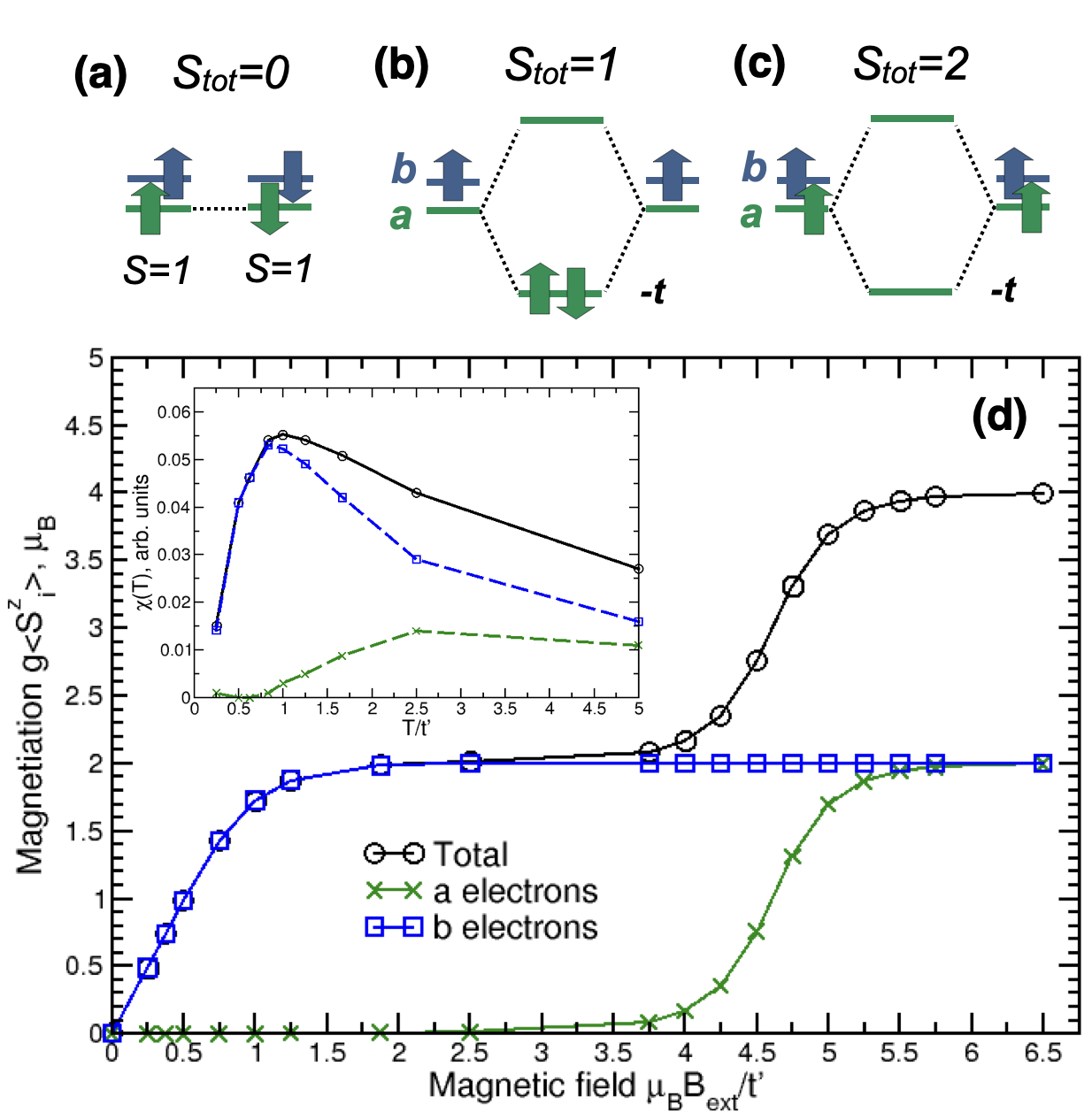}
  \caption{\label{Orbital-selective}(a--c) Three possible states of a dimer with two orbitals (shown by different colors) and two electrons per site. The hopping between $a$ orbitals is $t$, while between $b$ ones is $t'\ll t$. (d) Field dependence of magnetization as obtained in cluster DMFT for chain of such dimers. Inset shows magnetic susceptibility as a function of temperature. For detailed description see Ref.~\cite{Streltsov2014}.}
\end{figure}

Let us illustrate some implications of the presence of two types of orbitals. Consider first the situation with two electrons per site. In the usual treatment we expect that these two electrons at each site first form a state with spin $S=1$, according to the (first) Hund's rule. And then these $S=1$ states interact due to virtual $d$--$d$ hopping and finally form a state with opposite spins (more accurately, a total singlet state made of two antiferromagnetically coupled $S=1$ states at each site). This state thus gains the full Hund's energy, $-J_H$ at each site, and antiferromagnetic coupling adds the energy gain $\sim t^2/(U+J_H)$, see Fig.~\ref{Orbital-selective}(a).  As a result the energy of this state is
\begin{eqnarray}
\label{largeS-state}
E_{\rm I} = -2J_H - \frac {2t^2}{U+J_H}.
\end{eqnarray}

However, if the bonding-antibonding splitting $2t$ is larger than~$U$ and~$J_H$, one can form a different state,  shown in Fig.~\ref{Orbital-selective}(b). First one makes a singlet bond out of two orbitals with strong hopping. One gains by that the bonding energy ($-2t$ for two electrons).  The electrons at the other orbital remains localized, but the spin at a site is now not~1, but only~1/2. One loses in this state the Hund's energy (though not completely, as there is still some probability that the electron forming a singlet bonding state resides at each site and interacts with the remaining localized electron by the Hund's exchange). As a result the energy of this state is 
\begin{eqnarray}
\label{smallS-state}
E_{\rm II}=- 2t -J_H. 
\end{eqnarray}
Comparing these energies we see that the second state \eqref{smallS-state} (singlet bonding state of strongly overlapping orbitals, with the remaining localized electron with $S=\frac12$ per site) has lower energy than the state with $S=1$ per site \eqref{largeS-state}, if hopping $t \gtrsim J_H/2$. That is, for sufficiently strong intersite hopping for one of the orbitals the electrons at this orbital form a singlet bonding state and ``drop out of the game'', and the other electron(s) remain localized and can in particular form a magnetic ordering --- but with a strongly reduced magnetic moment (here $S=\frac12$ per site instead of $S=1$). We lose by this a part of the Hund's energy, but gain more on the bonding energy of ``itinerant'' electrons. This situation is actually rather typical for many magnetic systems made from $4d$ and $5d$ ions. These materials are often metallic --- metallic compounds are much more common among $4d$ and $5d$ compounds than among $3d$ ones. But if such $4d$ and $5d$ systems are insulating and become magnetic, typically they have magnetic moments much smaller than the nominal moments corresponding to the formal valence of respective ions. The mechanism described above is one possible mechanism explaining this behaviour. 

The model calculations of a chain consisting of such dimers using the cluster version of the Dynamical mean field theory (DMFT) confirm this picture~\cite{Streltsov2014}. Thus, in Fig.~\ref{Orbital-selective} we show how, in this situation, magnetization evolves under external magnetic field. One sees that we indeed start with the solution without any net magnetization. Then first it increases and reaches the value of  $2 \mu_B$ per dimer, which corresponds to polarization by a field of localized spins provided by localized ($b$) electrons ($S=\frac12$ and magnetic moment $1\mu_B$ per site). And only at much higher fields, sufficient to break the singlet bond, the other ($a$) electrons on the ``hopping'' orbitals (forming a singlet bonding state) become polarized, and the total moment reaches $4 \mu_B$ per dimer, i.e.\ $2 \mu_B$ per site, see Fig.~\ref{Orbital-selective}(c). Thus the system demonstrates orbital-selective behaviour, with different orbitals reacting differently to external perturbation depending on the ``strength'' of this perturbation. A very similar effect can be seen e.g.\ in the temperature dependence of magnetic susceptibility, when only the $b$ orbital is responsible for the low-temperature behaviour, while at higher temperature the $a$ orbitals starts to contribute, see inset in Fig.~\ref{Orbital-selective}.

\begin{figure}[t]
   \centering
  \includegraphics[width=0.49\textwidth]{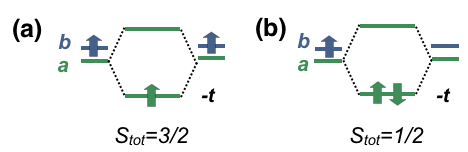}
  \caption{\label{DE}Illustration of two possible states of a dimer with two orbitals per site and three electrons. State (a) is an analogue of the double exchange in a concentrated case, while configuration (b) can be considered as a molecular low spin state. Green arrows are mobile electrons, blue ones are localized states.}
\end{figure}

This physics can lead to even stronger effects in the case of not~2, but 1.5 electrons per site (i.e.\ 3 electrons per dimer). This is the situation typically considered as leading to the famous double exchange (DE) --- the ferromagnetic coupling between localized electrons provided by the hopping of a more itinerant electron. This is the usually invoked mechanism of ferromagnetic exchange not only in metallic oxides such as colossal magnetoresistance  manganites La$_{1-x}$Sr$_x$MnO$_3$ and similar compounds~\cite{Izyumov2001} or in CrO$_2$ \cite{Korotin1998}, but also even in ferromagnetic metals such as Fe or Co. This mechanism is illustrated in Fig.~\ref{DE}(a): again two electrons are localized on ``non-hopping'' orbitals $b$, one at each site, and the mobile electron at the orbital $a$ hops from site to site, its spin being oriented parallel to the spin of localized electron at each site, i.e.\ all three spins are parallel, so that the total spin of a dimer is $S_{\rm tot}=\frac32$.  Thus the mobile electron ``forces''  localized spins to be parallel. This is the double exchange mechanism of ferromagnetism in conducting solids (and the presence of a small energy gap does not destroy this mechanism~\cite{Nishimoto2012}).  The energy of this state is 
\begin{eqnarray}
\label{DE-state}
E_{\rm DE} = -t - J_H,
\end{eqnarray}
while the competing state, shown in Fig.~\ref{DE}(b), with minimal possible spin would have energy $E_{\rm MO} = -2t - J_H/2$. Comparing these energies we see that such state with reduced magnetic moment is definitely favoured if the hopping, or bonding--antibonding splitting $\sim t$ is larger than the Hund's interaction~$J_H$.

As it happens, Zener first proposed this mechanism just on the example of a dimer which we consider now~\cite{Zener1951}. In the standard treatment of the double exchange mechanism of ferromagnetism described above one always considers the situation with the Hund's coupling much larger than all the hoppings, $J_H\gg t$. In fact $J_H$ is usually taken as infinite, and it does not even enter the final expressions in this field~\cite{Anderson1955,DeGennes1960}. This is a good approximation for $3d$ materials, but, as discussed above, it can break down for $4d$ and especially for $5d$ systems. In those it can happen that the hopping~$t$, or eventual bonding--antibonding splitting~$2t$ is of the order of or exceeds the Hund's energy~$J_H$. In this case we can form a different state: we redistribute the electrons, forming a bonding state of itinerant orbitals and putting two electrons in such bonding state, of course with antiparallel spins. The remaining electron with spin $\frac12$ may be localized at each site, or can be distributed between them due to the remnant weak hopping~$t'$, but in any case the total spin of a dimer with three electrons would be now not $\frac32$ as in the double exchange picture of Fig.~\ref{DE}(a), but only $S=\frac12$, Fig.~\ref{DE}(b). Thus, this orbital-selective formation of singlet bonding states actually suppresses the double exchange mechanism of ferromagnetism.
\begin{figure}[t]
   \centering
  \includegraphics[width=0.49\textwidth]{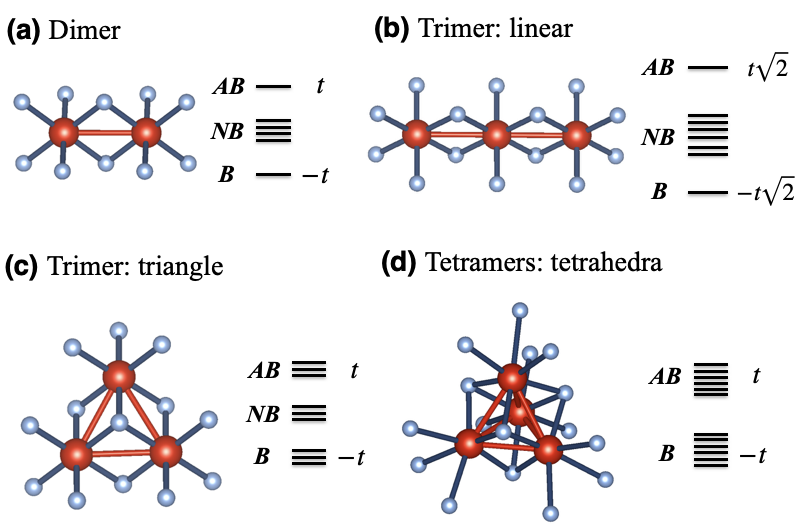}
  \caption{\label{MO-in-clusters}Formation of molecular orbitals in different types of clusters of transition metal ions (red balls) with non-completely filled $t_{2g}$ shell. In all situations only direct $d$--$d$ hopping (along red bonds), $t$, is taken into account. There is also hopping due to overlap between $d$ orbitals via ligand $p$ orbitals in real materials. This effective hopping can lift the degeneracy of corresponding molecular orbitals. Here, $B$ stands for bonding, $A$ for antibonding and $NB$ for nonbonding orbitals.}
\end{figure} 

In fact the competing states, described above, are nothing else but different spin states. However, these are different spin states not of a transition metal ion (such as the high spin and low spin states of Co$^{3+}$ or Fe$^{2+}$), but spin states or multiplets of the whole cluster. It is important to mention that the competition between these spin states affects not only the magnetic moment of a cluster, but may also change other magnetic characteristics, e.g.\ the temperature of magnetic ordering. It was shown in Ref.~\cite{Harland2019} that spin and orbital fluctuations in the vicinity of such spin state crossover substantially suppress N\'eel temperature in cluster Mott insulators.
\begin{table*}[t!]
\centering \caption{\label{Tab:Orbital-Selective}Examples of cluster Mott insulators showing orbital-selective behaviour with strong suppression of magnetic moments. Theoretical moments are calculated using formal valence of corresponding transition metals. $\mu$ and $\mu_{\it eff}$ stands for local and effective (in Curie-Weiss theory) magnetic moments.}
\vspace{0.2cm}
\begin{tabular}{llcccccc}
\hline
\hline
Material                            &Ionic    & Theor.                  & Exp.              \\                                                 
                                         &Conf.   & moment               & moment         \\
\hline
Y$_5$Mo$_2$O$_{12}$   & $4d^{1.5}$    & $\mu_{eff} = 2.3 \mu_{B}$/Mo & $\mu_{eff} = 1.7 \mu_{B}$/Mo \cite{Torardi1985}\\
Nb$_2$O$_2$F$_3$         & $4d^{1.5}$    & $\mu_{eff} = 3.9 \mu_{B}$/dimer & $\mu_{eff}\approx 2\mu_{B}$/dimer\cite{Tran2015}\\
$\alpha-$MoCl$_4$          & $4d^{2}$    & $\mu_{eff} = 2.8 \mu_{B}$/Mo & $\mu_{eff}\approx 0.9\mu_{B}$/Mo\cite{Kepert1968}\\
Ba$_3$YRu$_2$O$_9$ & $4d^{3.5}$   & $\mu=2.5 \mu_B/{\rm Ru}$  & $\mu =0.5 \mu_B/{\rm Ru}$ \cite{Senn2013a}\\
Ba$_3$LaRu$_2$O$_9$ & $4d^{3.5}$   & $\mu=2.5\mu_B/{\rm Ru}$  &$\mu =1.4 \mu_B/{\rm Ru}$ \cite{Senn2013a}\\
Ba$_5$AlIr$_2$O$_{11}$ & $5d^{4.5}$   & $\mu_{\it eff}= 3.3\mu_B/{\rm dimer}$  & $\mu_{\it eff} \approx 1 \mu_B/{\rm dimer}$ \cite{Terzic2015,Streltsov2017}\\
Ba$_4$NbRu$_3$O$_{12}$ & $4d^{4.7}$   & $\mu_{\it eff} \approx 4.3\mu_B/{\rm trimer}$  &$\mu =2.6 \mu_B/{\rm trimer}$ \cite{Nguyen2018a}\\
\hline
\hline
\end{tabular}
\end{table*}

While this effect of N\'eel temperature suppression has not yet been confirmed experimentally, there are many examples of strong reduction of magnetic moments (with respect to ionic values) in cluster Mott materials due to the orbitally-dependent formation of bonding singlets with the suppression of double exchange. This phenomenon is in fact rather common among  $4d$ and $5d$ compounds. A very clear example is presented by Y$_5$Mo$_2$O$_{12}$~\cite{Torardi1985}. This system contains linear chains of MoO$_6$ octahedra with common edge, see Fig.~\ref{Clusters-examples}(a), but this chain is very strongly dimerized (probably by the mechanism described in the previous section), so that actually it can be considered as consisting of almost isolated dimers. The valence of Mo here is $4.5+$, i.e.\ there are 3 electrons per dimer, as in the example considered above. If the system were in the double exchange state, the nominal effective magnetic moment per dimer (in susceptibility, $\mu_{\it eff} = 2 \sqrt {S(S+1)}\,$) would be $3.87\mu_B$, but experimentally it is only $0.8\mu_B$/dimer. The mechanism presented above explains this behaviour. {\it Ab initio} calculations indeed show that there appears in this system a bonding state containing two electrons in a singlet state, and the remaining moment is only $0.87\mu_B$~\cite{Streltsov2015MISM}. These calculations show that the bonding--antibonding splitting in Y$_5$Mo$_2$O$_{12}$ is $2.7\,\rm eV$ --- definitely larger than the Hund's energy, which for $4d$ elements is $\sim0.6$--$0.7\,\rm eV$. Thus, indeed here at least these bonding orbitals are  in the regime of metal--metal bonds, which overcomes the Hund's energy and suppresses double exchange. An even stronger effect is seen in the similar material Y$_5$Re$_2$O$_{12}$~\cite{Chi2003}. Here we have Re$^{4.5+}$($d^2/d^3$), which, if the double exchange picture applied here, would give total $S_{\rm tot}=\frac52$ per dimer. Experimentally we have here dimers with $S_{\rm tot}=\frac12$! That is, two electrons per Re (4 per dimers) make singlet MOs, the remaining one $d$ electron giving $S_{\rm tot}=\frac12$ per dimer.

There are in fact many examples of the orbital-selective behaviour in real materials, some of which are summarized in Table~\ref{Tab:Orbital-Selective}. A rather large variety of possible orbital-selective states is observed in compounds with the general formula Ba$_3AM_2$O$_9$, where the two transition metal ions $M$ form dimers sharing common faces, as shown in Fig.~\ref{Clusters-examples}(c), and the $A$ ion determines the number of electrons per this dimer (it can be a mono-, di-, tri-, or even four-valent  ion). One of the situations most studied in the recent years is when $A$ is 3+ and $M$ is a Ru ion. Since in this case Ru is $4.5+$ ($4d^{3.5}$) one would naively expect that the total spin of the dimer would be $S_{tot}=\frac52$ (six $t_{2g}$ electrons with spin up, one with spin down) and the magnetic moment would be $2.5 \mu_B/{\rm Ru}$. However, experimentally very different magnetic moments were observed in these materials: from 0.5 up to 1.4$\mu_B$~\cite{Senn2013a}. Further analysis has shown that two $d$ electrons are always in the lower-lying molecular orbital of the $a_{1g}$ symmetry, while the others can occupy the $e_g^{\pi}$ orbitals in different ways so that very different molecular spin states can be stabilized by slightly different local distortions (due to different size of the $A$ ions). For example  it was proposed that in Ba$_3$LaRu$_2$O$_9$ we have $S_{tot}=\frac32$ per dimer~\cite{Chen2020}, while in Ba$_3$YRu$_2$O$_9$ only $S_{tot}=\frac12$~\cite{Ziat2017}. The presence of several low-lying spin states strongly affects magnetic properties of cluster magnets. It was shown in Ref.~\cite{Li2020} that thermal population of these states may result in non-trivial temperature dependence of magnetic susceptibility with deviation from the Curie--Weiss law with $\mu_{\it eff}(T)$ (or, alternatively, one may consider this as temperate-dependent Van Vleck paramagnetism). 

Other Ba$_3AM_2$O$_9$ compounds also have much smaller magnetic moments than one would expect from the formal valence of the transition metal ion: in Ba$_3$CeRu$_2$O$_9$ with Ru$^{4+}$ ($S=1$) the total spin per dimer was also found to be not $S_{tot}=2$, but  $S_{tot}=1$~\cite{Chen2019}. In Ba$_3$TiRu$_2$O$_9$ with the same Ru$^{4+}$ the effective moment was measured to be $\sim 1.8 \mu_B/{\rm dimer}$, again very different from both what one would expect for isolated Ru$^{4+}$ ion or for a dimer with $S_{tot}=2$.  Ba$_3$CeIr$_2$O$_9$ with Ir$^{4+}$ ($5d^5$), where due to strong spin--orbit coupling we expect an effective total moment $j_{\it eff}=1/2$, see also Sec.~\ref{SOC}, turns out to have a nonmagnetic ground state, since both holes reside on antibonding $a_{1g}$ molecular orbitals~\cite{Revelli2019}. Strong bonding between Ir$^{5+}$ ions results in magnetic response in Ba$_3$ZnIr$_2$O$_9$ \cite{Nag2019}, whereas typically the isolated Ir$^{5+}$ ($t^4_{2g}$) ions are nonmagnetic having $J=0$, see Sec.~\ref{SOC}. Orbital-selective behaviour is probably responsible for charge ordering in Ba$_3$NaRu$_2$O$_9$, where two types of dimers: Ru$^{5+}$--Ru$^{5+}$ and Ru$^{6+}$--Ru$^{6+}$ were observed~\cite{Kimber2012}. This can be explained by the compensation of energy costs due to charge disproportionation by the energy gain due to formation of one very short and another slightly longer dimer (this gain is achieved because of strong non-linear dependence of electron hopping, i.e.\ bonding--antibonding splitting, on distance, similar to the situation in Nb$_2$O$_2$F$_3$~\cite{Tran2015,Gapontsev2016}).

Finally, it has to be mentioned that the orbital-selective behaviour is, of course, typical not only for dimerized, but also for other cluster magnets. In Fig.~\ref{MO-in-clusters} we present the examples of different possible types of transition metal clusters with active $t_{2g}$ orbitals, where one might expect the formation of molecular orbitals due to direct $d$--$d$ hopping ($e_g$ orbitals are directed to ligands and thus they are not prone to form molecular orbitals). Ba$_4$Ru$_3$O$_{10}$ and Ba$_4$Nb(Ir,Rh,Ru)$_3$O$_{12}$ are the examples of materials with linear trimers, see see Fig.~\ref{Clusters-examples}(b). In Ba$_4$Ru$_3$O$_{10}$ the formation of molecular orbitals results in a situation when the middle ion turns out to be nonmagnetic~\cite{Klein2011,Streltsov2012a}. In Ba$_4$Nb(Ir,Rh,Ru)$_3$O$_{12}$ it again leads to suppression of total magnetic moment~\cite{Nguyen2018a,Nguyen2019b}. Triangular clusters can be met e.g.\ in $A_xB_y$Mo$_3$O$_8$ systems, which will be considered in Sec.~\ref{Sec:PCO}, while tetramer clusters are met in square plaquettes, e.g.\ in CaV$_4$O$_9$~\cite{Starykh1996} but they are more typical for spinels (e.g.\ AlV$_2$O$_4$ discussed in Sec.~\ref{Sec:D-reduction}) or for lacunar spinels. On the example of these materials one can show that the Jahn--Teller physics discussed previously in Sec.~\ref{sec:JT} for isolated ions can be very important for cluster magnets as well.

\subsection{Jahn--Teller effect in cluster magnets}
\begin{figure}[b]
   \centering
  \includegraphics[width=0.5\textwidth]{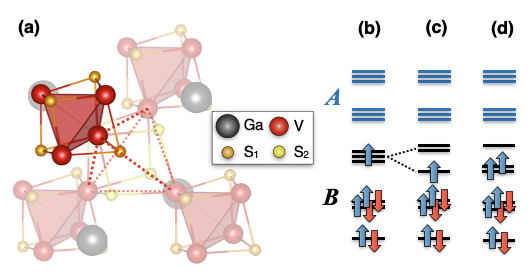}
  \caption{\label{Lacunar-spinel}(a) Crystal structure of lacunar spinel GaV$_4$S$_8$ (taken from \cite{Kim2018}). (b) Electronic structure of transition metal tetramers (or V$_4$S$_4$ ``cubanes''): due to hopping via ligands the initial bonding ($B$, shown in black) and antibonding ($A$, shown in blue) levels of Fig.~\ref{MO-in-clusters}(d) are split. Further splitting of a bonding triplet due to the ``cluster'' Jahn--Teller effect in GaV$_4$S$_8$ and GeV$_4$S$_8$ is shown in (c) and (d). Left part of the figure is reproduced with permission from Ref.~\cite{Kim2018}. Copyright 2020 American Physical Society.}
\end{figure}

Lacunar or $A$-site deficient spinels are $AB_2$L$_4$ spinels  in which half of $A$-ions are missing, so that in effect we have, e.g., Ga$_{1/2}$$\square_{1/2}$V$_2$S$_4$ = GaV$_4$S$_8$. The remaining $A$-ions and $A$-vacancies are ordered, and the structure is that of a cubic spinel but with the symmetry not Fd-3m as usual but F-43m~\cite{Pocha2000}.  In this structure the vanadium tetrahedra are not all equivalent: small and large tetrahedra alternate, see Fig.~\ref{Lacunar-spinel}(a). Thus if the V--V distance in small tetrahedra would be (significantly) smaller than in the large ones, one could consider these systems as containing tightly bound  clusters, in this case V$_4$S$_4$ ``cubane'' clusters, or V$_4$ tetrahedra, weakly coupled with each other. There  exist many materials of this type: GeV$_4$S$_8$, GaMo$_4$Se$_8$ etc. They can be visualized as breathing-type spinels, and there exist also similar real pyrochlores $A_2B_2$O$_7$ with breathing-type pyrochlore lattices, see e.g.~\cite{Kimura2014}.

When describing these systems, we again meet the problem discussed above: should we treat $B_4$, e.g.\ the V$_4$ tetramers, in the molecular-orbital picture, or do the TM ions in those still behave as ions with localized spins? Lacunar spinels give a very good opportunity to answer this question. Experience shows that in most of them the molecular-orbital picture works better. There are twelve $t_{2g}$ orbitals per tetrahedron. In the molecular-orbital scheme the energy levels with only direct $d$--$d$ hopping would be 6 bonding and 6 antibonding states (for 6 metal--metal bonds in the V$_4$ tetrahedra, see Fig.~\ref{MO-in-clusters}(d)). With inclusion of hoppings via ligands (which should be more important for such ligands as S, Se, Te) they are split into, from below, bonding singlet, doublet and triplet, and two antibonding triplets, see Fig.~\ref{Lacunar-spinel}(b)-(d). In GaV$_4$S$_8$ seven electrons per V$_4$ tetramer would then fill these levels as shown in this figure, with the upper, sevenths electron sitting on a  triplet, giving spin $S_{tot}=1/2$ per tetrahedron. The ordering of these spin-1/2 objects gives magnetic ordering. Similarly, in GeV$_4$S$_8$ with eight electrons per tetramer these electrons would, in the molecular picture, fill these levels, with two electrons on the upper triplet, which according to the Hund's rule should have parallel spins, i.e.\ in this scheme the V$_4$ tetrahedra in GeV$_4$S$_8$ would have spin~1.  And indeed experimentally this seems to be the case~\cite{Singh2014}. The example of GeV$_4$S$_8$ is very important in this respect. In the molecular-orbital picture, as explained above, the V$_4$ tetramers would be magnetic with $S_{tot}=1$. However if the localized picture worked here, then every V$^{3+}$ ($d^2$) would have $S=1$, and four V ions with $S=1$ each in a tetrahedron would have strong antiferromagnetic coupling, so that in effect we would have 
a nonmagnetic ground state with $S_{tot}=0$ for such tetrahedron.  Thus, in contrast to some cases considered before (V$^{3+}$ ($d^2$) triangles in LiVO$_2$, V tetrahedra in GaV$_4$S$_8$) in which at least the type of the ground state, its quantum numbers, would be the same in both pictures, that of molecular orbitals and that of localized electrons ($S_{tot}=0$ for LiVO$_2$; $S_{tot}=\frac12$ for GaV$_4$S$_8$), in GaV$_4$S$_8$ the very type of the ground state would be qualitatively different in these two limits: a magnetic $S_{tot}=1$ state in the molecular-orbital picture and a nonmagnetic $S_{tot}=0$ state in the picture with  localized electrons.

Yet another interesting and very important feature of lacunar spinels is seen from the molecular-orbital level scheme of  Fig.~\ref{Lacunar-spinel}(b)-(d): in both GaV$_4$S$_8$ and GeV$_4$S$_8$ there is an extra orbital degeneracy in the ground state, with one or  two electrons on a triply degenerate level. One should expect in this case the same Jahn--Teller distortion as for isolated  ions with orbital degeneracy, only here on degenerate levels of the whole cluster. And indeed such Jahn--Teller distortion occurs in both GaV$_4$S$_8$ and GeV$_4$S$_8$~\cite{Singh2014,Ruff2015,Hlinka2016}. The resulting level structures are shown in Fig.~\ref{Lacunar-spinel}(c) for GaV$_4$S$_8$ and in Fig.~\ref{Lacunar-spinel}(d) for GeV$_4$S$_8$. It is especially important because in this case it leads to multiferroic behaviour of these systems. The point is that in lacunar spinels, with the symmetry F-43m, there is no inversion symmetry, in contrast to the usual  cubic spinels with the symmetry Fd-3m. But in this state the material is  not ferroelectric: it has four equivalent polar [111] axes.  At the Jahn--Teller distortion this equivalence is broken, one of the [111] axes is ``selected'', and the system becomes polar, i.e.\ ferroelectric. Because of this extra degeneracy lacunar spinels of this type finally become multiferroic. And, moreover, they harbour skyrmions, of a novel type --- N\'eel  skyrmions,  in which spins rotate as in N\'eel domain walls~\cite{Kezsmarki2015,Nikolaev2019}, in contrast to the majority of skyrmions, which are of Bloch type~\cite{Muhlbauer2009}. Thus in this case the interplay of the existence of clusters and of orbital degeneracy leads to very diverse and interesting phenomena.  We stress that in these situations one can encounter very rich quantum effects, see e.g.\ \cite{Savary2016,Yan2020}. 

Thus we see that there can remain a substantial degeneracy in cluster magnets even after formation of molecular orbitals (see also the level scheme in Fig.~\ref{MO-in-clusters}). Often the account of hoppings via ligands also does not lift the degeneracy completely. Therefore, one may expect similar Jahn--Teller distortions in other cluster Mott insulators. In fact the Jahn--Teller effect, i.e.\ vibronic coupling, was shown to be very important for very different objects: isolated clusters of TM ions. In many situations it stabilizes some particular geometry and largely determines physical properties of e.g.\ Ni or Au clusters~\cite{Hakkinen2000,Xie2005}.

\subsection{Novel quantum states in cluster Mott insulators~\label{Sec:PCO}}

Quantum effects are typically connected with the presence of some degeneracy, e.g.\  due to a specific form of the energy surface in the Jahn--Teller physics (Sec.~\ref{sec:JT}), or they appear because of magnetic frustrations, which may result in such extraordinary phenomena as the formation of effective magnetic monopoles~\cite{Castelnovo2008,Bramwell2009,Jaubert2009}, of quantum spin liquids~\cite{Zhou2017,Savary2017,Broholm2020}, and order-by-disorder mechanism of spin ordering~\cite{Villain1980,Bergman2007}. As was shown in the previous section the presence of such  degenerate states is rather natural for cluster magnets. If Jahn--Teller distortions are suppressed (e.g.\ by spin--orbit coupling, see Sec.~\ref{JTplusSOC}, or by other competing mechanisms) or the resulting splitting of electronic bands is small (compared with the corresponding bandwidths), then one might expect an emergence of novel quantum states in this situation. Any frustrations (in the lattice formed by clusters) would support such states.

One of such emergent quantum states was proposed to explain the physical properties of $A_xB_y$Mo$_3$O$_8$ systems~\cite{Chen2016b,Chen2018a}. The $A$ and $B$ ions in these materials are chosen in such a way as to give a state $[A_xB_y]^{5+}$. Then each Mo is 3.6(6)+ and has electronic configuration $4d^{2.3(3)}$ (or we have nominally two Mo$^{4+}$ ($d^2$) and one Mo$^{3+}$ ($d^3$)). These crystals have layered structure. The Mo ions form triangular trimers as shown in Fig.~\ref{Clusters-examples}(d). If we now consider the lattice formed by these trimers, then this turns out to be a triangular lattice of trimers (while the lattice of Mo ions themselves is a breathing kagome lattice). There are 7 electrons per each cluster, 6 of which occupy 3 lowest-lying bonding states, see Fig.~\ref{MO-in-clusters}(d), and the remaining electron resides on the triplet non-bonding $d$ levels. (The better-studied ``undoped'' materials such as Zn$_2$Mo$_3$O$_8$ or Fe$_2$Mo$_3$O$_8$ have only 6 electrons per Mo$_3$ triangle, and the Mo sublattice in those is nonmagnetic. However e.g.\ Fe$_2$Mo$_3$O$_8$ presents significant interest as a polar magnet with extremely strong magnetoelectric response~\cite{Wang2015c,Kurumaji2017,Kurumaji2017a,Solovyev2019}).


\begin{figure}[t]
   \centering
  \includegraphics[width=0.5\textwidth]{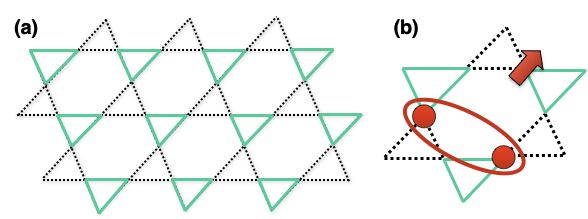}
  \caption{\label{PCO}(a) Kagome lattice formed by the Mo ions in A$_x$B$_y$Mo$_3$O$_8$ materials. See also Fig.~\ref{Clusters-examples}(d). Mo$_3$ trimers with short Mo-Mo bonds are shown by solid (green) lines. (b) The part of the wave function of the so-called plaquette charge order state is sketched. Two out of three electrons form a valence bond, while one electron with $S=1/2$ remains dangling. True many-body plaquette charge order state is constructed from such states taking into account the presence of inter-site Coulomb repulsion between electrons and exploring various possibilities to form valence bonds on the kagome lattice.}
\end{figure} 

The physical properties of materials $A_xB_y$Mo$_3$O$_8$  with one unpaired electron per Mo$_3$ triangle cluster are highly unexpected. LiZn$_2$Mo$_3$O$_8$ never orders magnetically, but there is a change in the slope of the temperature dependence of inverse magnetic susceptibility at $96\,$K instead~\cite{Sheckelton2012}. The effective magnetic moment decreases exactly by a factor of~3 at this temperature, signalling that 2/3 of the free spins turn out to be frozen, i.e.\ insensitive to the external magnetic field. Li$_2$InMo$_3$O$_8$, having the same crystal structure, shows a very different behaviour and becomes long-range ordered (with $120^{\circ}$ structure) at $T_N=12\,$K~\cite{Haraguchi2015}. The materials Li$_2$ScMo$_3$O$_8$ and Li$_2$In$_{1-x}$Sc$_x$Mo$_3$O$_8$ with $x \sim 0.6$ were claimed to be quantum spin liquids~\cite{Haraguchi2015,Akbari-Sharbaf2018,Iida2019}, again with the freezing of 2/3 of spins. ZnScMo$_3$O$_8$ is a ferromagnet (with $T_c=6\,$K~\cite{Sinclair2018}), which is rather atypical for insulating transition metal oxides with localized electrons, see Sec.~\ref{Sec:OO}.
 
Since the splitting between non-bonding and (anti)bonding states in Mo$_3$ trimers is $\sim2$eV~\cite{Nikolaev2020}, the simplest model to describe these surprising physical properties of $A_xB_y$Mo$_3$O$_8$ materials is the three-band model (three non-bonding states of the Mo$_3$ trimer) with a single electron. On-site Hubbard repulsion cannot localize electrons in such a dilute regime (one electron per three sites), and one needs to take into account inter-site repulsion and also the collective electron motion on the whole lattice (i.e.\ also between different trimers). If these non-bonding states are completely degenerate or the corresponding splitting is small, then the ground state is a quantum entangled state --- a so-called plaquette charge order (PCO)~\cite{Chen2016b,Chen2018a,Nikolaev2020}. This is a true many-electron state resembling a valence bond liquid, one  component of which is sketched in Fig.~\ref{PCO}(b). Two out of three electrons that we have per a hexagon (or rather star of David) on a kagome lattice form a spin singlet ($S_{tot}=0$), while the remaining electron still has an uncompensated spin. This explains the freezing of 2/3 of spins at low temperatures. The true many-electron state on the whole lattice is constructed from such states~\cite{Chen2016b,Chen2018a,Nikolaev2020}.

While the mechanisms of PCO formation can be very different,  see \cite{Chen2016b,Chen2018a,Nikolaev2020}, it seems that the splitting of non-bonding bands is the key parameter in this situation~\cite{Nikolaev2020}. Together with correlation effects --- both on-site and intersite Coulomb repulsion --- it suppresses collective charge fluctuations and localizes electrons on Mo$_3$ clusters (in the same way as crystal-field splitting helps Mott localization in the conventional Hubbard model, see e.g.\ \cite{Poteryaev2008}), so that the system can be described by a Heisenberg model with localized spins on a triangular lattice of Mo$_3$ trimers. {\it Ab initio} calculations show that this situation corresponds to Li$_2$ScMo$_3$O$_8$, Li$_2$InMo$_3$O$_8$ and most probably to ZnScMo$_3$O$_8$, while in LiZn$_2$Mo$_3$O$_8$ the PCO state is induced by the much smaller splitting of non-bonding bands~\cite{Nikolaev2020}. These examples show that the presence of degenerate states is rather typical for cluster insulators, and therefore one might expect similar physical effects in other insulating clusterized materials, especially those with small number of electrons (holes) and (partially) suppressed Jahn--Teller effect.


\subsection{Molecular orbitals without ``molecules''}

Typically one expects to see molecular orbitals in materials where some clusters (which one may consider as quasi ``molecules'') exist in a crystal structure or they are spontaneously formed by mechanisms discussed in  Sec.~\ref{Sec:D-reduction}. However, there can be features of the band
structure characteristic for the molecular orbitals even if there is no well-defined clusters (dimers, trimer etc.) in a crystal.
\begin{figure}[t]
   \centering
  \includegraphics[width=0.4\textwidth]{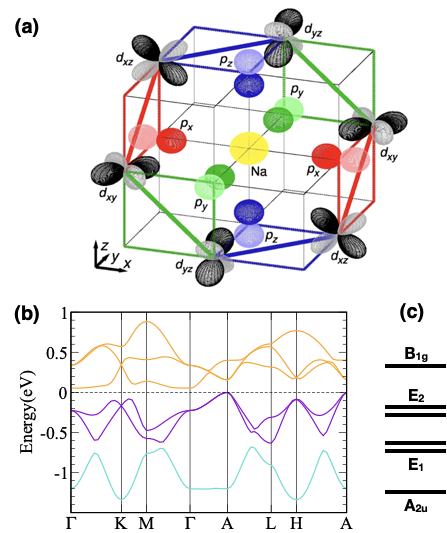}
  \caption{\label{Na2IrO3-MO} (a) Illustrates formation of quasimolecular orbitals in Na$_2$IrO$_3$.  Ir ions are in oxygen octahedra, which form honeycomb lattice. (b) Band structure of another honeycomb materials with similar crystal structure - SrRu$_2$O$_6$ (obtained in the nonmagnetic DFT calculations) and (c) Molecular orbitals in the benzene molecule.}
\end{figure} 

A good example are some materials with honeycomb lattice, such as Na$_2$IrO$_3$ or SrRu$_2$O$_6$.   Some of them are very popular nowadays Kitaev magnets, discussed below  in Sec.~\ref{sec:SOCexchange}. Transition metals in them have octahedral surrounding and these octahedra share edges in forming honeycombs, see  Fig.~\ref{Na2IrO3-MO}(a).

Suppose that the electrons hop between transition metal sites only via ligand $p$ orbitals. Then one can easily see that due to particular signs of $d$ orbitals in this geometry hoppings via ligands are orbital-specific. If we start e.g. with electron at the $xy$ orbital of site one, then it hops to the $xz$ orbital of the next, second site, and from it - to the $yz$ orbital of the third site etc, see Fig.~\ref{Na2IrO3-MO}(a). In effect an electron put on the $xy$ orbital of a leftmost site in Fig.~\ref{Na2IrO3-MO}(a) always remains on a corresponding hexagon, forming a state very similar to that of a benzene molecule, with the same molecular orbital. In fact as in a benzene molecule there are six such molecular orbitals, which in a solid of course form not levels, but bands, see Fig.~\ref{Na2IrO3-MO}(b). Such states, which were called in Ref.~\cite{Mazin2012,Foyevtsova2013} quasimolecular orbitals give an alternative description of Na$_2$IrO$_3$ to the standard one which uses the picture of localised electrons.  As was shown in Ref.~\cite{Streltsov2015a,Pchelkina2016} the same picture may explain unique magnetic properties of another system with hexagonal lattice - SrRu$_2$O$_6$, with a record value of N\'eel temperature of 560~K~\cite{Hiley2014,Hiley2015}. More Ru$^5+$ ion with $d^3$ electronic configuration is more prone to formation of such molecular orbitals~\cite{Streltsov2018b}. Thus even in a system with regular lattice without any visible geometrically-determined clusters, just the particular orbital structure can produce effectively molecular-like states. This is yet another manifestation of specific effects connected with the directional character of orbitals in TM compounds.

Obviously there are many factors, which can spoil formation of such quasimolecular orbitals. First of all, in real materials there is also a nonvanishing second nearest hopping, which puts antibonding states in SrRu$_2$O$_6$ nearly at the same energy range, see Fig.~\ref{Na2IrO3-MO}(b). Second, there are Coulomb correlations in case transition metal compounds, which tend to localize electrons at atomic sites and not on molecular orbitals. Then, quasimolecular orbitals are rather sensitive to local distortions of ML$_6$ octahedra and the spin-orbit coupling may also hinder their formation~\cite{Foyevtsova2013}. Nevertheless this notion can be quite useful in interpreting properties of these and similar materials.

\section{Role of the spin--orbit coupling \label{SOC}}

One more factor is extremely important in transition metal compounds with partially filled $d$-shells --- the real relativistic spin--orbit coupling. The very word ``orbital'' implies that the orbital moment can be an essential  factor --- and with it, the spin--orbit coupling. The orbital moment is $L=2$  for $d$ shells. As was discussed in Sec.~\ref{Sec:single-site-effect}, this moment can be partially quenched by crystal field, if the typical $t_{2g}$--$e_g$ splitting $\Delta_{CF} = 10Dq$ is larger than the strength of the spin--orbit coupling, characterized by $\lambda$ in~\eqref{eq:full-SOC}. This is always the case in $3d$ systems, and actually also in $4d$ and $5d$ ones. The SOC increases in going from $3d$ to $4d$ to $5d$: typical SOC for $3d$ is $\lambda_{3d} \sim 20$--$70\,$meV~\cite{Dunn1961}, for $4d$ we have $\lambda_{4d} \sim 0.1-0.2\,$eV~\cite{Dunn1961}, and $\lambda_{5d}$ may be $\sim0.3$--$0.5\,$eV~\cite{Yuan2017}. But concomitant to that the crystal field splitting also increases:  $10Dq$ for $3d$ ions is usually $\sim1.5$--$2\,$eV, for $4d$ it is 2.5--$3\,$eV and for $5d$ it is often $\gtrsim3\,$eV\null. Thus, in most TM compounds the dominant crystal-field splitting is larger than the SOC\null.  Then one can consider SOC for the $t_{2g}$ electrons separately. As was mentioned in Sec.~\ref{Sec:CFS} the orbital moment and SOC are quenched for the $e_g$ electrons, but SOC is operative for the $t_{2g}$ electrons, which can be described in the first approximation by the effective orbital moment $\tilde l=1$, which we will always use in what follows.  In this sense it is ``lucky'' that the $4d$ and $5d$ systems, for which one can expect a strong influence of SOC, are typically in the low spin state, i.e.\ very often for $4d$ and $5d$ ions we have only $t_{2g}$ levels partially occupied, so that the SOC for them is indeed instrumental and often crucial.

One more general point should be mentioned here. As is well known, there exist in the description of atoms two situations: one when for many-electron atoms or ions the $LS$ (or Russel--Saunders) coupling scheme is valid; and the other with the $jj$ coupling  scheme. The applicability of one or the other description is determined by the ratio between the Hund's rule coupling $J_H$ and the parameter $\lambda$, see Eq.~\eqref{eq:full-SOC}, characterizing the strength of the SOC. When the Hund's interaction dominates, the $LS$ coupling scheme is valid: to satisfy the intra-atomic Hund's interaction one should first form the state with the maximum possible spin $\mathbf S$ and with maximum total orbital moment $\mathbf  L$ consistent with it, and then the weaker SOC leads to the formation of multiplets with the total moment $\mathbf J =\mathbf  S + \mathbf L$. Note that these states are multi-electron states, they cannot be described by a single Slater determinant and therefore generally speaking one can hardly describe this situation by such techniques as e.g.\ DFT~\cite{Kohn1999}.

In the opposite situation with $\lambda>J_H$  the $jj$ coupling scheme is valid. It is essentially a one-electron description. In the $jj$ scheme we first form the state with the total moment $\mathbf  j$ for each electron, $\mathbf  j_i= \mathbf s_i + \mathbf l_i$, and then combine these to form the total moment $\mathbf  J=\sum_i \mathbf  j_i$. By this we gain the maximum energy of the SOC, but we lose a part of the Hund's energy.  (Note that the usual band structure calculation methods such as DFT actually are one-electron approaches, thus they in fact work with the $jj$ coupling scheme, even when they are used for $3d$ materials for which $LS$ scheme is more applicable).  

These two descriptions, $LS$ and $jj$ coupling, are actually just two limiting cases; the real materials may be somewhere in between. Numerical estimates show that for $3d$ systems the $LS$ scheme is applicable: for $3d$ ions the Hund's energy $J_H$ is $\sim0.8$--$1\,$eV~\cite{Vaugier2012}, but the SOC is much smaller (see above in this section). For $4d$ electrons $J_H \sim 0.6$--$0.7\,$eV~\cite{Vaugier2012} --- also larger than the typical values of SOC $\lambda_{4d} \sim 0.1$--$0.2\,\rm eV$~\cite{Dunn1961}. But here the system can already deviate from the pure $LS$ coupling and can acquire some features resembling the $jj$ case. For $5d$ ions $J_H \sim 0.5\,$eV --- actually of the same order as SOC~\cite{Yuan2017}. These systems are evidently not yet in a pure $jj$ coupling limit, but are already half-way to it. And the notions applied in this case can already be used for these materials.
  \begin{figure}[t]
   \centering
  \includegraphics[width=0.5\textwidth]{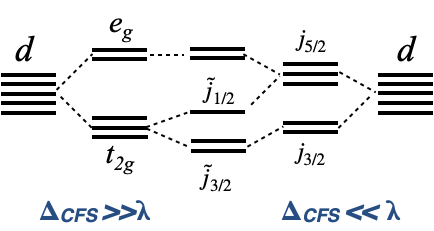}
  \caption{\label{CFS-SOC}One-electron level scheme in different regimes set up by the $t_{2g}$--$e_g$ crystal field splitting $\Delta_{CFS}=10 Dq$ and the strength of spin--orbit coupling given by corresponding constant~$\lambda$. Each level is doubly-degenerate (Kramers degeneracy).}
\end{figure}

Most often the qualitative pictures, the type of the ground state obtained in the $LS$ and $jj$ schemes are similar. The resulting state strongly depends on electron occupation~$n_d$. Let us shortly compare the results of these two approximations for different $n_d$ keeping in mind that only $t_{2g}$ states can be considered.

\begin{figure}[t]
   \centering
  \includegraphics[width=0.4\textwidth]{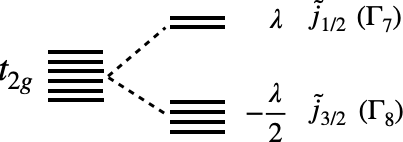}
  \caption{\label{t2g-eff}Splitting of the $t_{2g}$ subshell due to spin--orbit coupling.}
\end{figure}

{\bf $\mathbf {t_{2g}^1}$ configuration}.
In case of one $d$ electron the results of $LS$ and $jj$ descriptions do of course coincide. The spin is in this case $s=\frac12$, the effective orbital moment $\tilde l=1$. We remind that for $t_{2g}$ triplet we use the description in terms of an effective moment $\tilde l=1$. The difference from the real orbital moment is that the sign of the spin--orbit coupling $\lambda$ here is opposite to that of the real moment, so that in effect the second (or the third) Hund's rule changes to its opposite: for less-than-half-filled $t_{2g}$ subshell the order of multiplets is inverted, i.e.\ the lowest is the multiplet with the highest effective total moment~$\tildej$, and for more-that-half-filled case it is normal, and multiplets with smaller~$\tildej$ have lower energies, see e.g.~\cite{khomskii2014transition}. It this case we would get possible $\tildej=s+\tilde l$ equal to $\frac32$ and $\frac12$, the quartet $\frac32$ lying lower, see Fig.~\ref{t2g-eff}. The wavefunctions corresponding to the $\tildej=\frac12$ doublet can be obtained using corresponding Clebsch--Gordan coefficients:
\begin{eqnarray}
|\tildej_{1/2}, \tildej^z_{1/2} \rangle &=& -\frac 1 {\sqrt 3}|\tilde l^z_0, \uparrow \rangle + \sqrt{ \frac  23}\,| \tilde l^z_{1}, \downarrow \rangle  \nonumber \\ 
&=& -\frac 1 {\sqrt 3}\left( |xy, \uparrow \rangle +  {\rm i} | xz, \downarrow \rangle   + | yz, \downarrow \rangle  \right),  \nonumber  \\           
| \tildej_{1/2}, \tildej^z_{-1/2} \rangle  &=& \frac 1 {\sqrt 3}| \tilde l^z_0, \downarrow \rangle - \sqrt{ \frac  23}\,| \tilde l^z_{-1}, \uparrow \rangle  \nonumber \\ 
&=& \frac 1 {\sqrt 3}\left( |xy, \downarrow \rangle + {\rm i} |xz, \uparrow\rangle  - |yz, \uparrow \rangle     \right),  \label{1/2-states}
\end{eqnarray}
and for the $\tildej=\frac32$ quartet we have
\begin{eqnarray}
\label{3/2-states}
|\tildej_{3/2}, \tildej^z_{3/2}\rangle &=& |\tilde l^z_1, \uparrow \rangle = - \frac 1{\sqrt 2} ( {\rm i} |xz, \uparrow \rangle + |yz, \uparrow \rangle),\nonumber \\
|\tildej_{3/2}, \tildej^z_{-3/2}\rangle &=& | \tilde l^z_{-1}, \downarrow \rangle 
=  -\frac 1{\sqrt 2} ({\rm i} |xz, \downarrow \rangle - |yz, \downarrow \rangle  ),
\nonumber \\
|\tildej_{3/2}, \tildej^z_{1/2}\rangle &=& \sqrt{\frac 2 3}\,|\tilde l^z_0, \uparrow \rangle + \frac 1 {\sqrt{3}}|\tilde l^z_{1}, \downarrow \rangle  \nonumber \\ 
&=&  \sqrt{\frac 2 3}\,|xy, \uparrow\rangle - \frac 1 {\sqrt{6}}|{\rm i}xz + yz, \downarrow \rangle, \nonumber \\
|\tildej_{3/2}, \tildej^z_{-1/2}\rangle &=& \sqrt{\frac 23}\,|\tilde l^z_0, \downarrow \rangle + \frac 1 {\sqrt{3}}|\tilde l^z_{-1}, \uparrow \rangle  \nonumber \\ 
\quad \quad &=&  \sqrt{\frac 2 3}\,|xy, \downarrow \rangle - \frac 1 {\sqrt{6}}| {\rm i}xz - yz , \uparrow \rangle. \label{eq:j32}
\end{eqnarray}

In both $LS$ and $jj$ schemes we should put one electrons at a state of this quartet.

{\bf $\mathbf {t^2_{2g}}$ configuration}.  
In the $LS$ scheme the total spin is $S=1$, and $L=1$, so that the possible values of $J$ are $J=2, 1, 0$, with the quintet $J=2$ lying lowest (cubic crystal field further splits this quintet). One would get the similar configuration in the $jj$ scheme:  we now have to put two electrons on the lowest $\tildej=\frac32$ quartet in Fig.~\ref{t2g-eff}. To gain some extra energy from the remaining Hund's rule we can put for very large $\lambda$ these two electrons e.g.\ in the state $| \tildej^z_{3/2}, \tildej^z_{1/2} \rangle$, with the total moment $J^z=2$ (or $| \tildej^z_{-3/2}, \tildej^z_{-1/2}\rangle$ with the moment~$-2$).

{\bf $\mathbf {t^3_{2g}}$ configuration}.
In the $LS$ scheme we would have total spin $S=\frac32$, but the total orbital moment  would be quenched, $L=0$, i.e.\ the total moment would be a pure spin quartet $S=\frac32$. In the $jj$ scheme we have to put three electrons, or one hole  at the lowest one-electron quartet $\tildej=\frac32$ of Fig.~\ref{t2g-eff} --- i.e.\ we would also have a ground state quartet, but not a pure spin quartet with $S=\frac32$, but rather the total moment $J=\frac32$ quartet. 

{\bf $\mathbf {t^4_{2g}}$ configuration}.
This state is especially interesting. In the $LS$ scheme we have for the $t_{2g}^4$ configuration the total spin $S=1$ and total orbital moment $L=1$. From possible multiplets with $J=L+S =\{0, 1, 2\}$ here the singlet $J=0$ would be the ground state (more-than-half-filled case!). I.e.\ this state should be nonmagnetic, and without any orbital degeneracy (present in the absence of SOC). And we would get the same state in the $jj$ scheme: here we have to put 4 electrons on the lowest $\tildej=\frac32$ quartet of Fig.~\ref{t2g-eff}, so that all these states would be filled and the total moment would be zero.  This nonmagnetic state may be realized  in some (at least isolated) $4d^4$ ions, e.g Ru$^{4+}$,  and is very typical for $5d^4$ ions such as  Ir$^{5+}$. Thus Ir$^{5+}$  for example for the electron spin resonance (ESR) community is always considered as a typical  nonmagnetic ion, and it is even used sometimes for nonmagnetic dilution~\cite{Abragam}. Some cases in which Ir$^{5+}$ behaves at least partially as magnetic require special explanation~\cite{Terzic2017,Nag2016,Nag2019}.

\begin{figure}[t]
   \centering
  \includegraphics[width=0.3\textwidth]{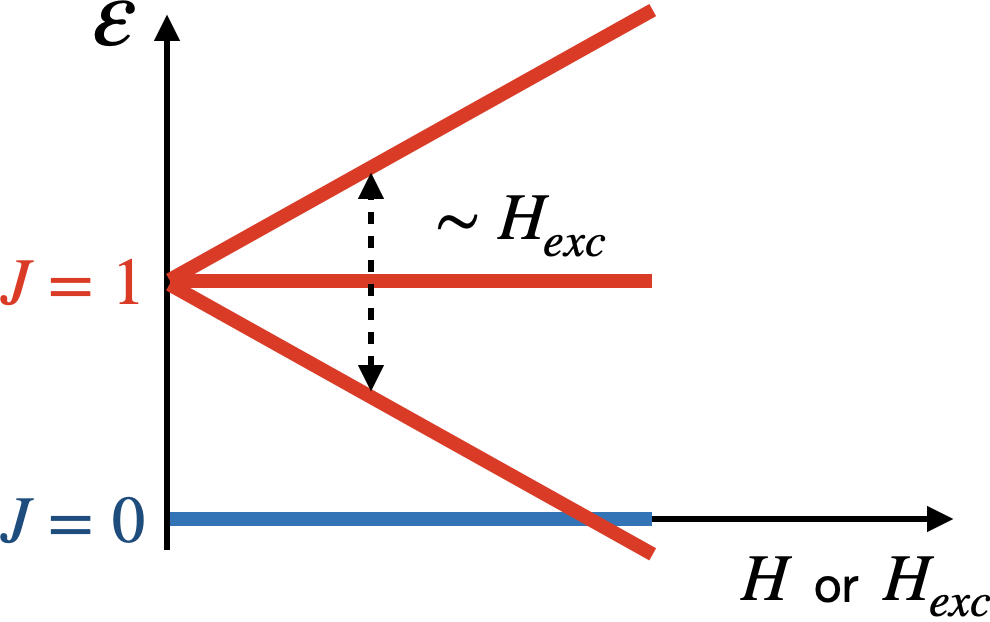}
  \caption{\label{Singlet-magnetism} Schematic illustration of the origin of magnetic state (singlet, or Van Vleck (excitonic) magnetism) in systems with the singlet ground state and triplet excited state. External or internal (exchange), $H_{exc}$, magnetic field leads to Zeeman splitting of the triplet state, and if it is strong enough, the resulting state becomes magnetically-ordered.
}
\end{figure}

The situation in concentrated systems with Ru$^{4+}$ ions is actually more interesting. In most cases Ru$^{4+}$ compounds are either metallic (SrRuO$_3$, Sr$_2$RuO$_4$)  --- in which case one cannot reliably use these notions; or, if they are insulating, they are usually magnetic (Ca$_2$RuO$_4$, etc.). Several factors can lead to this state. One is the possible  non-cubic crystal field (if it is sufficiently strong), which could split the $t_{2g}$ triplet and quench the orbital moment and spin--orbit coupling, and by that ``kill'' the nonmagnetic $J=0$ state. But the intersite interaction is usually more important. In this case we can meet the situation which is known as singlet magnetism, see e.g.\ Sec.~5.5 in~\cite{khomskii2014transition} (it was called excitonic magnetism in \cite{Khaliullin2013}). For $4d$ systems the excited $J=1$ triplet lies not very far from the ground state $J=0$ singlet. The intersite exchange interaction can lead to a Zeeman splitting of the excited triplet, so that if it is strong enough, one of the triplet levels would lie below the $J=0$ singlet, and the system would be magnetic, see Fig.~\ref{Singlet-magnetism}. But this is a special type of magnetism, with soft spins: not only direction, but also the very magnitude, the length of the magnetic moment may change, as it depends on the degree of admixture of the triplet $J=1$ to the singlet $J=0$~\cite{Khaliullin2013}. Consequently the character of spin excitations may be much richer compared to those in usual magnets: these could be not only ordinary spin waves, but also excitations in which the lengths of the spins change; these are actually similar to the famous Higgs excitations in high energy physics. These excitations seem indeed to be observed in Ca$_2$RuO$_4$~\cite{Jain2017}.

{\bf $\mathbf {t^5_{2g}}$ configuration}. The systems with this electronic occupation are especially popular nowadays. To these belong many Ir$^{4+}$ compounds with very intriguing properties, to be considered in the next section. These largely rely on the special type of the ground state of Ir$^{4+}$ ions. In the $LS$ scheme we have here $S=\frac12$ and $L=1$, similar to the case of $d^1$ occupation. But, in contrast to that case, here we have a more-than-half-filled situation, so that the order of multiplets is opposite to that for $d^1$: the ground state is a Kramers doublet $J=\frac12$. Thus in this case, as in the $d^4$ configuration, there remains no extra orbital degeneracy, in contrast to the original situation with $t_{2g}^5$ occupation in the absence of SOC, which would be triply-degenerate. We will get the same state also in the $jj$ scheme: we fill the $\tildej=\frac32$ quartet of Fig.~\ref{CFS-SOC} by four electrons, and the fifth electron goes to the upper-lying (in $jj$ scheme!)\ single-electron $\tildej=\frac12$ doublet. Thus due to strong spin--orbit coupling we actually got rid of orbital degeneracy, and the system becomes similar to that  of a non-degenerate Hubbard model (except the double Kramers degeneracy, which is present both for non-degenerate Hubbard model with $S=\frac12$ per electron and here with $\tildej=\frac12$). Thus this case seems to be conceptually simple. But in many respects the situation here is different and more complicated than in the usual non-degenerate Hubbard model, simply because of the completely different character of the wavefunction, with strong spin-orbit entanglement and anisotropic spin distribution. As we will see in the next two sections, it leads to drastic consequences, in particular to very specific quantum effects in respective systems. 

Below we present several examples of how the spin--orbit coupling can affect various physical properties of transition metal oxides.

\subsection{Electronic properties: Spin--orbit assisted Mott state\label{sec:SOCexchange}}
As was mentioned in Sec.~\ref{sec:el-in-solids}, it is the ratio between the on-site electron--electron repulsion~$U$ and the width of the corresponding band~$W$, which controls whether we are in insulating or metallic regime in strongly correlated materials. Another factor important for the case when we have several correlated bands is the crystal field splitting. It is clear that it leads to orbital polarization (i.e.\ one of the bands becomes more occupied than the other), and it suppresses orbital fluctuations and therefore favours an insulating regime~\cite{Poteryaev2008}. Spin--orbit coupling is also a single-site effect, which provides additional splitting and thus it acts in the same direction, and its contribution sometimes turns out to be the last straw that breaks the camel's back resulting in an insulating state. Such a state, realized e.g.\ in Sr$_2$IrO$_4$, is often called spin--orbit assisted Mott state. 

It was observed that going from Sr$_2$CoO$_4$ to Sr$_2$RhO$_4$ and finally to Sr$_2$IrO$_4$ we have  the same number of $t_{2g}$ electrons (five), but strongly increase the bandwidth, since we increase the principal quantum number and the corresponding extent of the respective wavefunctions. Therefore, going down in the Periodic Table we should go to a more metallic regime. But experimentally the first two compounds are metals~\cite{Matsuno2005,Kim2006}, while Sr$_2$IrO$_4$ is an insulator~\cite{Moon2009}. It was found in~\cite{Kim2008} that neither an account of correlation effects in a mean-field way (via the so-called $\rm DFT+U$ method~\cite{Anisimov1997}) nor the inclusion of the spin--orbit coupling ($\rm DFT+SOC$) alone could  give an insulating solution, but only their combination results in a formation of the band gap. The mechanism is the appearance of an additional splitting (between $\tildej=1/2$ and $\tildej=3/2$ states) in the $\rm DFT+U+SOC$ calculations due to spin--orbit coupling. More elaborate methods such as $\rm DFT+DMFT$ also give an insulating solution in case of Sr$_2$IrO$_4$~\cite{Arita2012,Martins2017}, but one has to keep in mind that the system must be in a critical regime even without spin--orbit coupling to realize this scenario. Thus e.g.\  in the 3D analogue of this material --- SrIrO$_3$ the bandwidth is much larger than in layered Sr$_2$IrO$_4$ and the first one is always metallic~\cite{Moon2008}.

There are two very important aspects when we speak about the spin--orbit assisted Mott state. First of all, the correlation effects result in localization of electrons, and this of course makes spin--orbit coupling more efficient. It was found that e.g.\ in Sr$_2$RhO$_4$ Coulomb correlations effectively enhance spin--orbit coupling  -- the spin-orbit coupling constant is increased by a factor of $\sim2$ with respect to its bare value~\cite{Liu2008}. Second, the critical value $U_c$ for the Mott transition increases with degeneracy  as $U_{c,N} \approx \sqrt{N}\, U_{c,N=1}$ (here $N$ is the orbital degeneracy)~\cite{Gunnarsson1996}, i.e. it decreases upon degeneracy lifting. In Sr$_2$IrO$_4$ the spin--orbit coupling splits $d$ levels and leads to the formation of a non-degenerate $\tildej=1/2$ band, separated from the other bands, and hence the orbital degeneracy and with it the critical $U_c$ strongly decrease (crudely by $\sqrt 3$) as compared with the initial triply-degenerate $t_{2g}$ bands. In effect this system  turns out to lie on the insulating side of the Mott transition, whereas Sr$_2$CoO$_4$ and Sr$_2$RhO$_4$ without such splitting are on the metallic side.

\subsection{Magnetic properties: Anisotropy of the exchange interaction, Kitaev materials.\label{sec:SOCexchange}}
Probably one of the most famous manifestations of the spin--orbit coupling influence on the exchange interaction is exchange anisotropy in the so-called Kitaev materials. We briefly discuss here the main ideas and the results obtained until now in this field, but for a more detailed description we recommend to turn to specialized reviews~\cite{Winter2017,Trebst2017,Hermanns2018,Takagi2019}.
\begin{figure}[t]
   \centering
  \includegraphics[width=0.5\textwidth]{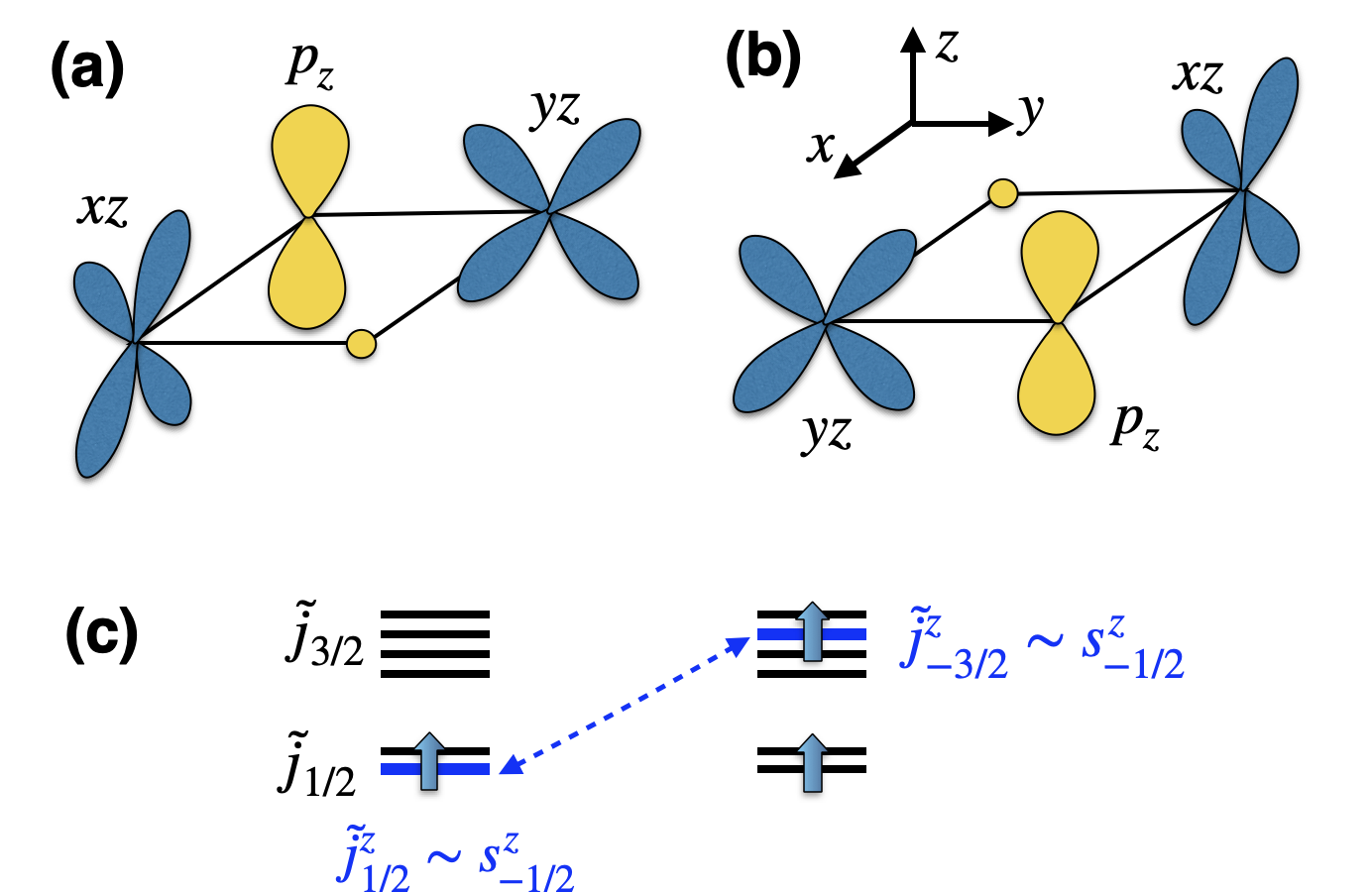}
  \caption{\label{Khaliullin}Illustration of the microscopic mechanism leading to a strongly anisotropic (Ising) exchange for a pair of two octahedra sharing  common edge. Only hopping process via ligand $p_z$ states shown in (a) and (b) are allowed. In the limit of a very strong spin--orbit coupling this leads to a situation when a hole can transfer from $|\tildej_{1/2} \tildej^z_{1/2}\rangle$ only to $|\tildej_{3/2} \tildej^z_{-3/2}\rangle$ state, see~(c). Similar process of transfer from $|\tildej_{1/2} \tildej^z_{-1/2}\rangle$ to $|\tildej_{3/2} \tildej^z_{3/2}\rangle$ state is not shown.}
\end{figure}

Let us start with a simple model system: a pair of transition metals $A$ and $B$ with $t_{2g}^5$ configuration (each) surrounded by octahedra having a common edge, see Fig.~\ref{Different-sharing}(a). We also assume that ($i$)~a direct $d$--$d$ hopping is absent and that ($ii$)~the spin--orbit coupling is large enough so that the ground state of such ions is the $\tildej=1/2$  Kramers doublet, with the excited $\tildej=3/2$ quartet.

In this situation only intersite hoppings via oxygens $t_{xz/yz}=t_{yz/zx}=t$ are non-zero, see Fig.~\ref{Khaliullin}(a,b), and a single hole resides on the $\tildej_{1/2}$ doublet~\eqref{1/2-states}. 
Having these hoppings one could easily calculate hoppings between $\tildej_{1/2}$ states and find the exchange interaction by the second order perturbation theory with respect to Hubbard~$U$ using~\eqref{2t-over-U}. We notice immediately that all hoppings between these $\tildej_{1/2}$ states via ligand $p$ orbitals are simply zero: $\langle j^z_{1/2,A} | \hat t | j^z_{-1/2,B} \rangle = \langle \tildej^z_{-1/2,A} | \hat t |\tilde  j^z_{1/2,B} \rangle= 0$ because an electron cannot change its spin during hopping from site to site, while $ \langle \tildej^z_{1/2,A} | \hat t | \tildej^z_{1/2,B} \rangle= \langle \tildej^z_{-1/2,A} | \hat t | \tildej^z_{-1/2,B} \rangle = {\rm i}t -{\rm i}t =0 $, due to the specific form of the $\tildej_{1/2}$ orbitals, given in~\eqref{1/2-states}. (In other words, due to the presence of the imaginary coefficient ``${\rm i}$'' in the expression \eqref{1/2-states} the hoppings from one $d$ site to another via two common oxygens exactly cancel). This fact, that the quantum interference suppresses conventional antiferromagnetic superexchange between the $\tildej_{1/2}$ orbitals in 90$^{\circ}$ geometry (this is not the case for  180$^{\circ}$ metal--oxygen--metal bonds), was first noticed by Jackeli and Khaliullin~\cite{Jackeli2009,Khaliullin2005}. 

Thus, as was explained in Sec.~\ref{Sec:OO}, one needs to take into account higher order terms determined by hoppings between half-filled $\tildej_{1/2}$ and empty (in hole representation) $\tildej_{3/2}$ orbitals (via ligand $p$ states). Similarly to \eqref{FMexchange} this contribution is ferromagnetic,
\begin{eqnarray}
\label{Kitaev-constant}
K \sim - \frac{t^2 J_H} {U^2} = - \frac{t_{pd}^4 J_H} {\Delta_{CT}^2 U^2},
\end{eqnarray}
where the second expression stresses that all hoppings occur via ligand $p$ states. Interestingly, most of the terms in the resulting exchange Hamiltonian (containing different components of the moment $\tildej=1/2$) are again suppressed by the symmetry of the $d$ orbitals; the only remaining component is the Ising interaction containing $\tildej^z$ components, where the local $z$ axis is perpendicular to the plane of the TM--O$_2$--TM plaquette, see Fig.~\ref{Khaliullin}.  

Using explicit form of the wave functions of $\tildej=1/2$ doublet and $\tildej=3/2$ quartet - Eqs. \eqref{1/2-states} and \eqref{3/2-states} - one can show that only virtual hoppings between $|\tildej_{1/2}, \tildej^z_{1/2}\rangle$ and $|\tildej_{3/2}, \tildej^z_{-3/2}\rangle$ (or states with reversed $\tilde j^z$) are nonzero. In effect the electron  (or hole)  transferred in such a way should return to its own site and hence there would be no real exchange of electrons, and, as explained  in Sec.~\ref{Sec:OO} after Eq.~\eqref{2t-over-U}, we would have only Ising-type interaction $\sim K j^z j^z$  (ferromagnetic because of the Hund's interaction in the intermediate state as it has been explained above), with the  $z$ axis perpendicular to the particular plaquette Ir$_2$O$_2$, see \cite{Jackeli2009,Matsuura2014}.

By itself this situation --- Ising ferromagnetism --- does not represent anything particularly interesting. However, if one could connect these pairs of edge-shared octahedra in a network with different common edges, i.e.\ with different orientations of Ir$_2$O$_2$ plaquettes, then each Ir pair would tend to orient spins (i.e.\ total moments) along its own local $z$ axis, and one would obtain strong magnetic frustrations.
\begin{figure}[t]
   \centering
  \includegraphics[width=0.5\textwidth]{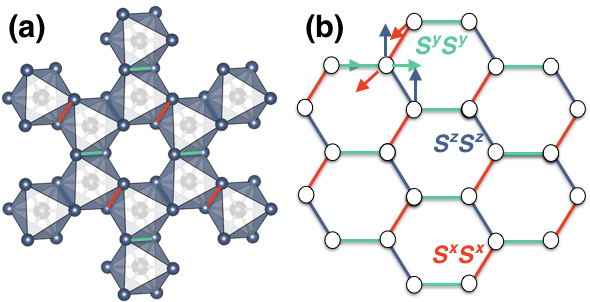}
  \caption{\label{Kitaev-lattice}(a)~Crystal structure of $\alpha$-RuCl$_3$ (and Na$_2$IrO$_3$). Every Cl$_6$ (O$_6$) octahedron surrounding Ru (Ir) ions shares three {\it different} edges with its neighbours. Thus every Ru--Ru bond has its own local exchange anisotropy axis. (b)~Spin lattice corresponding to the Kitaev model. There are three exchange bonds: $K \hat S_i^x \hat S_j^x$, $K \hat S_i^y \hat S_j^y$, and $K \hat S_i^z \hat S_j^z$, denoted by different colours.}
\end{figure}

Such materials do indeed exist. Jackeli and Khaliullin in their seminal paper \cite{Jackeli2009} noted that e.g.\ in Na$_2$IrO$_3$ the IrO$_6$ octahedra form a 2D honeycomb lattice with alternating common edges, see Fig.~\ref{Kitaev-lattice}(a), with Ir$_2$O$_2$ plaquettes in three orthogonal directions, the normals of which can be taken as $x$, $y$, and~$z$.  Ir here is 4+ ($t_{2g}^5$), and strong spin--orbit coupling localizes a single hole on the $\tildej=1/2$ orbital. Therefore one may expect that the spin subsystem can be described by the Hamiltonian~\cite{Jackeli2009}:
\begin{eqnarray}
\label{Kitaev}
H = \sum_{\langle ij \rangle_{\gamma}} K^{\gamma} \hat S^{\gamma}_i \hat S^{\gamma}_j,
\end{eqnarray}
where for different types of bonds $\gamma = \{x,y,z\}$ is different. This model is now called the Kitaev model. It has very much in common with the compass model \eqref{compass-model} discussed in Sec.~\ref{Sec:OO}: the interaction at each bond is of Ising type, but with different components of spin or pseudospin on different bonds. Kitaev has shown that this model for $S=1/2$ on the honeycomb lattice, see Fig.~\ref{Kitaev-lattice}(b), is exactly solvable~\cite{Kitaev2006}. The ground state of this model is a quantum spin liquid~\cite{Kitaev2006}. One can easily notice that even in a classical version of~\eqref{Kitaev}, where the spin operators are replaced by vectors, the ground state is highly degenerate~\cite{Baskaran2008}. The spin at any site of the honeycomb lattice does not know what to do: it is connected by three bonds (shown by different colors in Fig.~\ref{Kitaev-lattice}(b)) and each of them tends to orient the spin in its own way, but we are able to satisfy only one of these bonds. Therefore we can propose many spin configurations as candidates for the ground state, and all of them would have exactly the same total energy. Quantum effects result in tunnelling between these states giving rise to extremely rich physics, which can be described by Majorana fermions~\cite{Kitaev2006}. It is interesting to note that a very similar model (but with pseudospins instead of real spins) was introduced much earlier in \cite{KK-UFN}, see also \cite{Nussinov2015}, in the context of orbital ordering; it was called there the ``compass'' model. 
\begin{figure}[t]
   \centering
  \includegraphics[width=0.5\textwidth]{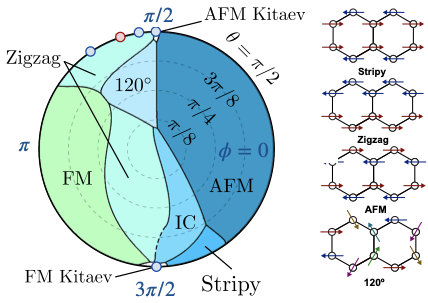}
  \caption{\label{Kitaev-PD}Phase diagram of the generalized Kitaev model \eqref{Kitaev-general} according to \cite{Winter2017a} (see also \cite{Rau2014,Katukuri2014,Winter2016,Yadav2016}), where coupling constants are parametrized via two angles $J=\cos \phi \sin \theta$, $K=\sin \phi \sin \theta$, and $\Gamma=\cos \theta$. $IC$ stands for incommensurate spiral order. Figures are reproduced from~\cite{Winter2017a}.}
\end{figure}

In fact in addition to Na$_2$IrO$_3$ there has been tested a number of compounds based on the Ir$^{4+}$, Ru$^{3+}$ and Rh$^{4+}$ ions as possible physical realizations of the Kitaev model. However, in real materials there is always direct $d$--$d$ exchange, distortions of ML$_6$ octahedra, and mixing with higher energy $e_g$ states. Because of that the interaction \eqref{Kitaev} turns out to be too simplified and one needs to use a more general expression for the exchange Hamiltonian, with (at least) isotropic ($J$) and anisotropic symmetric ($\Gamma$) terms:
\begin{eqnarray}
\label{Kitaev-general}
H = \sum_{\langle ij \rangle} K \hat S^{\gamma}_i \hat S^{\gamma}_j
+ J \hat {\mathbf S}_i \cdot \hat {\mathbf S}_j
+ \Gamma \left( \hat S_i^{\alpha} \hat S_j^{\beta} + \hat S_i^{\beta} \hat S_j^{\alpha}\right),
\end{eqnarray}
where $\alpha$ and $\beta$ index in-plane components, i.e.\ $x$ and $y$, if $\gamma = z$ for a given bond, but also with longer-range exchange.

 Analytic expressions for all these parameters can be found in~\cite{Rau2014,Winter2016}, while the phase diagram of \eqref{Kitaev-general} is shown in Fig.~\ref{Kitaev-PD}. One can see that Kitaev physics is realized in narrow regions at the very top and bottom of the phase diagram. This is the reason why most of the known until now materials listed in Table~\ref{Tab:Kitaev} show a long-range magnetic order instead of the quantum spin liquid ground state. However one can notice, first, rather large absolute values of the Curie--Weiss temperatures $\theta_{\rm CW}$: they are typically much larger than the N\'eel temperatures. This is the evidence of strong magnetic frustrations, which in a bipartite honeycomb lattice with isotropic (nearest neighbour) interactions would not be present. Second, there is typically a strong anisotropy of the magnetic susceptibility, which suggests bond-anisotropic character of the exchange interaction. All this speaks in favour of importance of  the Kitaev interaction in these materials.

Such a large field as the study of Kitaev materials of course cannot be discussed in details in this short section. But a few more aspects should be mentioned. First of all, 3D versions of the Kitaev model exist and, moreover, since they retain local geometry of the Kitaev model (and therefore retain strong frustration), their ground state in a pure form should also be  quantum spin liquids~\cite{Mandal2009}. Two modifications of Li$_2$IrO$_3$ with the so-called hyperhoneycomb, $\beta$-Li$_2$IrO$_3$, and stripy, or harmonic honeycomb,  $\gamma$-Li$_2$IrO$_3$, lattices were synthesized~\cite{Takayama2015,Modic2014}, but both have a spiral long range magnetic order~\cite{Takayama2015,Modic2014,Biffin2014,Ruiz2017,PhysRevLett.113.197201}. Second, while most of the Kitaev materials order at low temperature, one may suppress this order by a magnetic field, e.g.\ in the case of $\alpha$-RuCl$_3$ $B_c=7$--$8\,$T directed in the $ab$ plane is enough for this~\cite{Johnson2015}. Very unusual magnetic excitations characteristic for Kitaev physics were observed in this regime~\cite{Zheng2017,Wang2017,Baek2017,Kasahara2018,Wellm2018,Banerjee2018}.  Third, the only known Kitaev material which neither orders nor shows transition to a spin-glass state is H$_3$LiIr$_2$O$_6$, and Raman measurements seem to support the formation of a quantum spin-liquid state~\cite{pei2019}. Theoretical calculations and dielectric spectroscopy show that Kitaev exchange dominates in this material due to static, or more likely dynamic hydrogen disorder~\cite{Li2018,Wang2018a,Geirhos2020}. Quantum effects such as zero-point motion of protons are of course very important in this case~\cite{Li2018,Wang2018a}. The position of hydrogen strongly affects superexchange interaction, because there is a strong hybridization of H $1s$ and O $2p$ states, and fluctuations of protons in this case would lead the system to a spin-disordered state.
 \begin{figure}[t]
   \centering
  \includegraphics[width=0.4\textwidth]{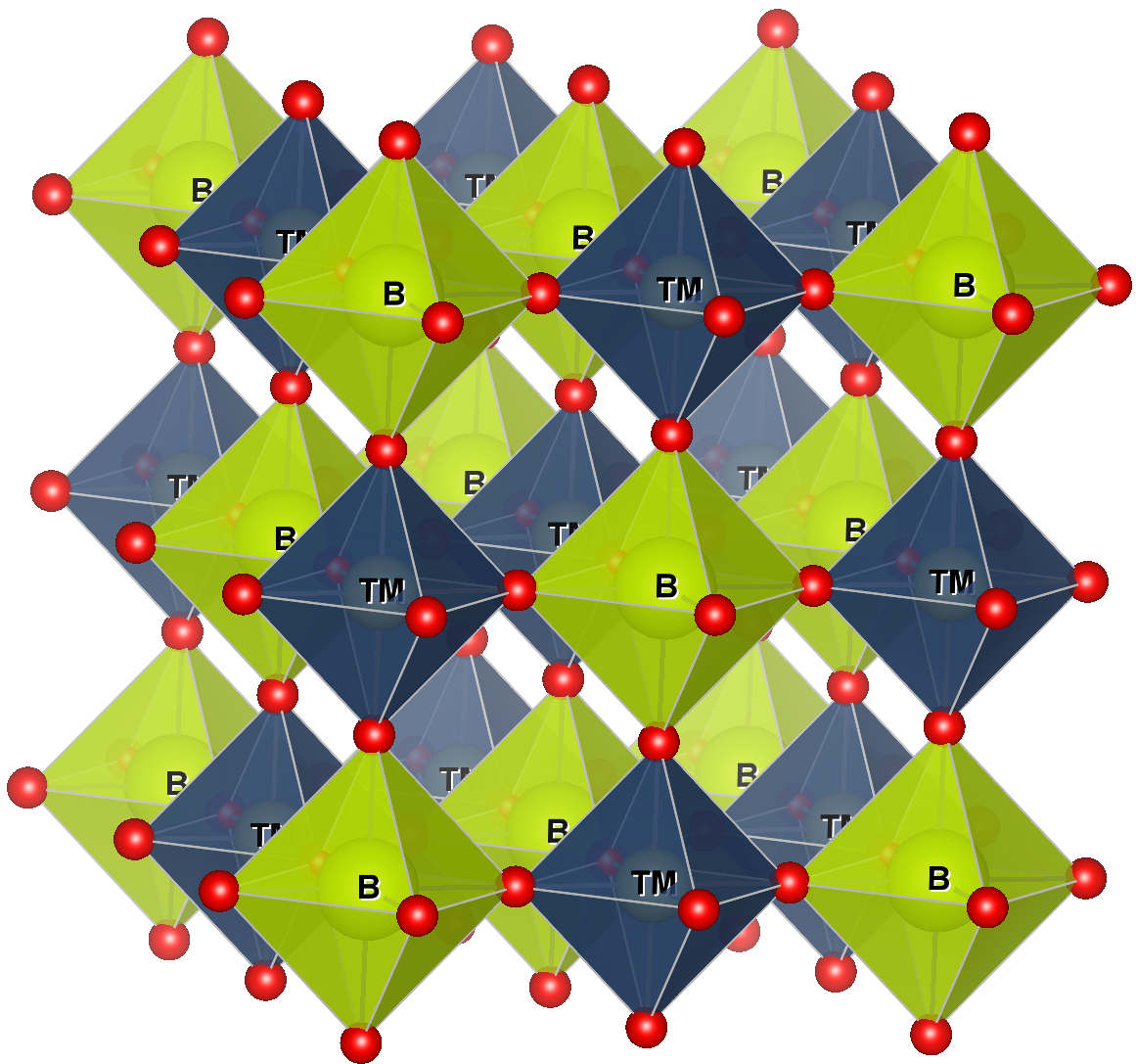}
  \caption{\label{Double-perovskite}Crystal structure of double perovskites $A_2B\,$(TM)$\,$O$_6$, where TM is a transition metal. Oxygen ions are shown by red balls, $A$ ions are not shown.}
\end{figure}
\begin{table*}[t]
\centering \caption{\label{Tab:Kitaev}Magnetic properties of Kitaev materials. N\'eel ($T_N$) and Curie--Weiss ($\theta_{\rm CW}$) temperatures are given in~K\null. If in-plane ($ab$) and out-of-plane ($c$) components of magnetic susceptibility were measured the corresponding values of $\theta_{\rm CW}$ are presented. Different types of magnetic order are shown in Fig.~\ref{Kitaev-PD}.}
\vspace{0.2cm}
\begin{tabular}{llcccccc}
\hline
\hline
Material & Mag. order  & $T_N$ & $\theta_{\rm CW}$                      \\
\hline
Na$_2$IrO$_3$               & Zigzag\cite{Liu2011,Ye2012}   & 13-18\cite{Singh2010,Liu2011,Ye2012}& $-176_{ab}$,$-40_c$\cite{Winter2017}  \\
$\alpha$-Li$_2$IrO$_3$ & Spiral\cite{Williams2016}              & 15\cite{Singh2012} & $5_{ab}$,$-250_c$\cite{Winter2017}     \\
$\alpha$-RuCl$_3$        & Zigzag\cite{Johnson2015,Sears2015}       & 8--14\cite{Johnson2015,Sears2015} & $40_{ab}$, $-216_c$\cite{Lampen-Kelley2018}\\ 
Li$_2$RhO$_3$             & Spin-glass\cite{Luo2013,Khuntia2017} & --       & -50\cite{Luo2013}                          \\
Cu$_2$IrO$_3$             & Spin-glass\cite{Abramchuk2017,Choi2019}  & 3\cite{Abramchuk2017}      & $-110$\cite{Abramchuk2017}                        \\
H$_3$LiIr$_2$O$_6$ & Spin-liquid?\cite{Kitagawa2018a}   & --       & $-105$\cite{Kitagawa2018a}                        \\
\hline
\hline
\end{tabular}
\end{table*}

Finally, at the end of this subsection we would like to stress that the strong anisotropy of exchange interaction due to a suppression of isotropic Heisenberg exchange is not a specific feature of $t_{2g}^5$ configuration only (i.e.\ Ir$^{4+}$, Rh$^{4+}$ or Ru$^{3+}$ ions). There exists a number of different combinations of geometries and electron fillings, for which one can observe very similar effects. Thus, the situation with the high spin Co$^{2+}$ ions having octahedral coordination and forming 2D honeycomb layers (octahedra again share their edges) looks promising for that. It has been known for a long time that Co$^{2+}$ in octahedra typically has unquenched orbital moment, which leads e.g.\ to a strong  magnetic anisotropy in such materials as CoO~\cite{Shull1951,Goodenough,Jo1998}, CoNb$_2$O$_6$~\cite{Coldea2010}, Ca$_3$Co$_2$O$_6$~\cite{Wu2005} and many others. The electronic configuration of Co$^{2+}$ is $t_{2g}^5e_g^2$, with the spin $S=\frac32$ and orbital moment $\tilde l=1$, which for this more-than-half-filled case  results in the formation of the lowest doublet with pseudospin~$\frac12$, similar in many respects to the $\tildej=\frac12$ state of Ir$^{4+}$~\cite{Abragam}. But, in contrast to Ir systems, the presence of both partially-filled $t_{2g}$ and $e_g$ levels leads here, in addition to the superexchange interaction between partially-filled $t_{2g}$ orbitals discussed above, to  the $e_g$--$e_g$ and $t_{2g}$--$e_g$ exchange channels. The first one occurs via two orthogonal $p$ orbitals and must be ferromagnetic, see Fig.~\ref{GKA-90} and discussions in Sec.~\ref{Sec:OO}, while the second one is antiferromagnetic. It was shown in Ref.~\cite{Liu2020} that their contributions to the isotropic Heisenberg exchange in case of perfect CoO$_6$ octahedra nearly compensate each other. Moreover, the $\Gamma$-term in \eqref{Kitaev-general} also vanishes, and we are left  only with the Kitaev interaction~\cite{Liu2020}. This idealized situation can be spoiled, however, by trigonal crystal splitting, which is inherent in layered systems, and by long-range exchange coupling (as the iridates story has taught us, the most important is the isotropic exchange between third nearest neighbours in honeycomb lattice~\cite{Winter2016,Yamaji2014}). There are many cobaltates having layered honeycomb structure e.g. Na$_2$Co$_2$TeO$_6$\cite{Viciu2007}, (Ag,Na,Li)$_3$Co$_2$SbO$_6$\cite{Viciu2007,Zvereva2016,Stratan2019}, CoTiO$_3$~\cite{Newnham1964,Yuan2020}, BaCo$_2$((As,P)O$_4$)$_2$\cite{Nair2018,Zhong2020}, which are under intense investigation now because of possible realization of Kitaev physics.

Another interesting situation is the $t_{2g}^1$ configuration on a honeycomb lattice in the presence of strong spin--orbit coupling. The exchange interaction turns out be again bond-dependent, and the final exchange Hamiltonian has very high SU(4) symmetry~\cite{Yamada2018,Natori2018,Yamada2019}. However, in real materials there is always a possibility for dimerization, which results in a trivial non-magnetic ground state~\cite{Ushakov2020}. Anisotropic bond-dependent exchange can also be expected in other lattices, e.g.\ in double perovskites with the general formula $A_2B\,$(TM)$\,$O$_6$, with the crystal structure shown in Fig.~\ref{Double-perovskite}.  Here the transition metal ions form FCC lattice and the nearest neighbour exchange could go only via the $p$ orbitals of two ligands. Moreover, these orbitals again have $90^{\circ}$ geometry as in the common edge situation. One can show that in the limit of very strong spin-orbit coupling the exchange interaction between a pair of metals lying in the $xy$ ($yz$, $zx$) plane will have Kitaev-like form with $z$ ($x$, $y$) components of spins coupled and now this coupling will be {\it antiferromagnetic}~\cite{Chen2010a,Revelli2019a}. Finally, one might expect bond-depending exchange not only for $d$, but also for $f$ systems, such as Li$_2$PrO$_3$ with honeycomb lattice~\cite{Jang2019b} or rare-earth double perovskites with general formula Ba$_2$LnSbO$_6$, where Ln is a rare-earth ion~\cite{Li2017b}.

\subsection{Lattice: Jahn--Teller effect and spin--orbit coupling\label{JTplusSOC}}
As has been mentioned in Sec.~\ref{sec:SOCexchange} and illustrated in Fig.~\ref{CFS-SOC}, the spin--orbit coupling does not split $e_g$, but  strongly affects $t_{2g}$ states. Thus, for the $t_{2g}$ orbitals one may expect an interplay between the spin--orbit coupling and the Jahn--Teller effect, which would also split $t_{2g}$ states, but in its own way. This fact was noticed long ago~\cite{Opik1957a}. The result of such an interplay depends on the number of electrons on the TM ion, and it can also  be affected by many-particle effects. In what follows we will consider the mutual interplay of the Jahn--Teller effect and the spin--orbit coupling, on the example of TM ions with partially filled $t_{2g}$ orbitals in octahedral coordination with the  vibronic coupling with tetragonal and orthorhombic distortions $E=\{Q_2,Q_3\}$, i.e. the $t \otimes E$ problem.
\begin{figure}[b]
   \centering
  \includegraphics[width=0.5\textwidth]{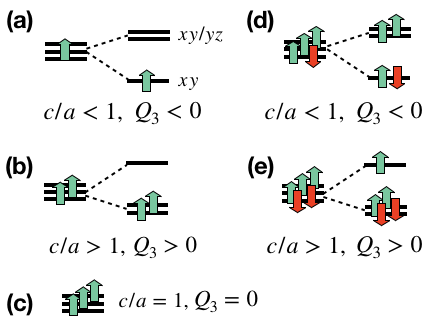}
  \caption{\label{JTt2g}Jahn--Teller effect for different number of $t_{2g}$ electrons. For the sake of simplicity only the tetragonal $Q_3$ mode is considered. $Q_3 > 0$ corresponds to elongation, and $Q_3 <0$ to contraction along one of the axes of metal-ligand octahedron, see Fig.~\ref{JT-Q}.}
\end{figure}

\begin{figure}[t]
   \centering
  \includegraphics[width=0.5\textwidth]{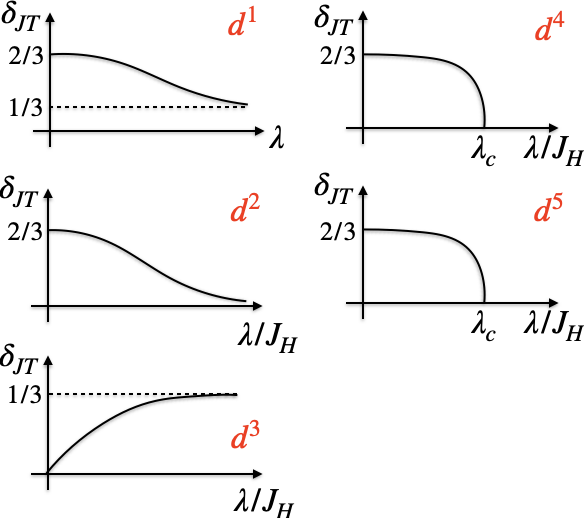}
  \caption{\label{SOCJT}Sketch illustrating how spin--orbit coupling changes the amplitude of the Jahn--Teller distortion, $\delta_{JT} = \sqrt{Q^2_2+Q^2_3}$ (measured in units of $g/B$, see \eqref{JTeg}), in case of different number of $t_{2g}$ electrons. The $t \otimes E$ problem is considered, $J_H$ is the intra-atomic Hund's exchange. See \cite{StreltsovPRX} for details.}
\end{figure}

In Fig.~\ref{JTt2g} we show how the Jahn--Teller effect would split $t_{2g}$ states in the simplest case of tetragonal distortions; for the definition of $Q_3$ see Fig.~\ref{JT-Q}(a,b). The results depend on the number of electrons on the ion, but in any case the Jahn--Teller effect tends to stabilize electrons on states with {\it real} wavefunctions (cubic harmonics): $xy$, $yz$, and~$zx$. Spin--orbit coupling would do this the other way round --- putting electrons on {\it complex} combinations, see Eqs.~\eqref{1/2-states} and~\eqref{3/2-states}. While these two mechanisms typically compete,  in certain cases they may even help each other. The results of many-particle calculations taking into account not only spin--orbit and vibronic interactions but also other terms, such as the intra-atomic Hund's exchange $J_H$, are schematically presented in Fig.~\ref{SOCJT}.

A very interesting situation develops in the case of the $t_{2g}^1$ configuration. While the Jahn--Teller effect would put an electron on the $xy$ orbital and compress the ML$_6$ octahedra ($Q_3<0$), see Fig.~\ref{JTt2g}(a), spin--orbit coupling wants to stabilize it on one of the $\tildej_{3/2}$ orbitals, which are very different from real cubic harmonics and are of course not optimal for the Jahn--Teller effect. Therefore increasing the strength of the spin--orbit coupling (via the spin--orbit coupling constant, $\lambda$, defined in \eqref{eq:full-SOC}) we suppress Jahn--Teller distortions as shown in Fig.~\ref{SOCJT}. In the limit of very large $\lambda$ the spin--orbit coupling finally wins, but there is still an orbital degeneracy in the $\tildej_{3/2}$ quartet (not one, but {\it two} Kramers doublets!), and vibronic interactions can be operative. I.e.\ in this case the Jahn--Teller effect still survives, albeit weakened, even for very strong spin--orbit coupling. It is interesting that in this limit it does not matter whether we put the electron on $\tildej^z_{\pm 3/2}$  (which would correspond to the local compression of ML$_6$ octahedra), $\tildej^z_{\pm 1/2}$ (local elongation) or on their linear combination --- in all cases we gain the same energy due to the Jahn--Teller effect, and the energy surface as a function of distortions has the form of the Mexican hat (note that only linear vibronic coupling \eqref{tau-Q} is considered here and the admixture of higher lying $e_g$ states is not taken into account)~\cite{StreltsovPRX}. 

\begin{figure}[t!]
   \centering
  \includegraphics[width=0.5\textwidth]{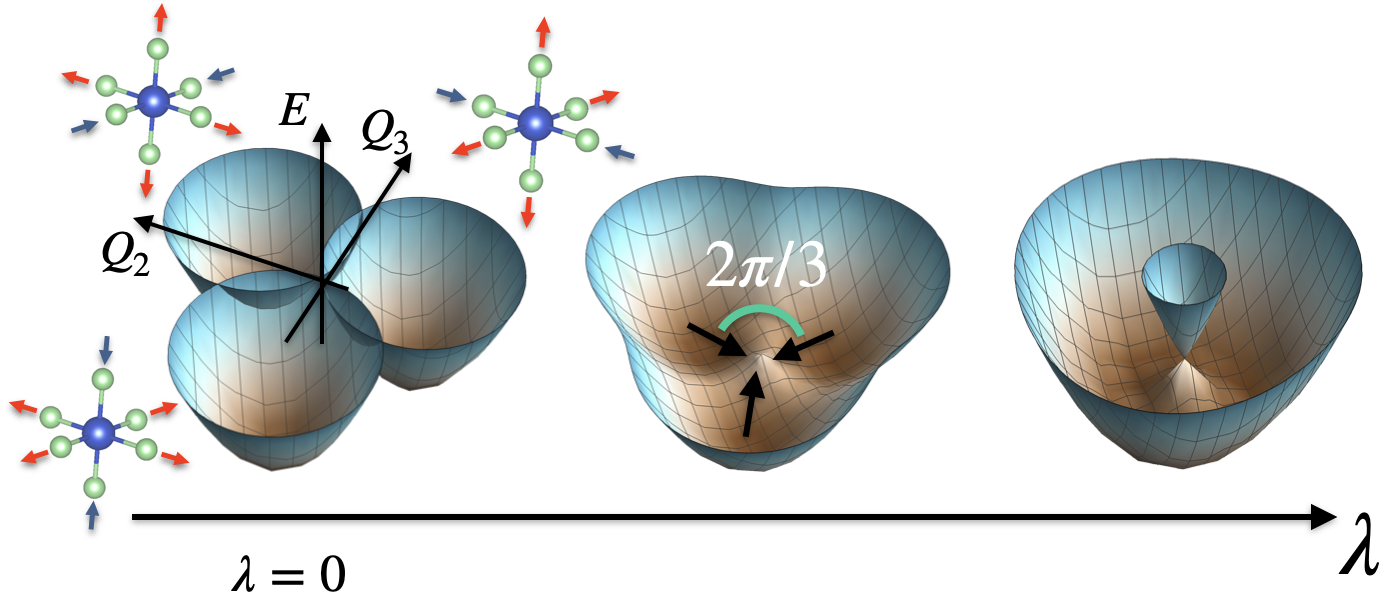}
  \caption{\label{t2g1SOC}Evolution of the energy surface for $d^1$ configuration as a function of $Q_2$ and $Q_3$ distortions with increasing strength of spin--orbit coupling (which is still assumed to be smaller than $10Dq$).}
\end{figure}

The transformation of the adiabatic potential energy surface with three minima (corresponding to octahedra compressed along three perpendicular directions) to the Mexican hat with the increase of SOC is shown in Fig.~\ref{t2g1SOC}. Thus, we see that strong spin--orbit coupling makes the situation with a single electron in the $t_{2g}$ shell qualitatively similar to that of the $e_g$ case without SOC\null. (Indeed, in this case we have one electron on a $\tildej_{3/2}$ quartet, i.e.\  besides Kramers degeneracy we have here an extra double degeneracy --- the same as for $e_g$ electrons). This result allows to explain recent experimental observations of elongated geometry in double perovskite K$_2$TaCl$_6$~\cite{Ishikawa2019}. It is nearly impossible to explain such type of distortion from the point of view of conventional Jahn--Teller effect, but if spin--orbit coupling makes all tetragonal distortions nearly equivalent, then anharmonic terms (which are expected to be rather strong for the relatively small K ions, compared with larger Rb, Cs) could stabilize elongated geometry, as they usually do in the $e_g$ case \cite{KhomskiiBrink2000}, see Sec.~\ref{sec:JT}. (Rb$_2$TaCl$_6$ and  Cs$_2$TaCl$_6$ have compressed octahedra~\cite{Ishikawa2019}.)

The situation with the $t_{2g}^2$ configuration is in some sense similar to the $t_{2g}^1$ case: for relatively small $\lambda$  the spin--orbit coupling suppresses Jahn--Teller distortions. But in contrast to the $d^1$ case, for two electrons the SOC gradually suppresses Jahn--Teller distortion down to zero due to the intra-atomic Hund's exchange coupling, which is always present in real systems. The reason is that for strong SOC and $jj$ coupling $J_H$ puts two electrons on two different orbital doublets, e.g.\ one on $\tildej^z_{3/2}$ and another one on $\tildej^z_{1/2}$ (or one on $\tildej^z_{-3/2}$ and another one on $\tildej^z_{-1/2}$), which have opposite Jahn--Teller distortions.

The case of $t_{2g}^3$ configuration is especially interesting and surprising. In this configuration spin--orbit coupling qualitatively changes the situation. This configuration in itself is Jahn--Teller inactive --- all three $t_{2g}$ orbitals are half-filled (no orbital degeneracy), see Fig.~\ref{JTt2g}(c). But spin--orbit coupling induces the splitting of the $t_{2g}$ shell into a $\tildej_{3/2}$ quartet and a $\tildej_{1/2}$ doublet, with the quartet lying lower. Three electrons now go to this $\tildej_{3/2}$ quartet, and we again have orbital degeneracy: as compared with the $d^1$ case we have here not one electron but one hole on the $\tildej_{3/2}$ quartet. Thus,  in this case the spin--orbit coupling {\it activates} or {\it induces} the Jahn--Teller effect, as shown in Fig.~\ref{SOCJT}.

The $t_{2g}^4$ and $t_{2g}^5$ configurations are similar in  some sense. Both are Jahn--Teller active at $\lambda =0 $, see Fig.~\ref{JTt2g}(d,e), but in both even a moderate spin--orbit coupling may completely suppress distortions. Moreover, this transition (from distorted to undistorted geometry) is rather drastic: for $\lambda > \frac 13 \frac{g^2} B$ the Jahn--Teller distortions vanish completely in an almost abrupt way. This is the reason why most iridates do not show any Jahn--Teller distortions. (Here we do not speak about other types of distortions; thus in perovskite-type compounds there may exist a GdFeO$_3$-distortion due to rotation and tilting of ML$_6$ octahedra, which is usually accompanied by certain distortion of octahedra themselves. Or the face sharing geometry by itself usually implies some distortions of ML$_6$ octahedra.) Another interesting example is the Cu$^{2+}$ ion in tetrahedral surrounding. There are also five $t_{2g}$ electrons in this situation, as in Ir$^{4+}$,  and this may explain the absence of Jahn--Teller distortions in the spinel CuAl$_2$O$_4$, where Cu occupies tetrahedral $A$ sites~\cite{Nirmala2017,Nikolaev2018a,Kim2019} (but one should also be aware of a rather large degree of disorder between Cu and Al in this system~\cite{Neill2005,Nirmala2017,Agzamova2020}, which may conceal Jahn-Teller distortions).

The mechanism behind the suppression of the Jahn--Teller effect in $d^4$ and $d^5$ systems is in fact very simple. For $t_{2g}^5$ and for very large $\lambda$ we must put a single hole at the $\tildej_{1/2}$ doublet, which does not have orbital degeneracy (only the Kramers degeneracy due to spin). The same is true for  $t_{2g}^4$, where four electrons completely fill the $\tildej_{3/2}$ quartet, leading to a singlet state $J=0$ without any degeneracy.

In real materials in addition to $E=\{ Q_2, Q_3\}$ distortions discussed above one needs to consider trigonal distortions $T=\{Q_3,Q_4,Q_5\}$ shown in Fig.~\ref{JT-Q}f. Indeed, it was recently demonstrated that e.g. in Ba$_2$YMoO$_6$ vibronic coupling constants for $T$ modes are just slightly smaller than for $E$ vibrations~\cite{Iwahara2018}. In case of trigonal models one needs to use 
 \begin{eqnarray}
\hat H_{JT}^{T} &=& -g \Big( (\hat l_y \hat l_z+\hat l_z \hat l_y)Q_4 + (\hat l_x \hat l_z+ \hat l_z \hat l_x)Q_5 \nonumber \\
 &+& (\hat l_y \hat l_x+\hat l_x \hat l_y) Q_6 \Big) + \frac {B^2}2 \left( Q_4^2 +Q_5^2 +Q_6^2\right),
\label{HJTT}
\end{eqnarray}
instead of \eqref{tau-Q} (here $\hat l_x, \hat l_y$, and $\hat l_z$ are effective orbital moment operators for $t_{2g}$ electrons), but qualitatively the results presented in Fig.~\ref{SOCJT} do not change. This can be easily rationalized. The most natural choice of electronic wave functions in case of trigonal distortions is the basis of the trigonal $a_{1g}$ and $e_g^{\pi}$ orbitals. Spin-orbitals can be expressed in terms of these orbitals in the same way as it was done for pure cubic harmonics in Eqs.~\eqref{1/2-states} and ~\eqref{eq:j32} with $a_{1g}$ playing role of $xy$ and $e_g^{\pi}$ -- of $xz/yz$ orbitals~\cite{Khomskii2016}. Then all the discussion presented above for the $t \otimes E$ problem can be repeated for the $t \otimes T$ situation.

One more remark is in order here. We stressed above  that the $\tildej=\frac32$ quartet is not one, but two Kramers doublets, i.e.\ similar to the case of $e_g$ electrons considered in Sec.~\ref{sec:JT}  it has an extra double degeneracy. This is actually the reason of Jahn--Teller activity of some of these states. It was noticed in Ref.~\cite{Jackeli2009a,Huang2014} that in this case one of these doublets has very peculiar  properties resembling somewhat those of $e_g$ electrons, discussed in Sec.~\ref{sec:JT}.  The Kramers doublet $|\tildej_{3/2}, \tildej^z_{\pm3/2}\rangle$  can be described by the effective spin $\sigma=\frac12$, but  different components of this effective spin have different meaning.  They are all time-reversal odd, i.e.\ are magnetic,  but have different transformation properties for some spatial transformations.  Thus $\sigma_z$ and $\sigma_x$ behave as usual magnetic dipoles, but $\sigma_y$ describes not magnetic dipoles but magnetic octupoles, similar to $\tau_y$ for the $e_g$ case.  This is easily seen when we use not the effective spin $\sigma=\frac12$ description, but go to the original description in terms of the operators of the total moment~$\hat J$. The states $|\tildej_{3/2}, \tildej^z_{\pm3/2}\rangle$ are eigenstates of $\hat J^z$, but in contrast to the usual spins $\frac12$ there is no off-diagonal matrix elements of $\hat J$, $\langle\tildej^z_{3/2}|\hat J^+|\tildej^z_{-3/2} \rangle=0$. Only the product of three operators $\hat J$ have such nonzero marix elements, e.g.\ $\langle \tildej^z_{-3/2} |(\hat J^+ \hat J^+ \hat J^+ - \hat J^- \hat J^- \hat J^-)| \tildej^z_{3/2} \rangle \neq 0$. But these products of three $\hat J$ operators correspond to magnetic octupoles. By projecting them to the effective $\sigma=\frac12$ operators describing this doublet one can indeed see that its $\sigma_y$ component describes magnetic octupoles.  This situation with the dipole--octupole character of such spin--orbit entangled states is most often met in some rare earth compounds, where it leads to very nontrivial effects \cite{Lhotel2015,Gaudet2019,Sibille2020}.  But in principle this can also be met in TM compounds such as e.g.\ defect elpasolites~\cite{Ishikawa2019,Hirai2020a}, for which also quadrupolar degrees of freedom --- actually our Jahn--Teller  ordering --- can exist.

Finally, it is worth mentioning that in this section only an interplay between  spin--orbit coupling and vibronic interactions was considered. However, we have seen in Sec.~\ref{Sec:OO} that there is also a pure exchange (Kugel--Khomskii) mechanism, which couples spin and orbital degrees of freedom and which may affect lattice distortions. Therefore, one might expect a third player --- spin --- in this intricate game, in which the final decision of the distortion is made. 

\section{Instead of conclusions: old and novel manifestations of quantum effects in orbital physics~\label{Sec:conclusions}}
Concluding this review we want once again to discuss possible interesting quantum effects which one encounters when dealing with orbitals in TM compounds. Some of them have been discussed in corresponding sections above; here we want not only to reiterate these points, paying main attention to the already existing and studied effects, but also discuss possible new directions in this field.

On a single-site level quantum effects are indeed well known and very strong; they constitute the large field of vibronic physics, very important and often crucial for molecules and magnetic impurities in solids. These effects arise due to nonadiabatic nature of single-site Jahn--Teller problem, where different electronic states  $|\psi_i \rangle$ are coupled to respective different states describing positions of nuclei $|\phi_i \rangle$, 
\begin{eqnarray}
|\Psi_i \rangle = |\psi_i\rangle |\phi_i\rangle.
\end{eqnarray}
Here we must treat both electronic and nuclear states quantum-mechanically. This leads to a number of nontrivial quantum effects in the ground state, as well as in excitation spectra, in thermodynamic and in dynamic properties of corresponding systems, such as  optical excitation spectrum, in the dynamics of some chemical reactions etc., see e.g.~\cite{Bersuker1989}. The most transparent visual manifestation of these effects is the appearance of the potential energy surface of the Mexican hat type, Fig.~\ref{JT-Mexican-hat}(a), with its highly degenerate (in the lowest approximation) ground state --- the trough in Fig.~\ref{JT-Mexican-hat}(a), with the conical intersection and the connected to it appearance of a geometric (Berry) phase, etc. This is an intrinsically quantum situation.

When going to concentrated systems, the situation may change significantly. At first glance one could also expect here very strong quantum effects. For example the mathematical description of systems with double ($e_g$) orbital degeneracy (e.g.\ in compounds of Cu$^{2+}$ or Mn$^{3+}$ ions with the configurations, respectively, $d^9=t_{2g}^6 e_g^3$ and $d^4= t_{2g}^3 e_g^1$) is very similar to that of magnetic systems with spin $S=1/2$, see Sec.~\ref{Sec:OO} above. Thus it looks as though we may have here the same strong quantum effects that are typical for $S=1/2$ magnets, especially in low-dimensional and frustrated systems. And indeed such suggestions were made many times in the literature \cite{Khaliullin2000,Ulrich2003,Saitoh2001}.
However there is a big problem, a hidden stone in these reasonings. Indeed a formally purely electronic, orbital system could develop similar states and display similar quantum effects as the corresponding spin systems, but there is one important difference between them. Whereas spins live more or less on their own and are often rather weakly coupled to other subsystems in the respective materials, notably to the lattice, the orbitals, in contrast, are charge degrees of freedom and as such have inherently strong coupling to the lattice. But the lattice in bulk crystals often behaves (quasi)classically: because the ions are much heavier, the lattice is ``slow'', and the usual adiabatic approximation is typically applicable, at least in insulating compounds. Indeed the inclusion of the lattice can strongly suppress the usual ``spin-like'' quantum effects. It can lead to long-range interaction between orbitals via elastic forces~\cite{Khomskii2001,Khomskii2003}, to the formation of orbital-glass or Jahn--Teller glass state in systems with random Jahn--Teller impurities~\cite{Ivanov1978}. It also gives an alternative interpretation\cite{Skoulatos2015}  of  neutron scattering results of \cite{Ulrich2003}, and a different explanation~\cite{Gruninger2002} of the orbital to the orbiton excitations claimed to be observed in~\cite{Saitoh2001}.

Thus typically one expects that the conventional quantum effects such as those for $S=1/2$ magnets can be strongly suppressed in orbital systems  by the electron (orbital)--lattice interaction. One can still hope to see such effects in some special situations --- e.g.\ for $t_{2g}$ systems, especially in the presence of strong spin--orbit interaction,  but these are rather exceptional situations.

On the other hand the question arises whether there may exist in concentrated systems quantum effects connected with vibronic effects for isolated Jahn--Teller centers. Such effects were probably first shortly discussed by Englman~\cite{englman1972} and were sometimes invoked for interpreting some experiments, e.g.\ the absence of Jahn--Teller distortion in Ba$_3$CuSb$_2$O$_9$~\cite{Katayama9305} --- although this interpretation is now questioned by new spectroscopic data~\cite{Takubo2020}. In any case, this is a very interesting but still quite open field --- can one, or should one combine the vibronic physics with the physics of cooperative Jahn--Teller systems, and if so --- in which cases, and what could be the consequences of that.

Another very important implication of orbital physics is that the directional character of orbitals often results in reduction of dimensionality, see Sec.~\ref{Sec:D-reduction}.  It is well known that quantum effects are typically much stronger in systems with low dimensionality -- especially in one-dimensional ones. Different transitions such as, for example, Peierls transition are basically of quantum nature. But if in these situations such transitions do not occur, we may have instead for example the formation of frustrated systems with their unusual states such as different types of spin liquids. Such effects can be especially enhanced by strong spin-orbit coupling, very important mainly for $4d$ and $5d$ systems. Kitaev spin-liquid state with the Majorana excitations is probably the most spectacular example of strong quantum effects dominating the behaviour of the systems with orbital freedom and strong spin-orbit interaction.  And even in cases when such Peierls-like transitions do occur and lead to the formation of e.g. breathing kagome and pyrochlore lattices with ``molecules'' in solids, some new quantum effects such as e.g. plaquette charge ordering may occur~\cite{Chen2016b,Chen2018a}, see Sec.~\ref{Sec:PCO}. 

Spin-orbit coupling and the resulting states with the strong spin-orbital entanglement is in general a very powerful source of different quantum effects. The extensive discussion of these effects goes far beyond the scope of the present review but some of examples of their  manifestations we still saw above. Kitaev physics, just mentioned,  is one such example. In Sec.~\ref{JTplusSOC} we also saw that spin-orbit coupling can ``activate'' quantum effects in some systems with Jahn-Teller effect, leading e.g. to the appearance of the energy surface of Mexican hat type, with its conical intersections, geometric (Berry) phase etc. 


Yet another example of quantum states in orbital physics is the appearance in some situations of rich quantum models, such as for example the Kugel-Khomskii model with not one but two coupled ``spins''  (spin $S$ and pseudospin $\tau$). In some cases it can even lead to intrinsically very quantum models: the compass model \eqref{compass-model}, the Kitaev model \eqref{Kitaev}, or the very symmetric SU(4) model \eqref{SU4} - the models  which, as we saw above, can be generated in many realistic situations.  And though in all these cases we again  have to worry about the role of the interaction with the lattice, which has a tendency  to suppress some of these quantum state and make system more classical, still all these interesting situations may have good potentials to reveal interesting quantum effects.

As was found in particular in \cite{Kitagawa2018a} an extra source of quantum effects can be the interaction of electron system with some extra degrees of freedom - in the case of H$_3$LiIrO$_6$ the protons introduced in the Kitaev-like material. Apparently  such substitution not only brings about some disorder in the Ir subsystem which can suppress long-range magnetic ordering and by that facilitate realization of the quantum spin-liquid state, but  most probably the  quantum nature of protons themselves, with their large zero-point vibrations, also plays important role. In general one can see that the coupling of different degrees of freedom  generically  enhances quantum effects. This we saw on many examples: vibronic effects in coupled orbital-lattice systems; coupling of spin and orbital degrees of freedom for example in Kugel-Khomskii systems;  coupling to extra quantum objects like hydrogen introduced in the material. In our review we mostly discussed insulating materials and  we practicality did not touch real electron motion existing for example in metallic systems.  But these of course can also bring about strong quantum effects. In many metallic systems orbital degrees of freedom also play very important role. This is a big special field requiring separate treatment; we only want to stress here that by thinking about quantum effects in orbital systems one has not to forget the eventual role of conducting electrons which can lead to completely new phenomena. 

Thus concluding this review we can say that the orbital physics, though not completely new subject, is still a very active field producing new and new surprises, and in particular it can lead to many novel and interesting quantum effects.

\tocless\section{Acknowledgements}
We thank for support the Deutsche Forschungsgemeinschaft (project number 277146847-CRC 1238), the Russian Foundation for Basic Researches (grant RFBR 20-32-70019) and the Russian Ministry of Science and High Education (program ``Quantum'' No. AAAA-A18-118020190095-4 and contract 02.A03.21.0006). The preparation of section ~\ref{SOC} was supported by the Russian Science Foundation (Project 20-62-46047).

\tocless\section{Author Contributions}
The manuscript was written with contributions of both authors. Both authors have given approval to the final version of the manuscript.

\tocless\section{Notes}
The authors declare no competing financial interest.

\tocless\section{Biographies}
 
Daniel Khomskii graduated from the Moscow University in 1962. From 1965, he worked in the Theoretical Department of the Lebedev Physical Institute of the Russian Academy of Sciences in Moscow. There, he defended his Ph.D. in 1969. In 1980 he obtained a second doctoral (Dr.)\ degree --- the Russian equivalent of the German {\it Habilitation} or a professorship in the US\null. From 1992 to 2003 he was a Professor at Groningen University in the Netherlands, and since 2003 he is a guest Professor in K\"oln (Cologne University) in Germany. His main research interests are the theory of systems with strongly correlated electrons, metal--insulator transitions, magnetism, orbital ordering (``Kugel--Khomskii model'') and superconductivity. In 2008 he was elected a Fellow of the American Physical Society.  Over the course of his career he has published more than 400 papers.

Sergey Streltsov graduated from the Ural Federal University in 2003 and subsequently obtained Ph.D. degree  in 2005 and then Dr.\ degree in condensed matter physics in 2015 at the Institute of Metal Physics, Ekaterinburg (Russia). In 2016 he was elected a Professor of the Russian Academy of Sciences (RAS) and in 2019 a Corresponding Member of the RAS\null. He is the Head of the laboratory of the ``Theory of Low-dimensional Spin Systems''  at the Institute of Metal Physics. His research interests encompass a range of topics in condensed matter physics with emphasis on theoretical study of interplay between orbital, charge, spin and lattice degrees of freedom and computational methods in application to solid state physics.

\tocless\section{References}
\bibliographystyle{achemso-my}
\bibliography{library}

\end{document}